\documentclass[%
 reprint,
amsmath,amssymb,
aps,
pra,
floatfix
]{revtex4-2}

\usepackage{graphicx}  
\usepackage{mathtools}                                          
\usepackage{amsthm,amssymb}                                     
\usepackage[colorlinks=true, allcolors=blue]{hyperref}          
\usepackage{cleveref}                                           
\usepackage{braket}                                             
\usepackage{dsfont}                                             
\usepackage{tikz}                                               
\usepackage{subcaption}                                         

\usetikzlibrary{positioning, calc, decorations.pathmorphing, decorations.pathreplacing, shapes.misc}


\newtheorem{theorem}{Theorem}

\newtheorem{corollary}[theorem]{Corollary}

\newtheorem{lemma}[theorem]{Lemma}

\begin{document}

\preprint{APS/123-QED}

\title{Quantum Wave Simulation with Sources and Loss Functions}

\author{Cyrill Bösch}
\thanks{These two authors contributed equally}
\email{cb7454@cs.princeton.edu}
\affiliation{Institute of Geophysics, ETH Zurich, 8092 Zurich, Switzerland}
\affiliation{Department of Computer Science, Princeton University, Princeton, NJ 08540, USA }
\author{Malte Schade}
\thanks{These two authors contributed equally}
\affiliation{Institute of Geophysics, ETH Zurich, 8092 Zurich, Switzerland}
\author{Giacomo Aloisi}
\affiliation{Institute of Geophysics, ETH Zurich, 8092 Zurich, Switzerland}
\author{Scott D. Keating}
\affiliation{Institute of Geophysics, ETH Zurich, 8092 Zurich, Switzerland}
\author{Andreas Fichtner}
\affiliation{Institute of Geophysics, ETH Zurich, 8092 Zurich, Switzerland}

\begin{abstract}

We present a quantum algorithmic framework for simulating linear, anti-Hermitian (lossless) wave equations in heterogeneous, anisotropic, and time-independent media. This framework encompasses a broad class of wave equations, including the acoustic wave equation, Maxwell’s equations and the elastic wave equation. 
Our formulation is compatible with standard numerical discretization schemes and allows for the efficient implementation of multiple practically relevant time- and space-dependent sources. Furthermore, we demonstrate that subspace energies can be extracted and wave fields compared through an $l_2$ loss function, achieving optimal precision scaling with the number of samples taken.
Additionally, we introduce techniques for incorporating boundary conditions and linear constraints that preserve the anti-Hermitian nature of the equations. Leveraging the Hamiltonian simulation algorithm, our framework achieves a quartic speed-up over classical solvers in 3D simulations, under conditions of sufficiently global measurements and compactly supported sources and initial conditions. 
This quartic speed-up is optimal for time-domain solutions, as the Hamiltonian of the discretized wave equations has local couplings.
In summary, our framework provides a versatile approach for simulating wave equations on quantum computers, offering substantial speed-ups over state-of-the-art classical methods.

\end{abstract}

\maketitle

\section{Introduction}\label{S:Introduction}

Wave-based inverse problems are prevalent in disciplines such as seismology \citep{fichtner2010full, cesca2015full, igel2017computational, arrowsmith2022big}, medical imaging \citep{pratt2007sound, gemmeke20073d, sanches2012ultrasound}, non-destructive testing \citep{naffa2002detection, kim2016non, lopez2018non} and metamaterial research \citep{bendsoe2013topology, christiansen2021inverse, sharma2022gradient, bordiga2024automated}. 
However, these fields are fundamentally limited by the current state of computational resources due to the excessive computational cost of the numerical wave simulation. 
In particular, resolution and uncertainty quantification in imaging problems remain heavily restricted by computational capacities \citep{yilmaz2001seismic, fichtner2011resolution, rawlinson2014seismic,betancourt2017conceptual, sen2017transdimensional, fichtner2019hamiltonian,ulrich2022analyzing}.

Quantum computers (QCs) offer promising runtime improvements for numerous computational problems \citep{nielsen2010quantum, montanaro2016quantum, cerezo2021variational, PRXQuantum.2.040203}. 
One of the most promising groups of quantum algorithms is Hamiltonian Simulation (HS).
It provides a framework for simulating quantum systems, i.e., the evolution of the Schrödinger equation, with up to exponential runtime advantages over classical solvers \citep{childs2012hamiltonian, low2017optimal, low2019hamiltonian, low2023complexity}.
A quantum state represented by $n$ two-state quantum particles or qubits has a state space dimension of $N = 2^n$.
Classical algorithms for simulating the evolution of a quantum state under a quantum Hamiltonian $\mathbf{H} \in \mathbb{C}^{N \times N}$ scale at least linearly with the state size $N$.
In contrast, under specific conditions utilizing HS algorithms, a QC is capable of performing quantum simulations with logarithmic scaling, thereby achieving an exponential speed-up compared to all classical computers. 
This can be formalized as:

\begin{theorem}[Optimal sparse Hamiltonian simulation, adapted \cite{low2017optimal} - Theorem 3] \label{TH: complexity}
A $d$-sparse Hamiltonian $\mathbf{H}$ acting on $n$ qubits with matrix elements specified to $m$ bits of precision can be simulated for time-interval $t$, $l_2$-error $\epsilon$, and success probability at least $1-2\epsilon$ with $\mathcal{O}(td||\mathbf{H}||_{\text{max}} + \frac{\log(1/\epsilon)}{\log\:\log(1/\epsilon)})$ queries to an oracle $\mathcal{H}$ that provides a description of $\mathbf{H}$ and a factor $\mathcal{O}(n+m\:\text{polylog}(m))$ additional quantum gates.
\end{theorem}

The apparent question is: Can HS algorithms be used to simulate classical wave equations with the same runtime advantages? 
A growing body of research answers this question affirmatively. 
These studies examine source-free wave equations in homogeneous media \citep{costa2019quantum, suau2021practical} or wave equations in media with varying parameters \citep{koukoutsis2023, sato2024hamiltonian,  sato2024quantum, schade2024quantum}. 
Furthermore, recent research has investigated solutions to mass-spring models \citep{babbush2023exponential, ito2023map} that yield exponential runtime advantages in certain settings, with particular types of source terms considered in \citep{danz2024calculating}, or wave simulation problems addressed through warped phase transformations \citep{jin2023quantum, jin2022quantum, ma2024schr}. 

While many existing studies tend to focus on simulation tasks, our research expands the scope to address the complete end-to-end stack by incorporating a wide range of practically relevant source scenarios and loss function measurements. 
We thoroughly analyze the conditions necessary to preserve quantum speedup throughout the entire algorithmic stack. 
Our primary result is to demonstrate that for many practically relevant sources that are either compact or rotationally symmetric, and when the measured quantities of the final state have sufficiently global characteristics relative to the system size, the proposed algorithmic framework achieves a quartic speed-up in the three-dimensional case compared to all classical wave equation solvers.

The rest of the paper is organized as follows.
In \Cref{s: General wave eq}, we demonstrate that for lossless wave equations, including the elastic, acoustic, and electromagnetic case, a natural quantum encoding can be employed, which enables a direct measurement of the wave field energy and $l_2$-norms.
We then analyze the runtime-advantage for time-domain solutions of wave equations (\Cref{S: Runtime advantage}).
In \Cref{S: Norms and energies} we introducing algorithms to measure $l_2$-related norms and subspace energies for extracting information from the simulated wave field (\Cref{S: Norms and energies}).
Then we propose a method for the implementation of sources on a QC (\Cref{s: sources}).
Finally, we analyze the limitations of our framework (\Cref{s: limitations}) and draw conclusions (\Cref{s: Conclusions}).


\section{A Natural Quantum Encoding for Linear, Anisotropic, Heterogeneous but Lossless Wave Equations} \label{s: General wave eq}

We consider continuous wave equations of the form
\begin{equation} \label{eq: general wave equation}
    \begin{aligned}
          \frac{d \hat{\mathbf{w}}(\mathbf{x}, t)}{dt} &= \hat{\mathbf{C}}(\mathbf{x})\hat{\mathbf{w}}(\mathbf{x}, t) + \hat{\mathbf{s}}(\mathbf{x}, t),\\
          \hat{\mathbf{w}}(\mathbf{x}, 0) &= \hat{\mathbf{w}}_0(\mathbf{x}),\\
          \hat{\mathbf{P}}_{\partial \Omega}\hat{\mathbf{w}}(\mathbf{x}, t) &= \mathbf{0}, \forall \mathbf{x} \in \partial\Omega,
    \end{aligned}
\end{equation}
where hats ( $\hat{}$ ) denote continuous fields and operators over $D$-dimensional space, $\mathbf{x} \in \mathbb{R}^D$, and time, $t$. 
The wave field $\hat{\mathbf{w}}$ is an element of a Hilbert space $\mathcal{H} = \{f: \mathbb{R}^D \rightarrow \mathbb{R}^C \}$ containing at least once-differentiable (in both space and time) and square-integrable $C$-dimensional vector fields over the domain $\Omega$, and $\hat{\mathbf{s}}$ is the source term.
For concreteness and without loss of generality, we have chosen Dirichlet boundary conditions (BC) on the domain boundary $\partial \Omega$, where $\hat{\mathbf{P}}_{\partial \Omega} \in \mathcal{B}(\mathcal{H})$ is a diagonal projector, which selects the elements of the wave field on which we impose homogeneous Dirichlet BC.

The wave operator $\hat{\mathbf{C}}$ considered here is assumed to have the following form
\begin{equation}
    \hat{\mathbf{C}} := \hat{\mathbf{B}}^{-1}(\mathbf{x})\hat{\mathbf{A}}(\mathbf{x}), 
\end{equation}
where $\hat{\mathbf{B}}: \mathcal{H} \rightarrow \mathcal{H}$ is any Hermitian (self-adjoint) and positive-definite operator, and $\hat{\mathbf{A}}: \mathcal{H} \rightarrow \mathcal{H}$ is any anti-Hermitian (anti-self-adjoint) operator with respect to the inner product
\begin{equation} \label{eq: quantum inner product}
    (\hat{\mathbf{w}} | \hat{\mathbf{w}}')_{\mathds{1}}(t) := \int_\Omega \braket{\hat{\mathbf{w}}(\mathbf{x}, t) | \hat{\mathbf{w}}'(\mathbf{x}, t)} \, d\mathbf{x},
\end{equation}
where $\braket{\hat{\mathbf{w}} | \hat{\mathbf{w}}'} = \sum_{l=1}^L \hat{w}_l^* \hat{w}'_l$, ($*$) denotes complex conjugation, and $\hat{\mathbf{w}},\hat{\mathbf{w}}' \in \mathcal{H}$ are any two vector fields.
Here, $\hat{\mathbf{A}}$ incorporates various types of spatial derivatives such as divergence, gradient or curl operations.
The operator $\hat{\mathbf{C}}$ is therefore anti-Hermitian with respect to the inner product
\begin{equation} \label{eq: B inner product}
    (\hat{\mathbf{w}} | \hat{\mathbf{w}}')_{\hat{\mathbf{B}}}(t) := \int_\Omega \braket{\hat{\mathbf{w}}(\mathbf{x}, t) | \hat{\mathbf{B}}(\mathbf{x}) | \hat{\mathbf{w}}'(\mathbf{x}, t)} \, d\mathbf{x},
\end{equation}
where
\begin{equation}
    \begin{aligned}
        &\braket{\hat{\mathbf{w}}(\mathbf{x}, t) | \hat{\mathbf{B}}(\mathbf{x}) | \hat{\mathbf{w}}'(\mathbf{x}, t)} \\ 
        &= \sum_{l=1}^L \sum_{m=1}^L \hat{w}_l^*(\mathbf{x}, t) \hat{B}_{l,m}(\mathbf{x}) \hat{w}'_m(\mathbf{x}, t).
    \end{aligned}
\end{equation}
As we will discuss shortly, the inner product in Eq. \eqref{eq: B inner product} gives rise to a notion of energy that can be estimated on a QC.
The anti-Hermiticity of $\hat{\mathbf{C}}$ with respect to $(\cdot | \cdot)_{\hat{\mathbf{B}}}$ is formalized by showing that
\begin{equation}
    (\hat{\mathbf{w}} | \hat{\mathbf{C}} \hat{\mathbf{w}}')_{\hat{\mathbf{B}}} = -(\hat{\mathbf{C}} \hat{\mathbf{w}} | \hat{\mathbf{w}}')_{\hat{\mathbf{B}}},
\end{equation}
using integration by parts and invoking the BC \citep{johnson2007notes, schade2024quantum}. 
Quantum states are naturally normalized with respect to a norm induced by the inner product \eqref{eq: quantum inner product}. 
Therefore, mapping Eq. \eqref{eq: general wave equation} to a Schrödinger equation amounts to transforming $\hat{\mathbf{C}} \rightarrow \hat{\mathbf{C}}_{Q}$, such that $\hat{\mathbf{C}}_{Q}$ is anti-Hermitian with respect to $(\cdot | \cdot)_{\mathds{1}}$ \cite{mostafazadeh2010pseudo, koukoutsis2023}.
Because $\hat{\mathbf{B}}$ is Hermitian and positive-definite, it has a unique square root $\hat{\mathbf{B}}^{1/2}$, which is also Hermitian and positive-definite. 
Transforming Eq. \eqref{eq: general wave equation} with $\hat{\mathbf{w}}_{Q} := \hat{\mathbf{B}}^{1/2} \hat{\mathbf{w}}$ and $\hat{\mathbf{C}}_{Q} := \hat{\mathbf{B}}^{-1/2} \hat{\mathbf{A}} \hat{\mathbf{B}}^{-1/2}$ leads to
\begin{equation}\label{eq: general schrödinger equation}
    \begin{aligned}
    \frac{d \hat{\mathbf{w}}_{Q}(\mathbf{x}, t)}{dt} &= \hat{\mathbf{B}}^{-1/2}(\mathbf{x}) \hat{\mathbf{A}}(\mathbf{x}) \hat{\mathbf{B}}^{-1/2}(\mathbf{x}) \hat{\mathbf{w}}_{Q}(\mathbf{x}, t), \\ \hat{\mathbf{w}}_{Q}(\mathbf{x}, 0) &= \hat{\mathbf{B}}^{1/2}(\mathbf{x}) \hat{\mathbf{w}}_0(\mathbf{x}), \\ \hat{\mathbf{B}}^{1/2}(\mathbf{x})&\hat{\mathbf{P}}_{\partial \Omega}\hat{\mathbf{w}}(\mathbf{x}, t) = \mathbf{0}, \, \forall \mathbf{x} \in \partial\Omega.
    \end{aligned}
\end{equation}
We ignore the source term here, as it can be incorporated into the initial conditions or handled separately, as we will show in \Cref{s: sources}. 
Clearly, because $\hat{\mathbf{A}}$ is anti-Hermitian with respect to $(\cdot | \cdot)_{\mathds{1}}$, it follows that $\hat{\mathbf{C}}_{Q} = \hat{\mathbf{B}}^{-1/2} \hat{\mathbf{A}} \hat{\mathbf{B}}^{-1/2}$ is also anti-Hermitian with respect to the same inner product.
Multiplying Eq. \eqref{eq: general schrödinger equation} by the imaginary unit $i$, we obtain a Schrödinger equation with the corresponding Hamiltonian
\begin{equation}\label{eq: continuous Hamiltonian}
    \hat{\mathbf{H}}(\mathbf{x}) = i \hat{\mathbf{C}}_{Q}(\mathbf{x}) = i \hat{\mathbf{B}}^{-1/2}(\mathbf{x}) \hat{\mathbf{A}}(\mathbf{x}) \hat{\mathbf{B}}^{-1/2}(\mathbf{x}).
\end{equation}
Finally, we observe that
\begin{equation}
    (\hat{\mathbf{w}} | \hat{\mathbf{w}})_{\hat{\mathbf{B}}} = (\hat{\mathbf{w}}_{Q} | \hat{\mathbf{w}}_{Q})_{\mathds{1}}.
\end{equation}
Therefore, if $(\cdot | \cdot)_{\hat{\mathbf{B}}}$ induces a norm related to the energy of the system, so does $(\cdot | \cdot)_{\mathds{1}}$ on the transformed states.


\subsection{The Acoustic Wave Equation} \label{s: acoustic wave eq}

For acoustics in $D$ space dimensions, we have that $\hat{\mathbf{w}}_{\text{acoustic}} = [\hat{u};\hat{\mathbf{v}}] \in  \{f: \mathbb{R}^D \rightarrow \mathbb{R}^{D+1}\}$, where $\hat{u}$ is the scalar pressure field and $\hat{\mathbf{v}}$ is the particle velocity field. 
The respective operators are given by \citep{Kaltenbacher2018}
\begin{equation}
    \hat{\mathbf{B}}_{\text{acoustic}}(\mathbf{x}) = \begin{bmatrix}
        \frac{1}{\rho(\mathbf{x})c^2(\mathbf{x})} & \mathbf{0}_{1 \times D} \\
        \mathbf{0}_{D \times 1} & \rho(\mathbf{x})\mathds{1}_{D \times D}
\end{bmatrix},
\end{equation}
where $c, \rho \in \{f: \mathbb{R}^D \rightarrow \mathbb{R}\}$ are the wave speed and density, respectively. 
The anti-Hermitian operator is given by
\begin{equation}
    \hat{\mathbf{A}}_{\text{acoustic}} = \begin{bmatrix}
        0 & -\nabla \cdot \\
        -\nabla & \mathbf{0}_{D \times D}
    \end{bmatrix}.
\end{equation}
The inner product relates to the sound energy as 
\begin{equation}
\begin{aligned}
    (\hat{\mathbf{w}}|\hat{\mathbf{w}})_{\hat{\mathbf{B}}}(t) &= \int_{\Omega} \frac{\hat{u}^2(\mathbf{x},t)}{\rho(\mathbf{x})c^2(\mathbf{x})} d\mathbf{x} \\ &+ \int_{\Omega} \braket{\hat{\mathbf{v}}(\mathbf{x},t)|\hat{\mathbf{v}}(\mathbf{x},t)}\rho(\mathbf{x}) d\mathbf{x} \\ &= 2E_{\text{potential}}(t) + 2 E_{\text{kinetic}}(t).
\end{aligned}
\end{equation}


\subsection{Maxwell's Equations}

For Maxwell's equations in $D$ dimensions, the state vector is defined as $\hat{\mathbf{w}}_{\text{Maxwell}} = [\hat{\mathbf{E}}; \hat{\mathbf{H}}] \in \{f: \mathbb{R}^D \rightarrow \mathbb{R}^{2D}\}$, where $\hat{\mathbf{E}}$ represents the electric field and $\hat{\mathbf{H}}$ represents the magnetic field. 
The operator containing the material properties is given by \citep{johnson2007notes, griffiths2023introduction}
\begin{equation}
    \hat{\mathbf{B}}_{\text{Maxwell}}(\mathbf{x}) = \begin{bmatrix}
        \boldsymbol{\epsilon}(\mathbf{x})& \mathbf{0}_{D\times D} \\
        \mathbf{0}_{D\times D} & \boldsymbol{\mu}(\mathbf{x})
    \end{bmatrix},
\end{equation}
where $\boldsymbol{\epsilon}$ and $\boldsymbol{\mu}$ are the generally complex but Hermitian dielectric permittivity and magnetic permeability tensors, respectively. 
These tensors reduce to a single coefficient multiplied with the identity for isotropic materials. 
For the presented framework to be applicable, both must be positive definite and Hermitian, thereby specifying a generally anisotropic but transparent medium without any losses \citep{johnson2001photonic}. 

We can further identify the anti-Hermitian operator
\begin{equation}
    \hat{\mathbf{A}}_{\text{Maxwell}} = \begin{bmatrix}
        \mathbf{0}_{D \times D} & \nabla \times \\
        -\nabla \times & \mathbf{0}_{D \times D}
    \end{bmatrix},
\end{equation}
where $\nabla \times$ is the curl operator.
Finally, the inner product induced norm, which is related to the electromagnetic energy, can be identified as \cite{jackson1999classical}
\begin{equation}
\begin{aligned}
    (\hat{\mathbf{w}}|\hat{\mathbf{w}})_{\hat{\mathbf{B}}} &= \int_{\Omega} \braket{ \hat{\mathbf{E}}(\mathbf{x},t)|\boldsymbol{\epsilon}(\mathbf{x})|\hat{\mathbf{E}}(\mathbf{x}, t)} d\mathbf{x}  \\
    &+ \int_{\Omega} \braket{\hat{\mathbf{H}}(\mathbf{x},t)|\boldsymbol{\mu}(\mathbf{x})|\hat{\mathbf{H}}(\mathbf{x}, t)} d\mathbf{x} 
    \\ &= 2E_{\text{electric}}(t) + 2E_{\text{magnetic}}(t).
\end{aligned}
\end{equation}


\subsection{The Elastic Wave Equation}

In the interest a of clear notation, we consider the linear elastic wave equation in 3D.
Here, the state vector is given by $\hat{\mathbf{w}}_{\text{elastic}} = [\hat{\mathbf{v}};\text{vec}(\boldsymbol{\hat{\sigma}})] \in  \{f: \mathbb{R}^3 \rightarrow \mathbb{R}^{9}\}$, where $\hat{\mathbf{v}}$ is the particle velocity field and $\text{vec}[\boldsymbol{\hat{\sigma}}] = [\sigma_{xx};\sigma_{yy};\sigma_{zz};\sigma_{xy};\sigma_{xz};\sigma_{yz}]$ denotes the six independent entries of the positive definite and symmetric stress tensor $\hat{\boldsymbol{\sigma}}$.
The corresponding strain tensor can be represented as $\text{vec}(\boldsymbol{\hat{\epsilon}})=[\epsilon_{xx}; \epsilon_{yy};\epsilon_{zz};2\epsilon_{xy};2\epsilon_{xz};2\epsilon_{yz}]$. 
This notation allows us to represent the constitutive relations in the Voigt notation as \citep{voigt1910lehrbuch}
\begin{equation}
    \text{vec}[\boldsymbol{\hat{\epsilon}}](\mathbf{x},t) = \mathbf{S}(\mathbf{x}) \text{vec}[\boldsymbol{\hat{\sigma}}](\mathbf{x},t),
\end{equation}
where $\mathbf{S} \in \{f: \mathbb{R}^3 \rightarrow \mathbb{R}^{6\times 6}\}$ is the compliance matrix; a positive definite and invertible matrix for regular elasticity (e.g., odd elasticity \citep{scheibner2020odd} cannot be covered with this formulation). 
Note that $\mathbf{S}$ has at most $21$ independent parameters \citep{aki2002quantitative}. 
We can now identify
\begin{equation}
    \hat{\mathbf{B}}_{\text{elastic}}(\mathbf{x}) = \begin{bmatrix}
        \rho(\mathbf{x})\mathds{1}_{3 \times 3} & \mathbf{0}_{3 \times 6} \\
        \mathbf{0}_{6 \times 3} & \mathbf{S}
    \end{bmatrix}.
\end{equation} 
To be specific, we further introduce the divergence operator that acts on the stress vector $\text{vec}[\boldsymbol{\sigma}]$ as
\begin{equation}
    \mathcal{D} = \begin{bmatrix}
        \partial_x & 0 & 0 & \partial_y & \partial_z & 0 \\
        0 & \partial_y  & 0 & \partial_x & 0  & \partial_z \\
        0  & 0 & \partial_z & 0 & \partial_x & \partial_y
    \end{bmatrix}.
\end{equation}

With these definitions, we can find 
\begin{equation}
    \hat{\mathbf{A}}_{\text{elastic}} = \begin{bmatrix}
        \mathbf{0}_{3 \times 6} & \mathcal{D} \\
        \mathcal{D}^T & \mathbf{0}_{6 \times 3} 
    \end{bmatrix}.
\end{equation}
Note that $\hat{\mathbf{A}}$ appears symmetric, but it is not: integration by parts, which shifts the derivative in the inner product introduces a minus sign, implying that $(\hat{\mathbf{A}}\hat{\mathbf{w}}|\hat{\mathbf{w}}') = - (\hat{\mathbf{w}}|\hat{\mathbf{A}}\hat{\mathbf{w}}')$. 
The elastic energy is computed as \cite{landau2012theory}
\begin{equation}
\begin{aligned}
    (\hat{\mathbf{w}}|\hat{\mathbf{w}})_{\hat{\mathbf{B}}} &= \int_\Omega  \braket{\hat{\mathbf{v}}(\mathbf{x},t)|\hat{\mathbf{v}}(\mathbf{x},t)}\rho(\mathbf{x}) d\mathbf{x} 
    \\ &+ \int_\Omega  \braket{\boldsymbol{\hat{\sigma}}(\mathbf{x},t)|\mathbf{S}(\mathbf{x})|\boldsymbol{\hat{\sigma}}(\mathbf{x},t)} d\mathbf{x} 
    \\ &= 2E_{\text{kinetic}}(t) + 2 E_{\text{potential}}(t).
\end{aligned}
\end{equation}


\subsection{Numerical Discretization and Boundary Conditions}

Our framework is compatible with any discretization method, provided that the (anti-)Hermitian properties of the operators are preserved.
We represent the numerical discretization of the wave field and operators to vectors and matrices as $\hat{\mathbf{w}} \rightarrow \mathbf{w}$, $\hat{\mathbf{H}} \rightarrow \mathbf{H}$, $\hat{\mathbf{B}} \rightarrow \mathbf{B}$, and $\hat{\mathbf{A}} \rightarrow \mathbf{A}$, where $\mathbf{w} \in \mathbb{R}^L$. 
The discretized equations of motion read
\begin{equation}\label{eq: discretized schrödinger equation}
\begin{aligned}
    \frac{d \mathbf{w}_{Q}(t)}{dt} &= -i\mathbf{H}\mathbf{w}_{Q}(t)= \mathbf{B}^{-1/2}\mathbf{A}\mathbf{B}^{-1/2}\mathbf{w}_{Q}(t) , \\ \mathbf{w}_{Q}(0) &= \mathbf{B}^{1/2}\mathbf{w}_0.
\end{aligned}
\end{equation}
where $\mathbf{A}$ and $\mathbf{B}$ incorporate the BC (\Cref{A: bcs implementation}).

Preserving the symmetry properties of these operators when discretizing requires that $\mathbf{B}$ is a Hermitian matrix and $\mathbf{A}$ an anti-Hermitian matrix.
In \Cref{A: FD discretization} we exemplify the symmetry-preserving discretization of the acoustic wave equation using finite differences with a staggered grid.
We further show that this discretization naturally gives Neumann or Dirichlet BC on a rectangular domain. 
Then, starting from the discretization of these particular BC that respect the symmetries, we discuss in \Cref{s: Constraints} how one can introduce a general class of BC that can be mapped to linear constraints \citep{hoepffner2007implementation}. 
This includes homogeneous or inhomogeneous Dirichlet or Neumann BC along non-rectangular domain boundaries.

Additionally, $\mathbf{B}$ must be a diagonal or block-diagonal matrix, with block sizes that are independent of the overall system size; otherwise, $\mathbf{B}^{1/2}$ becomes dense, and the initial quantum state $\mathbf{w}_{Q}(0)$ will lack sparsity.


\section{Runtime Advantage} \label{S: Runtime advantage}
Given that the discretized Hamiltonian $\mathbf{H}$ \eqref{eq: continuous Hamiltonian} has at most $d = \mathcal{O}(1)$ nonzero values in each row and column, which is ensured when $\mathbf{A}$ and $\mathbf{B}$ are $d$-sparse matrices, and if neither the bit precision $m$ nor the error $\epsilon$ is dominating, quantum simulation requires $\mathcal{O}(nt)=\mathcal{O}(\log Nt)$ operations (\Cref{TH: complexity}). 
In principle this offers an exponential speed-up over classical solvers, which all scale at least with $\mathcal{O}(Nt)$ \citep{banjai2009rapid, qian2009fast, wei2012fast, omar2015linear, smith2019compact}.
However, as we now show, the obtainable speed-up is usually only polynomial \citep{babbush2023exponential}. 
In wave propagation problems, increasing grid points typically aims to (i) enlarge the physical domain or (ii) increase the simulated frequencies. 
For systems with local couplings, such as those resulting from discretizing the considered wave equations, enlarging the domain requires a longer simulation time $t$ for waves to reach the boundaries. 
Similarly, a finer discretization to simulate higher frequencies raises $\|\mathbf{H}\|_{\max}$ due to the reduced grid spacing. 
As per \Cref{TH: complexity}, the number of query calls to $\mathcal{H}$ scales linearly with both $t$ and $\|\mathbf{H}\|_{\max}$. 
Consequently, the quantum speed-up over classical algorithms becomes a polynomial factor dependent on the spatial dimension $D$.
Specifically, let $t = \mathcal{O}(N^{1/D})$ represent the simulation time for the signal to reach the domain boundary in $D$ dimensions. 
For classical algorithms scaling linearly with $N$, the overall runtime complexity is $\mathcal{O}(N t) = \mathcal{O}(N^{1 + 1/D})$. 
In contrast, the HS algorithm has runtime $\tilde{\mathcal{O}}(t) = \tilde{\mathcal{O}}(N^{1/D})$, where the tilde hides logarithmic factors. 
Thus, we expect a quadratic speed-up in 1D, cubic in 2D, and quartic in 3D.
In contrast, \citep{babbush2023exponential} demonstrate an exponential speed-up for oscillator models with purely non-local couplings, where the simulation time does not scale with the system size.

Additionally, the algorithm's oracle $\mathcal{H}$ imposes constraints: to preserve quantum speed-up, the information transferred to the QC must be of order $\tilde{\mathcal{O}}(N^{1/D})$. 
This limits the number of independent material parameters to this order, necessitating functional or low-dimensional material representations.


\section{Estimating Subspace \texorpdfstring{$l_2$}{L2}-norms and Energies} \label{S: Norms and energies}

Reading the complete discrete wave field $\mathbf{w}_{Q}(t)$ from a quantum state comes with exponential and, hence, prohibitive costs. 
Consequently, it is essential to perform computations with the simulated wave field directly on the QC, e.g., by evaluating loss functions. 
We show that it is possible to efficiently measure $l_2$-norms of state subspaces containing multiple superposed quantum states, and that this measurement achieves optimal precision scaling with the number of experiment evaluations:

\begin{theorem}[Subspace $l_2$-norm, \Cref{A: Multi-state}] \label{TH: Multi-State}
    Let $\{\mathbf{w}_1, \mathbf{w}_2, \dots, \mathbf{w}_M\}  \in \mathbb{C}^L$. 
    Given a projector $\mathbf{P}_{\mathcal{S}} = \text{diag}(p_1, p_2, \dots, p_L)$, which projects onto a subspace $\mathcal{S} \subseteq \mathbb{C}^L$, where each $p_i \in \{0, 1\}$ and $d = \sum_{i=1}^L p_i$. 
    Let there be a state preparation oracle $\mathcal{U}$ that prepares the quantum state $\ket{\phi} = [\mathbf{w}_1; \mathbf{w}_2; \dots; \mathbf{w}_M] / ||[\mathbf{w}_1; \mathbf{w}_2; \dots; \mathbf{w}_M]||$.
    Then, if either $d = \mathcal{O}(\text{polylog}(L))$ or $L-d = \mathcal{O}(\text{polylog}(L))$, there exists a quantum algorithm that efficiently estimates the squared sum of $\{\mathbf{w}_1, \mathbf{w}_2, \dots, \mathbf{w}_M\}$ on $\mathcal{S}$ as
    \begin{equation} \label{eq: multi_state_qc_form}
    l_{\mathcal{S}}^2 = ||\mathbf{P}_{\mathcal{S}}\mathbf{w}_1 + \mathbf{P}_{\mathcal{S}}\mathbf{w}_2 + \dots + \mathbf{P}_{\mathcal{S}}\mathbf{w}_M ||^2,
    \end{equation}
    with probability at least $1-\delta$ within error $\epsilon$, which makes $\mathcal{O}(M\log(1/\delta)/\epsilon)$ calls to the oracle $\mathcal{U}$ (along with its inverse and controlled versions). 
\end{theorem}

We will now apply this result to compare two wave fields, such as a simulated and a target $\mathbf{B}^{1/2}$-transformed wave field and to measure subvolume energies (see \Cref{s: General wave eq}). 
By setting $M=2,\; \mathbf{w}_1=\mathbf{w}_{Q},\; \mathbf{w}_2=-\mathbf{w}_{Q}^{\text{target}}$, where $\mathbf{w}_{Q}$ is the simulated wave field and $\mathbf{w}_{Q}^{\text{target}}$ is a target or observed wave field which is only defined or known within subvolume $\mathcal{S}$, we can write
\begin{equation}
l_{\mathcal{S}}^2 = ||\mathbf{P}_{\mathcal{S}}\mathbf{w}_{Q} - \mathbf{P}_{\mathcal{S}}\mathbf{w}_{Q}^{\text{target}}||^2.
\end{equation}
This corresponds to an estimate of the $l_2$-distance between both fields on subspace $\mathcal{S}$. 
By setting $M=1,\; \mathbf{w}_1=\mathbf{w}_{Q}$ we obtain 
\begin{equation}
l_{\mathcal{S}}^2 = ||\mathbf{P}_{\mathcal{S}}\mathbf{w}_{Q}||^2,
\end{equation}
which estimates the energy of $\mathbf{w}$ on $\mathcal{S}$.
\Cref{TH: Multi-State} can also be used to compute energies and $l_2$-comparisons of a superposition of multiple wave fields, extending the work of \citep{babbush2023exponential} on measuring energies. 
These multi-state $l_2$- and energy estimates are key components of the source implementation algorithm, which we will introduce in \Cref{s: sources}.

We can utilize a weighted inner product to compare states in the standard, non-$\mathbf{B}^{1/2}$-transformed bases, as detailed in \Cref{s: weighted_l2}. 
However, as we discuss in \Cref{ss: limitation sources} this will result in globality issues in general. 
In practice, e.g., within the domain of inverse problems, it is usually sufficient to compare transformed wave fields.

Finally, given an efficient quantum algorithm for a unitary basis transformation, comparisons can also be performed in bases other than the computational basis.
This, for example, includes the spatial frequency domain, which can be efficiently accessed through the quantum Fourier transform \citep{nielsen2010quantum}.


\section{Implementing Sources} \label{s: sources}

Incorporating time-dependent sources into the HS of classical wave equations poses a significant challenge when aiming to maintain efficient runtime scaling, as in this framework the wave evolution is unitary and, hence, inherently source free. 
We consider a source term that is constituted of $S$ sources,
\begin{equation}\label{eq: source sum}
\hat{\mathbf{s}}(\mathbf{x}, t) = \sum_{s=1}^S \pmb{\chi}_s(\mathbf{x}) f_s(t),
\end{equation}
where $\pmb{\chi}_s \in \mathbb{R}^D \rightarrow \mathbb{R}^C$ is the spatial distribution of the $s$-th sources and $f_s(t)$ is its source time funcion.  

Assuming that $\pmb{\chi}_s$ is a point source and $f_s$ is a pulse function, whose support in time is compact and with a length either constant or inversely proportional to the maximum frequency, we will show that we can classically map the source to sparse initial conditions with an implementation complexity of $\mathcal{O}(1)$.
This already covers many practically relevant source scenarios.
We then show how to implement sources that lack compact support in time, but that give rise to rotationally covariant wave fields.


\subsection{Single Pulse Source}\label{s: n-independent}
Consider a single point-source defined by 
\begin{equation}
\hat{\mathbf{s}}(\mathbf{x}, t) = \hat{\pmb{\chi}}\delta(\mathbf{x} - \mathbf{x}_s) f(t),
\end{equation}
where $\hat{\pmb{\chi}}\in \mathbb{R}^C$ is an arbitrary vector. 
Its implementation complexity is tied to the number of values that need to be initialized on the QC.
We identify two optimal scenarios:


\subsubsection{Fixed Source Duration}

If the simulation domain grows while the source remains active over a fixed duration $\Delta T = T^\text{end} - T^\text{start}$, the implementation complexity naturally stays $\mathcal{O}(1)$. This is achieved by:

\begin{enumerate}
\item Classically solving the forced wave equation over $\Delta T$ within a fixed volume around the source, sufficient to contain all emitted waves during this interval.
\item Initializing the resulting sparse wave field on the QC.
\end{enumerate}
Since the required initialization volume does not scale with $N$, only a constant number of grid points needs to be initialized, ensuring $\mathcal{O}(1)$ runtime complexity.


\subsubsection{Increasing Grid Resolution}

When refining the grid to simulate higher-frequency wave pulses, typically the source time function $ f(t) $ includes higher frequencies, resulting in a duration inversely proportional to the increased frequency bandwidth $ \Delta \omega $:
\begin{equation}
\Delta T \propto \frac{1}{\Delta \omega} = \frac{1}{\omega^\text{max} - \omega^\text{min}},
\end{equation}
where $[\omega^\text{min}, \omega^\text{max}] $ is the support of the Fourier transform $ \tilde{f}(\omega) $ of $ f(t) $.
For example, increasing the grid resolution by a factor of 2 (i.e., halving the spatial grid spacing $ \Delta x \rightarrow \tfrac{\Delta x}{2} $) allows us to resolve higher spatial frequencies. 
If we scale the frequency bounds of $ f(t) $ as $ \omega^\text{max} \rightarrow 2\omega^\text{max} $ and $ \omega^\text{min} \rightarrow \alpha\,\omega^\text{min} $ with $ \alpha \leq 2 $,
then this results in at least a doubling of $ \Delta \omega $ and thus halving $ \Delta T $.
Consequently, the classically simulated volume decreases proportionally across all dimensions, and the number of initialized grid points remains constant, maintaining $\mathcal{O}(1)$ complexity.


\subsection{Multiple Pulse Sources} \label{S: multi_source}

We extend our analysis to sources composed of multiple point sources
\begin{equation}
\hat{\mathbf{s}}(\mathbf{x}, t) = \sum_{s=1}^S \hat{\pmb{\chi}}_s\delta(\mathbf{x}-\mathbf{x}_s) f_s(t).
\end{equation}


\subsubsection{Synchronous Activation}

If all sources are active over the same time interval, i.e., $T_s^\text{start} = T^\text{start}$ and $T_s^\text{end} = T^\text{end}$ for all $s$, the previous approach applies directly. Each source can be classically simulated within its respective volume, and the resulting wave fields are initialized on the QC as a single initial condition. 
The runtime scaling remains $\mathcal{O}(1)$ provided that the number of sources $S$ does not scale with $N$.


\subsubsection{Asynchronous Activation} \label{S: asynchronous activiation}

When activation times vary among different point sources, an additional step is required to synchronize them. 
Let $\mathbf{w}_s$ for $s = 1, \dots, S$ denote the initial conditions from classically simulating each source over their respective durations $\Delta T_s = T_s^{\text{end}} - T_s^{\text{start}}$. The combined and energy-transformed initial state is given by
\begin{equation}
    \phi(0) = (I^{\otimes \log_2(S)} \otimes \mathbf{B}^{1/2}) [\mathbf{w}_1; \dots; \mathbf{w}_S],
\end{equation}
where
\begin{equation}
    I = \begin{bmatrix} 1 & 0 \\ 0 & 1 \end{bmatrix}.
\end{equation}
The initial quantum state then is
\begin{equation}
    \ket{\psi(0)} = \frac{\phi(0)}{\|\phi(0)\|},
\end{equation}
incorporating initial conditions at different global times. 
To synchronize these states to a common global time $T^{\text{sync}} = \max\{T_1^{\text{end}}, \dots, T_S^{\text{end}}\}$, we evolve the state using a time-dilating Hamiltonian $\mathbf{H}^{\text{sync}}$:
\begin{equation}
    \ket{\psi(T^{\text{sync}})} = e^{-i \mathbf{H}^{\text{sync}}} \ket{\psi(0)},
\end{equation}
where
\begin{equation}
    \mathbf{H}^{\text{sync}} = \begin{bmatrix}
        (T^{\text{sync}} - T_1^{\text{end}}) \mathbf{H} & \dots & \mathbf{0} \\
        \vdots & \ddots & \mathbf{0} \\ 
        \mathbf{0} & \mathbf{0} & (T^{\text{sync}} - T_S^{\text{end}}) \mathbf{H}
    \end{bmatrix}.
\end{equation}
This effectively time-dilates each component, aligning them to $T^{\text{sync}}$. 
Subsequently, the synchronized state is evolved under the multi-state wave equation Hamiltonian $\mathbf{H}^{\text{mult}} = I^{\otimes \log_2 (S)} \otimes \mathbf{H}$ to the final simulation time $T$:
\begin{equation}
    \ket{\psi(T)} = e^{-i (T - T^{\text{sync}}) \mathbf{H}^{\text{mult}}} \ket{\psi(T^{\text{sync}})}.
\end{equation}

Due to the linearity of the wave equation, the superposition of partial wave fields in $\ket{\psi(T)}$ corresponds to the solution of Eq. \eqref{eq: discretized schrödinger equation} at time $T$. 
Consequently, we can estimate an $l_2$-distance to a target field or subspace energies using \Cref{TH: Multi-State}.
Since the addition operation is inherently non-unitary, it must be integrated within the measurement process. 
Furthermore, as the loss function is nonlinear, we cannot compute losses for individual sub-fields separately and then combine them classically to derive the joint loss; therefore, we have to rely on \Cref{TH: Multi-State} to estimate the loss function.

This approach remains efficient under the conditions that (a) the number of point sources $S$ is independent of $N$, and (b) each individual source meets the previously outlined constraints.


\subsection{Rotationally symmetric sources without compact support in time} \label{s: n-dependent}

Implementing a source time function $f(t)$ that lacks compact support in time is inefficient with the above approach, as it results in initial conditions lacking compact support in space. 
Such initial fields are prohibitively expensive to compute classically and initialize on the QC. 

To target this, we introduce an algorithm that efficiently initializes rotationally covariant wave fields as quantum states: 
\begin{lemma}[Vector Field Initialization,  \Cref{app: proof of lemma 4}] \label{TH:rot_covariant_initialization}
Let $\mathbf{w}^{sym} : \mathbb{R}^D \to \mathbb{R}^C$ be a vector field. 
Assume that $\mathbf{w}^{sym}$ is rotationally covariant around a point $\mathbf{x}_0 \in \mathbb{R}^D$.
Consider an isotropic radially symmetric grid centered at $\mathbf{x}_0$, discretized into $N = A^D$ points, where $A$ is the number of radial divisions. 
The grid points are specified by radial indices $a \in \{1, \dots, A\}$ and angular indices $\boldsymbol{\theta}$.
Define the quantum state
\begin{equation}\label{eq:psi_def}
    \ket{\psi} = \frac{1}{\sqrt{\mathcal{N}}} \sum_{c=1}^C \sum_{a=1}^A \sum_{\boldsymbol{\theta}} w^{sym}_c\big( \mathbf{x}_{a, \boldsymbol{\theta}} \big) \ket{c} \otimes \ket{a} \otimes \ket{\boldsymbol{\theta}},
\end{equation}
where $w^{sym}_c\big( \mathbf{x}_{a, \boldsymbol{\theta}} \big)$ is the $c^{\text{th}}$ component of $\mathbf{w}^{sym}$ at grid point $\mathbf{x}_{a, \boldsymbol{\theta}}$, and $\mathcal{N} = \sum_{c,a,\boldsymbol{\theta}} \left| w^{sym}_c\big( \mathbf{x}_{a, \boldsymbol{\theta}} \big) \right|^2$ ensures normalization.
This state can be initialized with $\mathcal{O}(N^{1/D})$ classical evaluations of $\mathbf{w}^{sym}$ and $\tilde{\mathcal{O}}(CN^{1/D})$ quantum gates, where the tilde hides polylogarithmic factors.
\end{lemma}

We consider a spatial source distribution $\pmb{\chi} = \pmb{\chi}^{sym}$ for which an arbitrary rotation in space and the same rotation of the source vector yield the same result.
Formally, let there be a $\mathbf{R}(\mathbf{S}) \in \mathrm{SO}(C)$, where $\mathbf{S}\in \mathrm{SO}(D)$; then, rotationally covariant sources satisfy $\mathbf{R}(\mathbf{S})\pmb{\chi}^{sym}(\mathbf{x}) = \pmb{\chi}^{sym}(\mathbf{S}\mathbf{x}) \; \forall \mathbf{x}, \mathbf{S}$. 
We further denote the rotationally covariant Green's function that solves the wave equation \eqref{eq: general wave equation} in a homogeneous medium for the source $\hat{\mathbf{s}}_\delta(\mathbf{x},t) = \pmb{\chi}^{sym}(\mathbf{x}) \delta(t)$ as $\hat{\mathbf{G}}^{sym}$.
In this case, the wave field arising from the source time function $f$ is given by $\hat{\mathbf{w}}^{sym}=\hat{\mathbf{G}}^{sym} * f$.
It can be discretized as $\hat{\mathbf{w}}^{sym}\rightarrow \mathbf{w}^{sym}$ and efficiently initialized on a spherical grid centered around the source location using \Cref{TH:rot_covariant_initialization}.

\begin{figure*}
    \includegraphics[width=1\linewidth]{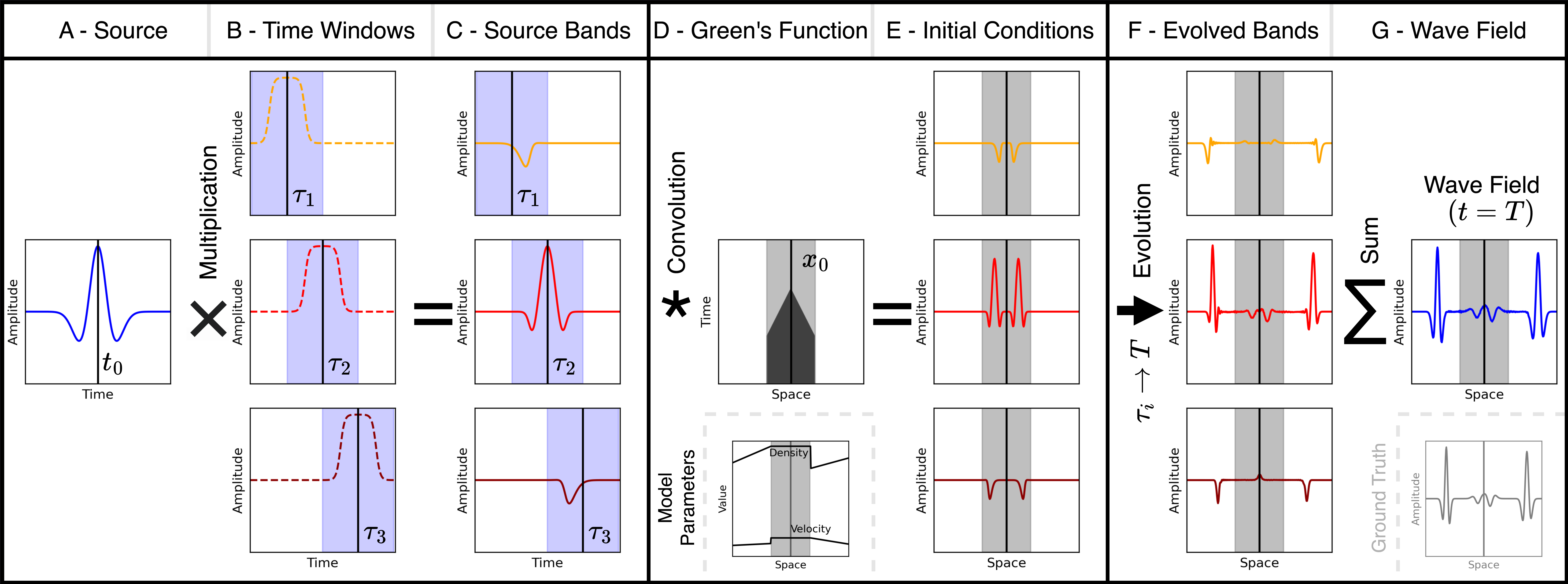}
    \caption{Algorithm implementing a point source function on a QC for a 1D acoustic problem. $A$ shows the source-time function for a single point location. We decompose the source-time function in time ($B$), which gives a set of functions that, in superposition, equate to the original source-time function ($C$). The inset at the bottom shows the acoustic parameters profiles with the homogenous volume around the source. By convolving them in time with the homogeneous Green's function ($D$), we obtain multiple sets of initial conditions at different global times ($E$). We evolve these initial conditions over different time intervals to the final simulation time $T$ ($F$). Here, the superposition of the source slices is equal to the solution of the wave equation ($G$). We can utilize this to efficiently estimate an $l_2$-distance to a target state using \Cref{TH: Multi-State}.}
    \label{fig: sources}
\end{figure*}

Now let us assume that the medium is generally heterogeneous but there exists a small homogeneous volume around the source with radius $r_s$. 
This may hold true in certain experimental setups and can otherwise be taken as an approximation.
Let the travel time in the volume be $T^{hom}_s = r_s/c^{hom}$ where $c^{hom}$ is the wave speed in the homogeneous medium. 
If the source time function is active for time period $T_s>T^{hom}_s$ then we can split the source in $J \approx T_s/T^{hom}_s$ contributions:
\begin{equation} \label{EQ: source_windows}
    f(t) \approx \sum_{j=1}^{J} W_{j}(t-\tau_{j}) f(t),
\end{equation}
where $W_j$ are sigmoid window functions (see \Cref{ap:windowing} for details).
All $J$ source terms for $f(t)$ can be recognized as individual sources that satisfy $\Delta T_j\leq T_{j}^{hom}$, where $\Delta T_j$ is the duration of the $j^{th}$ source. 
This means that for none of them, the corresponding wave leaves the homogeneous domain before it terminates.
Therefore, the resulting $J$ initial wave fields are all rotationally covariant
\begin{equation}
\begin{aligned}
    \hat{\mathbf{w}}^{sym}_{j}(\mathbf{x},t-\tau_{j}) &= \hat{\mathbf{G}}^{sym}(\mathbf{x},t) * (W_{j}(t) f(t+\tau_{j})),
    \\ &|\mathbf{x}| < r, \quad j = 1,...,J.
\end{aligned}
\end{equation}
By using the same discretization as in the homogeneous case, we can initialize the state $[\mathbf{w}^{sym}_{1};...;\mathbf{w}^{sym}_{J}]$ with a runtime complexity of $\tilde{\mathcal{O}}(J C N^{1/D})$.
This state can then be synchronized and measured as introduced in \Cref{S: asynchronous activiation}.
Hence, if $J = \tilde{\mathcal{O}}(\text{polylog}(N))$ the runtime advantage remains.
The whole workflow is depicted in \Cref{fig: sources}.

Note, that the rotationally covariant initial wave field must be pre-computed classically along a reference direction on the spherical grid with a runtime complexity of at worst $\tilde{\mathcal{O}}(CN^{1/D})$.
This is, for example, possible via analytic Green's functions for acoustic and elastic isotropic (explosive) point sources.


\section{Limitations}\label{s: limitations}

The limitations considered here are those that either impact the algorithmic runtime in terms of big-$\mathcal{O}$ complexity, or which are fundamentally not implementable in this framework.


\subsection{Loss functions}\label{ss: limitation sources}

We identify two limitations within the loss function estimations introduced:

(i) Estimating time-integrated loss functions: Currently, our algorithm estimates the loss at a single measurement time step. 
Computing time-integrated loss functions involves separate simulation experiments for each measurement time, with the overall loss metric computed across all steps using either \Cref{TH: Multi-State} or classical post-processing. 
However, this approach becomes inefficient for long comparison/integration times, where the number of comparison time steps scales worse than $\mathcal{O}(\text{polylog}(N))$. 
Non-destructive estimation of the loss function \citep{harrow2020adaptive, rall2023amplitude} could provide a solution by preserving the state throughout the measurement operation, enabling iterative evolution and loss estimation of the wave field for multiple time points.

(ii) Globality of measurement quantities: A limitation of QC loss estimation via observable measurements is the requirement for sufficiently global measurement quantities. 
Suppose a state is measured in a subspace with only a small probability amplitude fraction compared to the entire quantum state. 
In that case, more samples are needed to estimate the norm accurately, as the expected value becomes vanishingly small. 
E.g., ensuring the subspace size is large with respect to $N$ can mitigate this issue in practice. 


\subsection{Source Terms}

We identify three limitations: 

(i) The number of sources in space must not scale worse than $\mathcal{O}(\text{polylog}(N))$. 
For instance, it is inefficient to have a source everywhere on the surface of a 3D body.

(ii) Situations where the number of initialized nonzero values grows polynomially with $N$ must be avoided to prevent restricting the overall runtime.
An example of this could be a continuously active source within an inhomogeneous medium.

(iii) For symmetric sources without compact time support, the requirement for radially symmetric grids may be practically limiting as domains can be non-symmetric around the source location.


\subsection{Damping}

A significant limitation of the presented framework is its inability to implement damping due to its non-unitary nature. 
However, damping in the context of HS has been recently studied, demonstrating that quantum speed-up can be preserved under certain conditions \citep{krovi2024quantum, koukoutsis2024quantum}. 
An alternative approach is to map a damped classical wave equation to an open quantum system. 
Simulating such systems on QC is a recent research focus \citep{hu2020quantum}. 
We believe that our work can eventually be combined with these approaches.


\subsection{Time-Dependent Media} \label{S: time_dependet}

A final remark concerns time-dependent media, such as those in time-modulated meta-materials \citep[e.g.,][]{nassar2018quantization, chen2019mechanical, galiffi2022photonics}. 
In these cases, $\mathbf{B} = \mathbf{B}(t)$ becomes time-dependent, implying that it no longer commutes with the time derivative. 
Consequently, a time-dependent symmetric problem is mapped to a time-dependent non-Hermitian Schrödinger equation, rendering the introduced framework inapplicable \citep{bosch2023differences}.

\section{Code}
We present an educational Git repository, where we implement all the presented methods and concepts for the acoustic wave equation and Maxwell's equations. 
The numerical implementation is based on a staggered grid finite difference method as detailed in \Cref{A: FD discretization}.
The repository can be found here: \href{https://github.com/malteschade/Quantum-Wave-Simulation-with-Sources-and-Loss-Functions}{github.com/malteschade/Quantum-Wave-Simulation-with-Sources-and-Loss-Functions}.


\section{Conclusions} \label{s: Conclusions}

In this work, we have demonstrated that we can efficiently simulate forced, linear, anti-Hermitian wave equations with arbitrary heterogeneous and anisotropic material properties within the HS framework. 
While the theoretical speed-up of this approach can be exponential, practical 3D applications are expected to see a quartic speed-up due to the scaling of the simulation time with the system size. 
This quartic speed-up is maintained throughout our algorithm if the state read-out is sufficiently global compared to the overall system size.

Our method is compatible with any numerical approach that respects the anti-Hermitian nature of the operator and results in a diagonal or block diagonal representation of the material properties. 
The introduced representation of the wave equations naturally leads to a quantum encoding that enables the measurement of energies in sub-volumes of the model with optimal precision scaling in the number of measurements taken. 
Additionally, we have shown that the $l_2$-distance of the final simulated wave field to a target or observed wave field can be measured with optimal precision scaling and that a broad class of source terms can be implemented efficiently. 

Recent studies have indicated that HS could be feasible on noisy QC, potentially preceding the advent of noise-free quantum hardware \citep{kikuchi2023realization, wright2024noisy}.
Furthermore, new advances in quantum error correction hint that noise-free QC might be realized soon \citep{acharya2024quantum}. 
These findings suggest that our proposed framework may have practical applications in the near future, thereby underscoring the importance of continued research in this field. 


\begin{acknowledgments}
We would like to thank Marion Dugué for insights on the source decompositions. Furthermore, we are thankful to Patrick Marty and Dr. Ines Ulrich for sharing insights on acoustic wave physics in the context of medical imaging and to Dr. V\'aclav Hapla for initial brainstorming sessions.
\end{acknowledgments}

\noindent \textbf{Grant Acknowledgment}: C.B. was supported by the Swiss National Science Foundation through a Postdoc.Mobility Fellowship. S.K. was partially supported by the project DT-GEO, grant number 101058129. Furthermore, we gratefully acknowledge the support provided by Google Quantum AI, which has significantly contributed to the advancement of this research.

\noindent \textbf{Data statement}: No data have been used for this research.

\noindent \textbf{Author contribution statement}: study conception: C. Bösch; sources: M. Schade and C. Bösch; loss functions and energy estimation: M. Schade and C. Bösch, S. Keating; 
acoustic FD implementation: G. Aloisi; general boundary conditions and constraints: G. Aloisi and C. Bösch; manuscript preparation: M. Schade, C. Bösch, G. Aloisi, S. Keating, A. Fichtner. All authors reviewed the results and approved the final version of the manuscript.


\appendix


\section{Estimating Loss Functions} \label{s: loss fcts}

To prove \Cref{TH: Multi-State} we first introduce the following auxiliary result:

\begin{lemma}[Two-state subspace $l_2$-norm, \Cref{A: Subdomain norm}] \label{TH: Substate L2}
    Let $\mathbf{a}, \mathbf{b} \in \mathbb{C}^L$. Given a projector $\mathbf{P}_{\mathcal{S}} = \text{diag}(p_1, p_2, \dots, p_L)$, which projects onto a subspace $\mathcal{S} \subseteq \mathbb{C}^L$, where each $p_i \in \{0, 1\}$ and $d = \sum_{i=1}^L p_i$. Let there be a state preparation oracle $\mathcal{U}$ that prepares the quantum state $\ket{\phi} = [\mathbf{a}; \mathbf{b}]/||[\mathbf{a}; \mathbf{b}]||$.
    Then, if either $d = \mathcal{O}(\text{polylog}(L))$ or $L-d = \mathcal{O}(\text{polylog}(L))$, there exists a quantum algorithm that efficiently estimates the $l_{2,\mathcal{S}}$-norm between $\mathbf{a}$ and $\mathbf{b}$ on $\mathcal{S}$ as
    \begin{equation} \label{eq: l2_qc_form}
    l_{2,\mathcal{S}}^2 = ||\mathbf{P}_{\mathcal{S}} \mathbf{a} - \mathbf{P}_{\mathcal{S}} \mathbf{b}||^2,
    \end{equation}
    with probability at least $1-\delta$ within error $\epsilon$, which makes $\mathcal{O}(\log(1/\delta)/\epsilon)$ calls to the oracle $\mathcal{U}$ (along with its inverse and controlled versions).
\end{lemma}

\subsection{Proof of \texorpdfstring{\Cref{TH: Substate L2}}{Lemma \ref{TH: Substate L2}}} \label{A: Subdomain norm}

The squared $l_2$-norm on this subspace is defined as
\begin{equation} \label{eq: l2}
    l^2_{2,\mathcal{S}} = \sum_{i}^{L}\big|
    (\mathbf{P}_{\mathcal{S}}\mathbf{a})_i-(\mathbf{P}_{\mathcal{S}}\mathbf{b})_i\big|^2.
\end{equation}
By introducing $\ket{\phi} = [\mathbf{a};\mathbf{b}]/||[\mathbf{a};\mathbf{b}]||$, we can write Eq. \eqref{eq: l2} as an expected value estimation 
\begin{equation} \label{eq: partial_l2}
    l_{2,\mathcal{S}}^2 = \braket{\phi|\mathbf{O}|\phi}||[\mathbf{a};\mathbf{b}]||^2,
\end{equation} 
where the Hermitian observable $\mathbf{O}$ is given by
\begin{equation}
    \mathbf{O}_{{2L \times 2L}}=\begin{bmatrix}
        \mathbf{P}_{\mathcal{S}} & -\mathbf{P}_{\mathcal{S}} \\
        -\mathbf{P}_{\mathcal{S}} & \mathbf{P}_{\mathcal{S}}\\
    \end{bmatrix}.
\end{equation}
The observable $\mathbf{O}$ can be recognized as a difference operator that estimates the $l_2$-norm between $\mathbf{a}$ and $\mathbf{b}$ on the subspace $\mathcal{S}$ up to normalization and can, in principle, be implemented on a QC with sub-optimal precision scaling \citep{nielsen2010quantum}.

To estimate $l_{2,\mathcal{S}}^2$ with optimal precision scaling beyond the central limit, we will introduce an auxiliary qubit to double the state space dimension.
In this augmented space we can find a unitary decomposable observable $\tilde{\mathbf{O}}$, whose expectation value with respect to the augmented state $\ket{\psi} = [\mathbf{P}_{\mathcal{S}}\mathbf{a}; \mathbf{P}_{\mathcal{S}}\mathbf{b}; (\mathds{1}-\mathbf{P}_{\mathcal{S}})\mathbf{a}; (\mathds{1}-\mathbf{P}_{\mathcal{S}})\mathbf{b}]/||[\mathbf{a};\mathbf{b}]||$ is equal to $l_{2,\mathcal{S}}$, such that
\begin{equation}
    l_{2,\mathcal{S}}^2 = \braket{\psi|\tilde{\mathbf{O}}|\psi}||[\mathbf{a};\mathbf{b}]||^2.
\end{equation}
We first demonstrate that there exists a unitary initialization procedure for $\ket{\psi}$ given an initialization oracle for $\ket{\phi}$ and discuss the efficiency of its implementation. 
We then explicitly construct $\tilde{\mathbf{O}}$, which can be decomposed into $4$ commuting unitary observables independent of $L$ and $\mathcal{S}$. We conclude by analyzing the overall complexity of our algorithm.

To start, let us combine $\mathbf{U}_\mathbf{a}$ preparing $\mathbf{a}$ and $\mathbf{U}_\mathbf{b}$ preparing $\mathbf{b}$ into a single unitary, $\mathbf{U}$, that prepares $\ket{\phi}$ as
\begin{equation}
    \ket{\phi}  = \mathbf{U}\ket{0^{n+1}} =\begin{bmatrix}
        \mathbf{U}_\mathbf{a} & 0 \\ 0 & \mathbf{U}_\mathbf{b}
    \end{bmatrix}\ket{0^{n+1}} 
\end{equation}
where $n=\log_2 (L)$.
We further define the unitary matrix $\tilde{\mathbf{P}}_\mathcal{S}\in \mathbb{C}^{4L\times4L}$ constructed from $\mathbf{P}_\mathcal{S}$ and its complement subspace $\mathds{1}-\mathbf{P}_{\mathcal{S}}$ as 
\begin{equation} \label{eq: p-hat}
    \tilde{\mathbf{P}}_\mathcal{S} =
    \begin{bmatrix}
        \mathbf{P}_{\mathcal{S}} & 0 & (\mathds{1}-\mathbf{P}_{\mathcal{S}}) & 0 \\
        0 & \mathbf{P}_{\mathcal{S}} & 0 & (\mathds{1}-\mathbf{P}_{\mathcal{S}}) \\
        (\mathds{1}-\mathbf{P}_{\mathcal{S}}) & 0 & \mathbf{P}_{\mathcal{S}} & 0 \\
        0 & (\mathds{1}-\mathbf{P}_{\mathcal{S}}) & 0 & \mathbf{P}_{\mathcal{S}}
    \end{bmatrix}.
\end{equation}
By adding one auxiliary qubit,
which is left invariant during the application of $\mathbf{U}$, 
the target state $\ket{\psi}$ can finally be obtained as
\begin{equation}
\begin{aligned}
    \ket{\psi} &=
    \tilde{\mathbf{P}}_{\mathcal{S}} (I \otimes \mathbf{U}) \ket{0^{n+1}} \\
    &=
    [\mathbf{P}_{\mathcal{S}}\mathbf{a}; \mathbf{P}_{\mathcal{S}}\mathbf{b}; (\mathds{1}-\mathbf{P}_{\mathcal{S}})\mathbf{a}; (\mathds{1}-\mathbf{P}_{\mathcal{S}})\mathbf{b}]/||[\mathbf{a};\mathbf{b}]||,
\end{aligned}
\end{equation}
where 
\begin{equation}
    I = \begin{bmatrix}
    1 & 0 \\ 0 & 1
    \end{bmatrix}.
\end{equation}

As $\mathbf{P}_{\mathcal{S}}$ is diagonal, $\tilde{\mathbf{P}}_{\mathcal{S}}$ can be recognized as a permutation matrix, which is decomposable into $\tilde{\mathbf{P}}_{\mathcal{S}}=\prod_j^d\tilde{\mathbf{P}}_{\mathcal{S}_j}$. 
Each $\tilde{\mathbf{P}}_{\mathcal{S}_j}$ can be implemented with one multi-controlled X-gate (MCX).
As MCX gates can be synthesized with $\mathcal{O}(n)$ controlled-not (CNOT) gates with a circuit depth of $\mathcal{O} (\log_2 (n))$ using $\mathcal{O}(n)$ auxiliary qubits \citep{iten2016quantum,he2017decompositions,zindorf2024efficient}, the overall gate complexity when implementing $\tilde{\mathbf{P}}_{\mathcal{S}}$ is $\mathcal{O}(dn)$ with a circuit depth of $\mathcal{O}(d\log_2 (n))$.
Hence, if either $d = \mathcal{O}(\text{polylog}(L))$ or $L-d = \mathcal{O}(\text{polylog}(L))$, the implementation of $\tilde{\mathbf{P}}_{\mathcal{S}}$ is guaranteed to be efficient.

As future QC might allow for implementing hardware native MCX-gates, further reductions in gate complexity are possible \citep{martinez2016compiling,goel2021native,kim2022high}.
Additionally, regular patterns in $\tilde{\mathbf{P}}_{\mathcal{S}}$ can be exploited to implement multiple $\tilde{\mathbf{P}}_{\mathcal{S}_j}$ at once. 
Even for large $d$, this can allow for an efficient implementation of $\tilde{\mathbf{P}}_{\mathcal{S}}$ in some instances, e.g., for projections to the first half of the domain.

We now introduce the augmented observable 
\begin{equation}
    \tilde{\mathbf{O}}_{4L \times 4L}=
    \begin{bmatrix}
        \mathds{1} & - \mathds{1} & 0 & 0\\
        -\mathds{1} &  \mathds{1} & 0 & 0\\
        0 & 0 & 0 & 0 \\
        0 & 0 & 0 & 0
    \end{bmatrix}.
\end{equation}
Using the definition of $\ket{\psi}$, it is easy to check that
\begin{equation}
    l_{2,\mathcal{S}}^2 = \braket{\psi|\tilde{\mathbf{O}}|\psi}||[\mathbf{a};\mathbf{b}]||^2.
\end{equation}
The augmented observable $\tilde{\mathbf{O}}$ is always - independent of the state dimension $L$ and the subspace $\mathcal{S}$ - represented by the 4 Pauli matrices
\begin{equation}
    \begin{split}
        \tilde{\mathbf{O}}_1 &= I \otimes I \otimes I \otimes \cdots \otimes I\\
        \tilde{\mathbf{O}}_2 &= I \otimes X \otimes I \otimes \cdots \otimes I\\
        \tilde{\mathbf{O}}_3 &= Z \otimes I \otimes I \otimes \cdots \otimes I\\
        \tilde{\mathbf{O}}_4 &= Z \otimes X \otimes I \otimes \cdots \otimes I
    \end{split},
\end{equation}
 where
\begin{equation}
        X = \begin{bmatrix}
            0 & 1 \\ 1 & 0
        \end{bmatrix}, \quad
        Z = \begin{bmatrix}
            1 & 0 \\ 0 & -1
        \end{bmatrix},
\end{equation}
and with coefficients $\mathbf{c}=[0.5,-0.5,0.5,-0.5]$,
such that
\begin{equation} \label{eq: l2-obs}
    \tilde{\mathbf{O}} = \sum_{j=1}^4 \mathbf{c}_j\tilde{\mathbf{O}}_j.
\end{equation}
Hence, we conclude that $l_{2,\mathcal{S}}^2$ can be estimated as
\begin{equation}
    l_{2,\mathcal{S}}^2 = ||[\mathbf{a};\mathbf{b}]||^2 \sum_{j=1}^4 \mathbf{c}_j\braket{\psi|\tilde{\mathbf{O}}_j|\psi}.
\end{equation}

As proven in \citep{knill2007optimal}, the measurement of the expected value of a quantum state with respect to an observable that is unitary using high-fidelity amplitude amplification techniques has optimal precision scaling with a query complexity of $\mathcal{O} (\log(1/\delta)/\epsilon)$. As all Pauli observables are unitary and the number of Pauli observables that represent $\tilde{\mathbf{O}}$ is independent of $L$ and $\mathcal{S}$, it follows that our algorithm has the same query complexity.


\subsection{Proof of \texorpdfstring{\Cref{TH: Multi-State}}{Theorem \ref{TH: Multi-State}}}

\label{A: Multi-state}
Building on \Cref{TH: Substate L2}, we can estimate general $l_2$-norms of arbitrary many sub-states. 
Consider a quantum state 
\begin{equation}
    \ket{\phi} = [\mathbf{w}_1; \mathbf{w}_2; \dots; \mathbf{w}_M]/||[\mathbf{w}_1;\mathbf{w}_2; \dots; \mathbf{w}_M]||,
\end{equation}
with $\mathbf{w}_i \in \mathbb{C}^L$, where $M$ is the number of sub-states. 
We aim to estimate a squared $l_2$-norm, $l^2_{\Sigma}$, on a subspace $\mathcal{S}$ defined by the projector $\mathbf{P}_{\mathcal{S}}$, such that
\begin{equation}
    l_{\Sigma,\mathcal{S}}^2 = ||\mathbf{P}_{\mathcal{S}}\mathbf{w}_1 + \mathbf{P}_{\mathcal{S}}\mathbf{w}_2 + \dots + \mathbf{P}_{\mathcal{S}}\mathbf{w}_M ||^2.
\end{equation}
It remains to generalise $\tilde{\mathbf{P}}_\mathcal{S}$ (see Eq. \eqref{eq: p-hat}) and $\tilde{\mathbf{O}}$ (see Eq. \eqref{eq: l2-obs}).

We extend $\tilde{\mathbf{P}}_\mathcal{S}$ by recognizing that it corresponds to the unitary matrix
\begin{equation}
    \tilde{\mathbf{P}}_{\mathcal{S}, A} = 
    \begin{bmatrix}
        \mathbf{P}_{\mathcal{S}} & (\mathds{1}-\mathbf{P}_{\mathcal{S}})\\
        (\mathds{1}-\mathbf{P}_{\mathcal{S}}) & \mathbf{P}_{\mathcal{S}}
    \end{bmatrix},
\end{equation}
which is then tensor-multiplied with a number of $I$-gates depending on $M$, such that 
\begin{equation}
    \tilde{\mathbf{P}}_{\mathcal{S}, \Sigma} = 
    \tilde{\mathbf{P}}_{\mathcal{S}, A}
    \otimes I^{\otimes \log_2 (M)}.
\end{equation}

Similarly, the specific structure of $\tilde{\mathbf{O}}$ lets us define the generalized observable $\tilde{\mathbf{O}}_\Sigma$ as 
\begin{equation} \label{eq: obs_sum_iz}
    \tilde{\mathbf{O}}_\Sigma = \frac{1}{2}(I \otimes\tilde{\mathbf{O}}_A + Z \otimes \tilde{\mathbf{O}}_A) \otimes I^{\otimes \log_2 (L)},
\end{equation} 
where 
\begin{equation} \label{eq: obs_sum_a}
    \tilde{\mathbf{O}}_A =  \sum_{\mathbf{A} \in \{I, X\}^{\otimes \log_2 (M)}} \mathbf{A}.
\end{equation}
In this context, Eq. \eqref{eq: obs_sum_iz} represents the half-domain measurement of $\ket{\psi}$ through
\begin{equation}
    \frac{1}{2}(I + Z) = 
    \begin{bmatrix}
        1 & 0 \\ 0 & 0
    \end{bmatrix},
\end{equation}
and the dimension of each sub-state $\mathbf{w}_i$ as $\dim\left(I^{\otimes \log_2 (L)}\right) = L \times L$, while Eq. \eqref{eq: obs_sum_a} describes the summation of the sub-states.

The number of Pauli observables in Eq. \eqref{eq: obs_sum_a} depends on $M$, as $|\{I,X\}^{\otimes \log_2 (M)}| = M$. 
Therefore, the number of unitary observables that need to be estimated is $2M$.
Consequently, the overall query complexity of the algorithm is $\mathcal{O} (M\log(1/\delta)/\epsilon)$.
For verification, with $M=2$, $\mathbf{w}_1=\mathbf{a}$, and $\mathbf{w}_2=-\mathbf{b}$, we see that $\tilde{\mathbf{O}}_\Sigma = \tilde{\mathbf{O}}$ and $\tilde{\mathbf{P}}_{\mathcal{S}, \Sigma} = \tilde{\mathbf{P}}_\mathcal{S}$, confirming that $l_{\Sigma,\mathcal{S}}^2 = l_{2,\mathcal{S}}^2$.

\subsection{\texorpdfstring{Weighted $l_2$-norms}{Weighted L2-norms}} \label{s: weighted_l2}
To compare the states in the standard basis, we utilize a weighted inner product which reverses the transformation with $\mathbf{B}^{1/2}$:
\begin{corollary}[Weighted $l_2$-norm] \label{CO: Weighted L2}
    Given \Cref{TH: Multi-State}, we can estimate a weighted $l_2$-norm with $\mathcal{O}(V\log(1/\delta)/\epsilon)$ calls to the oracle $\mathcal{U}$ (along with its inverse and controlled versions), as \begin{equation} \label{eq: all_l2}
    l_{v}^2 = \sum_j^V v_j l_{\mathcal{S}_j}^2,
    \end{equation}
    where $\mathbf{v} \in \mathbb{R}_+^V$ is a weight vector consisting of all $V$ unique weights. The estimation is efficient if the number of unique weights satisfies $ V = \mathcal{O}(\text{polylog}(L))$, and if for every $\mathbf{P}_{\mathcal{S}_j}$ either $d_j = \mathcal{O}(\text{polylog}(L))$ or $L-d_j = \mathcal{O}(\text{polylog}(L))$.
\end{corollary}
Applying this to the comparison between two wave fields we obtain
\begin{equation}
\begin{aligned}
    l_{B}^2 &= \sum_j^{\bar{B}} B^{-1/2}_j ||\mathbf{P}_{\mathcal{S}_j}\mathbf{w}_{Q} - \mathbf{P}_{\mathcal{S}_j}\mathbf{w}_{Q}^{\text{target}}||^2 
    \\ &= \sum_j^{\bar{B}} ||\mathbf{P}_{\mathcal{S}_j}\mathbf{w} - \mathbf{P}_{\mathcal{S}_j}\mathbf{w}^{\text{target}}||^2,
\end{aligned}
\end{equation}
where $\bar{B}$ is the number of unique weights in the diagonal matrix $\mathbf{B}$, i.e. unique material parameters, and $S_j$ is the subspace only containing grid points with material parameters $B_j$.
Here, we need to assume that $\mathbf{B}$ is diagonal in $\mathcal{S}$, which holds true for the acoustic wave equation or the isotropic Maxwell's equation.
While this, in principle, allows for comparing two wave fields in the standard basis, we expect prohibitive sampling costs if the respective subspaces are small due to a loss of estimation globality (see \Cref{ss: limitation sources}).


\section{Sources}\label{S: general initial}

\subsection{Proof of \texorpdfstring{\Cref{TH:rot_covariant_initialization}}{Theorem \ref{TH:rot_covariant_initialization}}} \label{app: proof of lemma 4}

\paragraph{Rotational Covariance}
Since $\mathbf{w}^{sym}$ is rotationally covariant around $\mathbf{x}_0=0$, we have
\begin{equation}\label{eq:H_rot_covariance}
    \mathbf{R}(\mathbf{S}) \mathbf{w}^{sym}(\mathbf{x}) = \mathbf{w}^{sym}( \mathbf{S}\mathbf{x} ), \quad \forall \mathbf{x},\mathbf{S},
\end{equation}
where $\mathbf{S} \in \mathrm{SO}(D)$ uniquely decomposes $\mathbf{R}(\mathbf{S}) \in \mathrm{SO}(C)$.

\paragraph{Grid Construction}
We discretize the space around $\mathbf{x}_0$ into a grid as follows:
\begin{itemize}
    \item \textbf{Radial Coordinates}: The radial distance $r$ is discretized into $A$ divisions $r_a$, where $a = 1, 2, \dots, A$.
    \item \textbf{Angular Coordinates}: Each angular coordinate $\theta_i$ is discretized into $\Theta = A$ divisions, for $i = 1, 2, \dots, D-1$.
    \item \textbf{Grid Points}: Each grid point is specified by
    \begin{equation}
        \mathbf{x}_{a, \boldsymbol{\theta}} = \mathbf{x}_0 + r_a \, \tilde{\mathbf{u}}_{\boldsymbol{\theta}},
    \end{equation}
    where $\tilde{\mathbf{u}}_{\boldsymbol{\theta}}$ is a unit vector determined by the angular indices $\boldsymbol{\theta} = (\theta_1, \dots, \theta_{D-1})$.
    \item \textbf{Total Number of Points}: $N = A \Theta^{D-1} = A^D$ (for $D>1)$.
\end{itemize}

\paragraph{Efficient State Preparation Strategy}
We proceed with the following steps:

\subparagraph{Step 1: Compute Reference Field Values}
Select a reference direction, such as along the first coordinate axis, and compute $\mathbf{w}^{sym}( \mathbf{x}_{a, 0} )$ for $a = 1, 2, \dots, A$, where
\begin{equation}
    \mathbf{x}_{a, 0} = \mathbf{x}_0 + r_a \mathbf{e}_1,
\end{equation}
and $\mathbf{e}_1$ is the unit vector along the first coordinate axis. 
This requires $\mathcal{O}(A)=\mathcal{O}(N^{1/D})$ classical evaluations of $\mathbf{w}^{sym}$.

\subparagraph{Step 2: Prepare Reference Quantum State}
Prepare the quantum state
\begin{equation}\label{eq:phi_def}
    \ket{\phi} = \frac{1}{\sqrt{\mathcal{N}'}} \sum_{c=1}^C \sum_{a=1}^A w^{sym}_c\big( \mathbf{x}_{a, 0} \big) \ket{c} \otimes \ket{a},
\end{equation}
where $\mathcal{N}' = \sum_{c,a} \left| w^{sym}_c\big( \mathbf{x}_{a, 0} \big) \right|^2$. 
Since the number of terms is $CA$, this step requires $\tilde{\mathcal{O}}(CA)=\tilde{\mathcal{O}}(CN^{1/D})$ quantum gates \citep{gleinig2021efficient, malvetti2021quantum, de2022double, zhang2022quantum}.

\subparagraph{Step 3: Create Superposition Over Angular Indices}
Apply Hadamard gates to the angular index qubits to create  the equal superposition
\begin{equation}\label{eq:Phi_def}
    \ket{\Phi} = \ket{\phi} \otimes \frac{1}{\sqrt{\Theta^{D-1}}} \sum_{\boldsymbol{\theta}} \ket{\boldsymbol{\theta}}.
\end{equation}

\begin{figure*}
    \centering
    \begin{subfigure}[b]{0.9\textwidth}
        \centering
        \includegraphics[width=\linewidth]{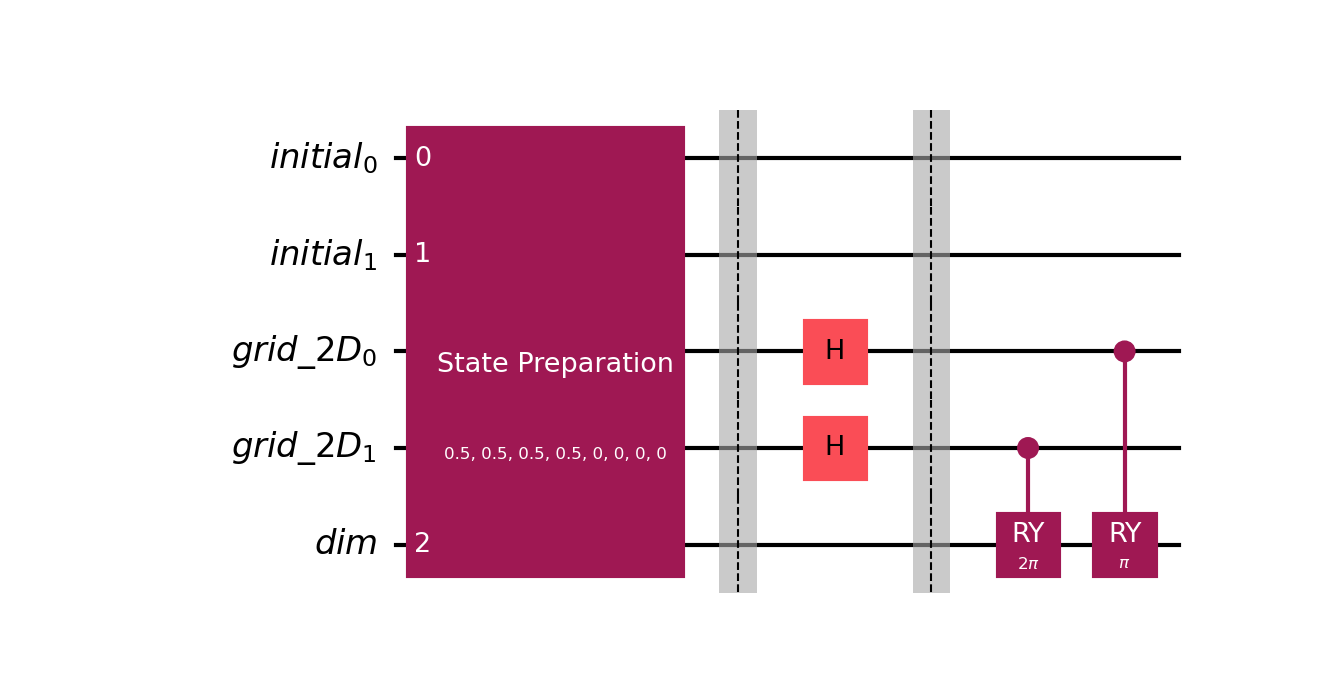}
        \caption{2D Initialization circuit. The full vector field is constructed from a single ray in $[1,0]$ direction.}
        \label{fig:circuit_2d}
    \end{subfigure}
    
    \begin{subfigure}[b]{0.9\textwidth}
        \centering
        \includegraphics[width=\linewidth]{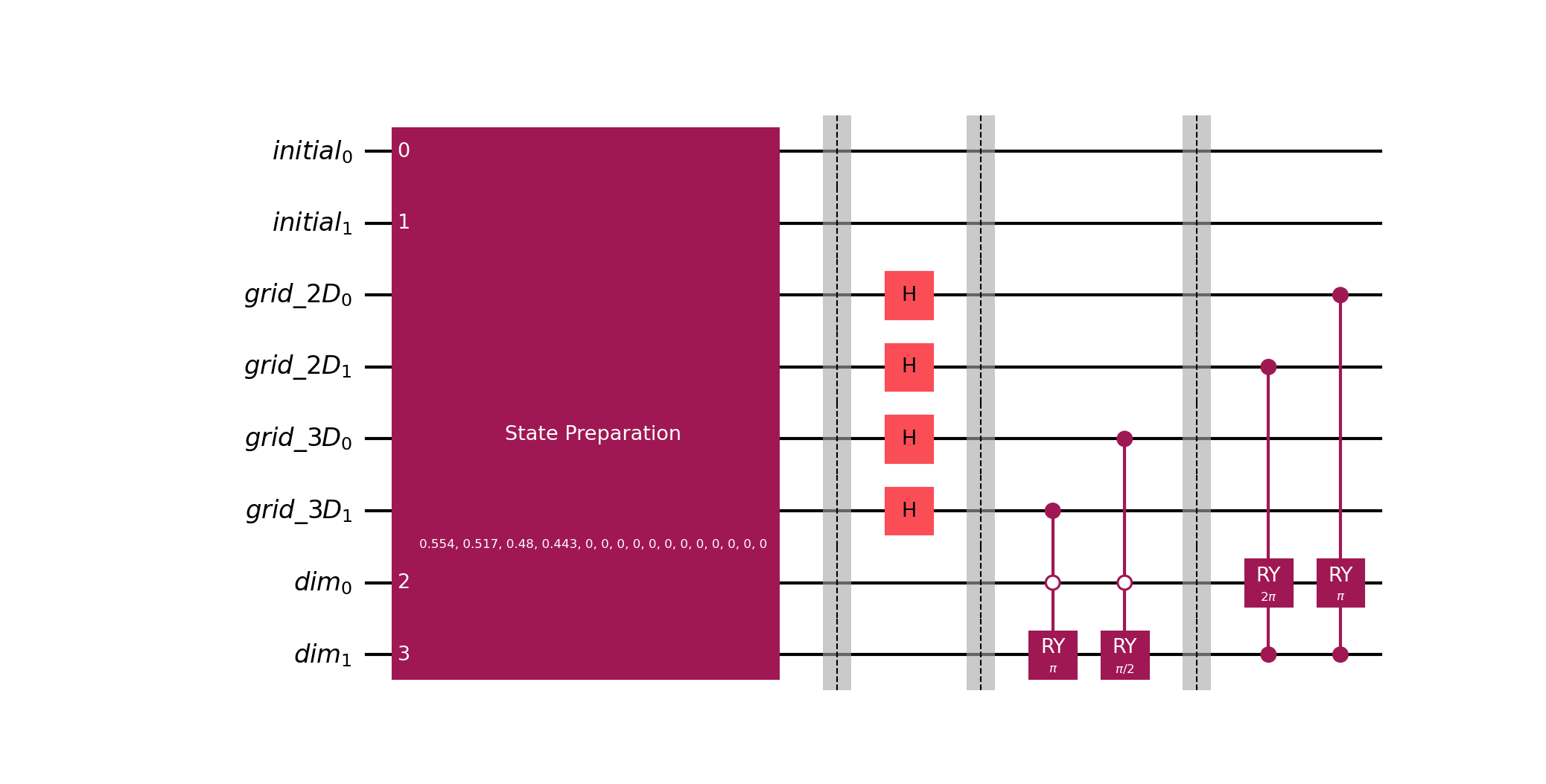}
        \caption{3D Initialization circuit. The full vector field is constructed from a single ray in $[1,0,0]$ direction.}
        \label{fig:circuit_3d}
    \end{subfigure}
    \caption{Quantum circuits for efficient vector field initialization in 2D and 3D. The initial ray is prepared in the quantum register $initial$ with vector components saved in the register $dim$. The registers $grid_{2D}$ (and $grid_{3D}$) are transformed into a state of equal superposition. Afterwards (multi-)controlled RY rotations are applied with decreasing angles, which gives the rotationally symmetric vector field in the combined registers.}
    \label{fig:init_circuits}
\end{figure*}

\subparagraph{Step 4: Apply Controlled Rotations}
Our goal is to rotate the field components in $\ket{\Phi}$ to their correct orientations corresponding to each angular index $\boldsymbol{\theta}$.
This is achieved by applying a sequence of controlled rotations, where the rotation angles are determined by the angular indices $\boldsymbol{\theta}$, and the rotations are performed in the planes spanned by pairs of coordinate axes.
Specifically, we proceed as follows:

\begin{enumerate}
    \item \textbf{Binary Representation of Angular Indices}:
    Each angular coordinate $\theta_i$ is discretized into $\Theta$ divisions and represented using $k = \log_2 (\Theta)$ qubits. 
    For $\Theta = 2^k$, the total number of angular qubits is $(D - 1)k$.
    
    \item \textbf{Rotation Decomposition}: 
    To rotate the initial vector $\mathbf{w}^{sym}( \mathbf{x}_{a, 0} )$ to all possible directions $\tilde{\mathbf{u}}_{\boldsymbol{\theta}}$, we decompose the rotation $\mathbf{R}_{\boldsymbol{\theta}} \in \mathrm{SO}(C)$ into a sequence of $C - 1$ rotations.
    Specifically, we perform rotations in the $(i,i+1)$-planes for $i = 1, \dots, C-1$:
    \begin{equation}
        \mathbf{R}_{\boldsymbol{\theta}} = \prod_{i=1}^{C-1} R_{i,i+1}(\theta_i),
    \end{equation}
    where $R_{i,i+1}(\theta_i)$ is a rotation in the plane spanned by the $i^{\text{th}}$ and $(i+1)^{\text{th}}$ coordinates, parameterized by the angle $\theta_i$ determined by $\boldsymbol{\theta}$.
    
    \item \textbf{Sequence of Controlled Rotations}:
    For each plane $(i,i+1)$, we apply controlled rotation gates $\mathrm{CR}_{i,i+1}(\theta_i^{(m)})$ in the real plane, where $m$ indexes the bits in the binary representation of the angular coordinate $\theta_i$. The rotation angles $\theta_i^{(m)}$ are defined as:
    \begin{equation}
        \theta_i^{(m)} = \frac{\pi}{2^{m}},
    \end{equation}
    so that the cumulative rotation angle $\theta_i = \sum_{m=1}^k b_{i,m} \theta_i^{(m)}$, where $b_{i,m} \in \{0,1\}$ is the $m^{\text{th}}$ bit of $\theta_i$.
    
    The controlled rotations are defined as:
    \begin{equation}
    \begin{aligned}
        \mathrm{CR}_{i,i+1}(\theta_i^{(m)}) &= \ket{0}\bra{0}_{\theta_{i,m}} \otimes I^{\otimes C} \\&+ \ket{1}\bra{1}_{\theta_{i,m}} \otimes R_{i,i+1}(\theta_i^{(m)}),
        \end{aligned}
    \end{equation}
    where $\theta_{i,m}$ is the $m^{\text{th}}$ qubit of the angular index corresponding to the rotation in the $(i,i+1)$-plane.
    As the control qubits are in a state of equal superposition, all binary combinations of rotation angles are realized.
    
    \item \textbf{Control and Target Qubits}:
    The control qubits are the angular index qubits $\ket{\boldsymbol{\theta}}$, and the target qubits are the field component qubits $\ket{c}$.
\end{enumerate}

\paragraph{Gate Complexity}
The total number of controlled rotation gates required is $\mathcal{O}\left( C k \right)$, since for each of the $C - 1$ planes, we apply $k$ controlled rotations corresponding to the bits of the angular indices. 
Since $k = \log (\Theta) = \log (A)$, the total number of controlled rotations is $\mathcal{O}\left( C \log (A) \right) = \mathcal{O}\left( \frac{C}{D} \log (N) \right)$. 
However, for fixed dimensions $C$ and $D$ the gate complexity of implementing these rotations \citep{vale2023circuit} is dominated by the complexity of initializing $\ket{\phi}$, which requires $\tilde{\mathcal{O}}(CA)=\tilde{\mathcal{O}}(CN^{1/D})$ quantum gates \citep{gleinig2021efficient, malvetti2021quantum, de2022double, zhang2022quantum}.

\paragraph{Final State Preparation}
After applying the sequence of controlled rotations, the state becomes
\begin{equation}
\begin{aligned}
    U_{\mathrm{rot}} \ket{\Phi} =& \frac{1}{\sqrt{\mathcal{N}' \Theta^{D-1}}} \sum_{c'=1}^{C} \sum_{a=1}^{A} \sum_{\boldsymbol{\theta}} \left( \prod_{i=1}^{C-1} R_{i,i+1}(\theta_i) \right)_{c c'} \\&w^{sym}_{c'}(\mathbf{x}_{a, 0}) \ket{c} \otimes \ket{a} \otimes \ket{\boldsymbol{\theta}} \nonumber \\
    =& \frac{1}{\sqrt{\mathcal{N}' \Theta^{D-1}}} \sum_{c=1}^{C} \sum_{a=1}^{A} \sum_{\boldsymbol{\theta}} \\ &w^{sym}_c(\mathbf{x}_{a, \boldsymbol{\theta}}) \ket{c} \otimes \ket{a} \otimes \ket{\boldsymbol{\theta}},
\end{aligned}
\end{equation}
where we used the fact that $w^{sym}_c(\mathbf{x}_{a, \boldsymbol{\theta}}) = \left( \mathbf{R}_{\boldsymbol{\theta}} \mathbf{w}^{sym}(\mathbf{x}_{a, 0}) \right)_c$ due to the rotational covariance property.
This is the desired state, $\ket{\psi}$, which proves \Cref{TH:rot_covariant_initialization}.

\paragraph{Practical Considerations}
Applying $U_{\mathrm{rot}}$ requires implementing rotations with angles that decrease exponentially with the number of qubits used to represent the angular coordinates. 
In practice, this poses challenges due to the finite precision of quantum gates. 
Specifically, the minimal rotation angle scales inversely with the state size $A$, resulting in a gate precision requirement that scales linearly with $A$. 
However, the precision requirement remains independent of $C$ and $D$. 
This independence allows for the implementation of a larger total number of grid points $N = A^D$ without necessitating additional precision.

\paragraph{Numerical Implementation}
We provide numerical implementations for the 2D and the 3D case in the supplements to the publication. 
\Cref{fig:init_circuits} shows the quantum circuits for the 2D and 3D case that construct the radially symmetric vector field from a single ray.

\subsection{Windowing}\label{ap:windowing}

While different choices of windows are possible, in our implementation we use a symmetric double sigmoid function
\begin{equation}
    W_{j}(t) = \frac{1}{1 + e^{-z \cdot t}} - \frac{1}{1 + e^{-z \cdot (t - (\tau_{{j+1}} - \tau_{j}))}},
\end{equation}
where $z$ defines the steepness of the sigmoid functions.
This function realizes overlapping windows that satisfy Eq. \eqref{EQ: source_windows} with adaptable accuracy.
In contrast to, e.g., box window functions ($z \rightarrow \infty$), sigmoid functions lead to smooth initial conditions. 
However, because the sigmoid functions do only converge to $0$ at the window boundary in the limit where $z \rightarrow \infty$, choosing a finite $z$ provides a trade-off between source approximation accuracy and smoothness of the sections.

Note, that if the windows are chosen to be adjacent block functions with no overlap ($z = \infty$), their appliance to a source time function $f(t)$ does not increase the number of nonzero values, which need to be initialized. 
In this case we would obtain the same initialization complexity as for a non-windowed state.
However, as we reduce the steepness of the windows, they begin to overlap.
This increases the amount of initialized values and is, in the worst case, a multiplicative complexity increase with the number of windows $J$.


\section{Finite Difference Discretization for the Acoustic Wave equation}\label{A: FD discretization}

As an example, in this Appendix, we discuss the discretization of the acoustic wave equation using the FD method on a uniform, rectangular, and staggered grid \citep{virieux1984}.
We show how to build a pair of discrete gradient and divergence operators to ensure that $\hat{\mathbf{A}}^T = -\hat{\mathbf{A}}$ is anti-Hermitian (see \Cref{s: General wave eq}).
We also discuss the implementation of Dirichlet and Neumann BCs on arbitrary boundaries of the domain by employing the constraints reduction introduced in \Cref{s: Constraints}.

Recall the acoustic wave equation in pressure-velocity formulation introduced in \Cref{s: acoustic wave eq}
\begin{equation}
    \begin{aligned}
        \hat{\mathbf{B}}_{\text{acoustic}} \frac{\partial \hat{\mathbf{w}}_{\text{acoustic}}}{\partial t} &= \hat{\mathbf{A}}_{\text{acoustic}} \hat{\mathbf{w}}_{\text{acoustic}} \\
        \begin{bmatrix}
            \frac{1}{\rho(\mathbf{x})c^2(\mathbf{x})} & \mathbf{0}_{1 \times D} \\
            \mathbf{0}_{D \times 1} & \rho(\mathbf{x})\mathds{1}_{D \times D}
        \end{bmatrix} \frac{\partial}{\partial t} \begin{bmatrix}
            u\\
            \mathbf{v}
        \end{bmatrix} &= \begin{bmatrix}
            0 & -\nabla \cdot \\
            -\nabla & \mathbf{0}_{D \times D}
        \end{bmatrix} \begin{bmatrix}
            u\\
            \mathbf{v}
        \end{bmatrix},
    \end{aligned}
\end{equation}
The FD method aims to construct the mapping $\nabla \rightarrow \mathbf{G} \in \mathbb{R}^{N_v \times N_u}$ and $\nabla \cdot \rightarrow \mathbf{D} \in \mathbb{R}^{N_u \times N_v}$ by defining the discrete versions of the gradient and divergence operators that will be applied to the discretized pressure fields $\mathbf{u}$ and $\mathbf{v}$ respectively.
The gradient and divergence operators are constructed independently of the BCs and follow from the definition of the FD operator used to approximate derivatives in space.
Without loss of generality, we use the first-order accurate central-FD stencil to compute derivatives in the $i^{\text{th}}$ dimension
\begin{equation}
    \frac{\partial f}{\partial x_i}(\mathbf{x}) \approx \frac{f(\mathbf{x} + \frac{\Delta x_i}{2}\mathbf{e}_i) - f(\mathbf{x} - \frac{\Delta x_i}{2}\mathbf{e}_i)}{\Delta x_i},
\end{equation}
where $\mathbf{e}_i$ is the canonical vector and $\Delta x_i$ is the grid step size in the $i^{\text{th}}$ direction.
The gradient and divergence operators are then discretized accordingly, such that
\begin{equation}
    \begin{aligned}
        (\nabla u(\mathbf{x}, t))_i &= \frac{\partial u}{\partial x_i}(\mathbf{x}, t) \\ &\approx \frac{u(\mathbf{x} + \frac{\Delta x_i}{2}\mathbf{e}_i, t) - u(\mathbf{x} - \frac{\Delta x_i}{2}\mathbf{e}_i, t)}{\Delta x_i},
    \end{aligned}
\end{equation}
and
\begin{equation}
    \begin{aligned}
        \nabla \cdot \mathbf{v}(\mathbf{x}, t) &= \sum_{i \in D} \frac{\partial v_i}{\partial x_i}(\mathbf{x}, t) \\ &\approx \sum_{i \in D} \frac{v_i(\mathbf{x} + \frac{\Delta x_i}{2}\mathbf{e}_i, t) - v_i(\mathbf{x} - \frac{\Delta x_i}{2}\mathbf{e}_i, t)}{\Delta x_i}.        
    \end{aligned}
\end{equation}

We now need to replace the continuous fields $u(\mathbf{x}, t)\colon \mathbb{R}^D \times [0, T] \to \mathbb{R}, \mathbf{v}(\mathbf{x}, t)\colon \mathbb{R}^D \times [0, T]$ with their discretized counterparts $\mathbf{u}(t)\colon [0, T] \to \mathbb{R}^{N_u}, \mathbf{v}(t)\colon [0, T] \to \mathbb{R}^{N_v}$ by employing a \emph{staggered} grid where the velocities are staggered in between pressure points.
We introduce the set of numbers of grid points $\mathcal{N} = \{N_i \mid i \in \{1, \ldots, D\}, N_i = \text{\# of grid points in the $i^{\text{th}}$ direction}\}$.
Since the velocity is a vector field, we need to distinguish between velocities in different dimensions: we will use $\mathbf{v}_i$ from now on to denote the velocity acting in the $i^{\text{th}}$ direction, such that $\mathbf{v}_i(t)\colon [0,T] \to \mathbb{R}^{N_{v_i}}$ where $N_{v_i} = (N_i - 1) \prod_{j \neq i} N_j$. Notice that this discretization requires the discretized velocity field in the $i^{\text{th}}$ direction to have one less DOF in that direction.
The pressure field $\mathbf{u}$ is discretized in $N_u = \prod_i N_i$ DOFs and the full velocity field $\mathbf{v} = [\mathbf{v}_1, \ldots, \mathbf{v}_D]$ in $N_v = \sum_i ((N_i - 1) \prod_{j \neq i} N_j)$ DOFs.


\subsection{2D Staggered Grid}

\begin{figure}
\centering
\begin{tikzpicture}

\tikzstyle{dot}=[circle, fill, inner sep=0pt, minimum size=5pt]
\tikzstyle{redcross}=[cross out, draw=red, minimum size=5pt, inner sep=0pt, outer sep=0pt]
\tikzstyle{bluecross}=[cross out, draw=blue, minimum size=5pt, inner sep=0pt, outer sep=0pt]

\coordinate (O) at (-0.5,3.5);
\draw[thick,->] (O) -- ++(.5,0) node[anchor=south east]{$x$};
\draw[thick,->] (O) -- ++(0,-.5) node[anchor=south east]{$y$};

\foreach \x in {0, ..., 3}
{
    \draw (\x,0) -- (\x,3);
}
\foreach \y in {0, ..., 3}
{
    \draw (0,\y) -- (3,\y);
}

\foreach \x in {0, ..., 3}
{
    \foreach \y in {0, ..., 3}
    {
        \draw (\x, \y) node[dot] {};
    }
}

\foreach \x in {0.5, ..., 2.5}
{
    \foreach \y in {0, ..., 3}
    {
        \draw (\x, \y) node[redcross] {};
    }
}

\foreach \x in {0, ..., 3}
{
    \foreach \y in {0.5, ..., 2.5}
    {
        \draw (\x, \y) node[bluecross] {};
    }
}

\draw [decorate,decoration={brace,amplitude=5pt,mirror,raise=0.2cm}]
  (2,0) -- (3,0) node[midway,yshift=-0.6cm]{$\Delta x$};

\draw [decorate,decoration={brace,amplitude=5pt,mirror,raise=0.2cm}]
  (3,0) -- (3,1) node[midway,xshift=0.65cm]{$\Delta y$};

\matrix [draw,below left] at (-1,2.5) {
  \node [dot,label=right:{$\mathbf{u}$}] {}; \\
  \node [redcross,label=right:{$\mathbf{v}_x$}] {}; \\
  \node [bluecross,label=right:{$\mathbf{v}_y$}] {}; \\
};

\end{tikzpicture}
\caption{Staggered grid visualization for 2D pressure-velocity formulation.}
\label{fig: staggered-grid}
\end{figure}
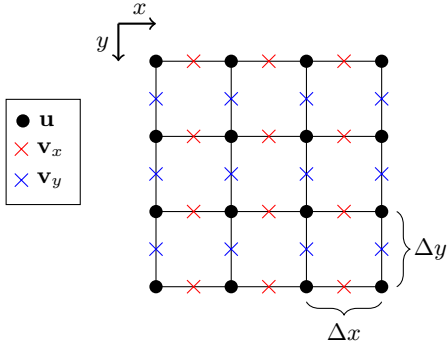

We now demonstrate how the staggered grid is constructed in practice for the $D = 2$ case, though the same procedure can be extended to the general $D$-dimensional case.

In 2D, we have three fields: $\mathbf{u}$, $\mathbf{v}_x$, and $\mathbf{v}_y$, discretized on different grid points within a rectangular domain $\Omega = [x_0, x_1] \times [y_0, y_1]$. \Cref{fig: staggered-grid} provides a visual representation of the staggered grid used. Further details on its construction are described below.

We introduce the set of all grid points for the pressure field $\mathcal{X}_u$ as follows
\begin{equation}
\begin{aligned}
\mathcal{X}_u =& \{(x, y) \in \Omega \mid \\ &x = x_0 + (i - 1) \Delta x, \quad y = y_0 + (j - 1) \Delta y, \\ &\forall i \in \{1, \ldots, N_x\}, \quad \forall j \in \{1, \ldots, N_y\} \},
\end{aligned}
\end{equation}
where $\Delta x = (x_1 - x_0) / (N_x - 1)$, $\Delta y = (y_1 - y_0) / (N_y - 1)$, $N_x = N_1$, and $N_y = N_2$. By staggering these points, we define the set of all grid points for the velocity field in the $x$-direction $\mathcal{X}_{v_x}$ and the $y$-direction $\mathcal{X}_{v_y}$ as
\begin{equation}
\mathcal{X}_{v_x} = \{ (x + \Delta x / 2, y) \in \Omega \mid (x, y) \in \mathcal{X}_u \},
\end{equation}
and
\begin{equation}
\mathcal{X}_{v_y} = \{ (x, y + \Delta y / 2) \in \Omega \mid (x, y) \in \mathcal{X}_u \}.
\end{equation}
Notice that $|\mathcal{X}_u| = N_u = N_x N_y$, $|\mathcal{X}_{v_x}| = N_{v_x} = (N_x - 1) N_y$, and $|\mathcal{X}_{v_y}| = N_{v_y} = N_x (N_y - 1)$.

The mappings from the linear indices of the discretized fields to the grid points are given by $\Phi_u = \Phi_u(\ell) \colon \{1, \ldots, N_u\} \to \mathcal{X}_u$, $\Phi_{v_x} = \Phi_{v_x}(k) \colon \{1, \ldots, N_{v_x}\} \to \mathcal{X}_{v_x}$, and $\Phi_{v_y} = \Phi_{v_y}(k) \colon \{N_{v_x} + 1, \ldots, N_{v_x} + N_{v_y}\} \to \mathcal{X}_{v_y}$. These mappings are constructed such that each index $k$ corresponds to one and only one grid point. For example, to map the pressure points, we introduce the Cartesian indices $(i, j)$ referring to the grid point at some position $(x_i, y_j) \in \mathcal{X}_u$, and then linearize the Cartesian index using $k = i + (j-1) N_x$, so that $\Phi_u(\ell) = (x_i, y_j)$. A similar linearization can be performed for the velocity mappings $\Phi_{v_x}$ and $\Phi_{v_y}$.

We now have all the necessary components to explicitly formulate the semi-discretization of pressure and velocities
\begin{equation}
\begin{aligned}
    [\mathbf{u}]_\ell(t) =& u(\Phi_u(\ell), t),\\ 
    &\forall \ell \in \{1, \ldots, N_u\},\\
    [\mathbf{v}_x]_k(t) =& v_x(\Phi_{v_x}(k), t), \\
    &\forall k \in \{1, \ldots, N_{v_x}\},\\
    [\mathbf{v}_y]_{k - N_{v_x}}(t) =& v_y(\Phi_{v_y}(k), t),\\
    &\forall k \in \{N_{v_x} + 1, \ldots, N_{v_x} + N_{v_y}\}.
\end{aligned}
\end{equation}


\subsection{Discrete Gradient and Divergence Operators, and their Anti-Transpose Relationship}

We construct the discretized gradient and divergence operators in 2D based on the first-order accurate central FD scheme introduced earlier. Generalization to FD schemes with higher orders of accuracy is straightforward since the symmetry of the FD operators is retained. Therefore, the results from this section remain unchanged when using higher-order schemes.

Let $(\mathbf{G})_{k, \cdot}$ denote the $k^{\text{th}}$ row of the gradient operator $\mathbf{G}$. When taking the dot product of this row with the discretized pressure field $\mathbf{u}$, the FD approximation of a component of the gradient $\nabla u$ is computed for the semi-discrete equation concerning the discretized velocity field component $[\mathbf{v}]_k$. If $k$ refers to a discretized velocity component in the $x$-direction (i.e., $k \in \{1, \ldots, N_{v_x}\}$), the $x$-component of the gradient is computed; otherwise, the $y$-component is computed.

This can be expressed by the following FD approximation
\begin{equation}
\begin{aligned}
    (\nabla u)_x(\Phi_{v_x}(k)) &= \frac{\partial u}{\partial x}(\Phi_{v_x}(k)) \\
    &\approx \frac{u(\Phi_{v_x}(k) + \frac{\Delta x}{2} \mathbf{e}_x)}{\Delta x} \\
    & - \frac{u(\Phi_{v_x}(k) - \frac{\Delta x}{2} \mathbf{e}_x)}{\Delta x} \\
    &\approx \frac{u(\Phi_{u}(\ell_{+})) - u(\Phi_{u}(\ell_{-}))}{\Delta x} \\
    &\approx \frac{[\mathbf{u}]_{\ell_{+}} - [\mathbf{u}]_{\ell_{-}}}{\Delta x},
\end{aligned}
\end{equation}
for all $k \in \{1, \ldots, N_{v_x}\}$. Similarly, for the $y$-component
\begin{equation}
\begin{aligned}
    (\nabla u)_y(\Phi_{v_y}(k)) &= \frac{\partial u}{\partial y}(\Phi_{v_y}(k)) \\
    &\approx \frac{u(\Phi_{v_y}(k) + \frac{\Delta y}{2} \mathbf{e}_y)}{\Delta y} \\
    & - \frac{u(\Phi_{v_y}(k) - \frac{\Delta y}{2} \mathbf{e}_y)}{\Delta y} \\
    &\approx \frac{u(\Phi_{u}(\ell_{+})) - u(\Phi_{u}(\ell_{-}))}{\Delta y} \\
    &\approx \frac{[\mathbf{u}]_{\ell_{+}} - [\mathbf{u}]_{\ell_{-}}}{\Delta y},
\end{aligned}
\end{equation}
for all $k \in \{N_{v_x} + 1, \ldots, N_{v_x} + N_{v_y}\}$.

For each equation $k \in \{1, \ldots, N_{v_x}\}$ (respectively, $k \in \{N_{v_x} + 1, \ldots, N_{v_x} + N_{v_y}\}$), there will be only two non-zero coefficients in the row $(\mathbf{G})_{k, \cdot}$, specifically those associated with discretized pressure components matching $\Phi_u(\ell_{\pm}) = \Phi_{v_x}(k) \pm \frac{\Delta x}{2} \mathbf{e}_x$ (respectively, $\Phi_u(\ell_{\pm}) = \Phi_{v_y}(k) \pm \frac{\Delta y}{2} \mathbf{e}_y$). Here, $\ell_{\pm}$ represents the index of the discretized pressure component $[\mathbf{u}]_{\ell_{\pm}}$. These coefficients are $\pm 1 / \Delta x$ (respectively, $\pm 1 / \Delta y$).

With these observations, we can explicitly write each component of the gradient operator as follows
\begin{equation}
\begin{aligned}
    \nabla \to \mathbf{G} &= (\mathbf{G})_{k, \ell} \\&= \begin{cases}
        \pm 1 / \Delta x &  \begin{aligned}
            &\text{if }k \in \{1, \ldots, N_{v_x}\}\\
            &\text{and }\Phi_u(\ell) = \Phi_{v_x}(k) \pm \frac{\Delta x}{2} \mathbf{e}_x,
        \end{aligned}\\
        \pm 1 / \Delta y &  \begin{aligned}
            &\text{if }k \in \{N_{v_x} + 1, \ldots, N_{v_x} + N_{v_y}\} \\
            &\text{and } \Phi_u(\ell) = \Phi_{v_y}(k) \pm \frac{\Delta y}{2} \mathbf{e}_y,
        \end{aligned}\\
        0                &\text{otherwise},
    \end{cases}
\end{aligned}
\end{equation}
for all $k \in \{1, \ldots, N_v\}, \ell \in \{1, \ldots, N_u\}$.

We can apply a similar procedure for the divergence operator using the FD approximation
\begin{equation}
\begin{aligned}
    \nabla \cdot \mathbf{v}(\Phi_u(\ell))
    &= \frac{\partial v_x}{\partial x}(\Phi_u(\ell)) + \frac{\partial v_y}{\partial y}(\Phi_u(\ell)) \\
    &\approx \frac{v_x(\Phi_{u}(\ell) + \frac{\Delta x}{2} \mathbf{e}_x) - v_x(\Phi_{u}(\ell) - \frac{\Delta x}{2} \mathbf{e}_x)}{\Delta x} \\
    &+       \frac{v_y(\Phi_{u}(\ell) + \frac{\Delta y}{2} \mathbf{e}_y) - v_y(\Phi_{u}(\ell) - \frac{\Delta y}{2} \mathbf{e}_y)}{\Delta y} \\
    &\approx \frac{v_x(\Phi_{v_x}(k_{x+})) - v_x(\Phi_{v_x}(k_{x-}))}{\Delta x} \\
    &+       \frac{v_y(\Phi_{v_y}(k_{y+})) - v_y(\Phi_{v_y}(k_{y-}))}{\Delta y} \\
    &\approx \frac{[\mathbf{v}_x]_{k_{x+}} - [\mathbf{v}_x]_{k_{x-}}}{\Delta x} \\
    &+       \frac{[\mathbf{v}_y]_{k_{y+}} - [\mathbf{v}_y]_{k_{y-}}}{\Delta y},
\end{aligned}
\end{equation}
resulting in
\begin{equation}
\begin{aligned}
    \nabla \cdot \to \mathbf{D} &= (\mathbf{D})_{\ell, k} \\&= \begin{cases}
        \pm 1 / \Delta x &  \begin{aligned} 
        &\text{if }k \in \{1, \ldots, N_{v_x}\} \\ 
        &\text{and }\Phi_{v_x}(k) = \Phi_{u}(\ell) \pm \frac{\Delta x}{2} \mathbf{e}_x,
        \end{aligned}\\
        \pm 1 / \Delta y &  \begin{aligned}
        &\text{if }k \in \{N_{v_x} + 1, \ldots,N_{v_x} + N_{v_y}\}\\
        &\text{and }\Phi_{v_y}(k) = \Phi_{u}(\ell) \pm \frac{\Delta y}{2} \mathbf{e}_y,\end{aligned}\\
        0                &\text{otherwise}.
    \end{cases}
\end{aligned}
\end{equation}

The anti-transpose property $\mathbf{G} = -\mathbf{D}^T$ can be verified by showing that $(\mathbf{G})_{k, \ell} = -(\mathbf{D})_{\ell, k}$ for all $\ell, k$. This is easily checked by observing that the only difference in the definition of the operators $\mathbf{G}$ and $\mathbf{D}$ is the position of the mappings $\Phi_u$, $\Phi_{v_x}$ (respectively, $\Phi_u$, $\Phi_{v_y}$) in the conditions $\Phi_u(\ell) = \Phi_{v_x}(k) \pm \frac{\Delta x}{2} \mathbf{e}_x$ and $\Phi_{v_x}(k) = \Phi_{u}(\ell) \pm \frac{\Delta x}{2} \mathbf{e}_x$ (respectively, the conditions $\Phi_u(\ell) = \Phi_{v_y}(k) \pm \frac{\Delta y}{2} \mathbf{e}_y$ and $\Phi_{v_y}(k) = \Phi_{u}(\ell) \pm \frac{\Delta y}{2} \mathbf{e}_y$), which is the cause of sign flip of the components.
This shows that the FD discretization using a uniform, rectangular, staggered grid produces a pair of gradient and divergence operators leading to an anti-Hermitian matrix $\mathbf{A} = \begin{bmatrix}
        \mathbf{0}_{N_u \times N_u} & -\mathbf{D} \\
        -\mathbf{G} & \mathbf{0}_{D \times D}
    \end{bmatrix}$.


\subsection{Implementation of Boundary Conditions} \label{A: bcs implementation}

\begin{figure*}
\centering

\begin{subfigure}[b]{0.49\textwidth}
\centering

\begin{tikzpicture}

\tikzstyle{dot}=[circle, fill, inner sep=0pt, minimum size=5pt]
\tikzstyle{reddot}=[circle, draw=red, inner sep=0pt, minimum size=5pt]
\tikzstyle{bluedot}=[circle, draw=blue, inner sep=0pt, minimum size=5pt]
\tikzstyle{redcross}=[cross out, draw=red, minimum size=5pt, inner sep=0pt, outer sep=0pt]
\tikzstyle{bluecross}=[cross out, draw=blue, minimum size=5pt, inner sep=0pt, outer sep=0pt]

\coordinate (O) at (-1,4);
\draw[thick,->] (O) -- ++(.5,0) node[anchor=south east]{$x$};
\draw[thick,->] (O) -- ++(0,-.5) node[anchor=south east]{$y$};

\foreach \x in {-0.5, 3.5}
{
    \draw[densely dashed] (\x, -0.5) -- (\x, 3.5);
}
\foreach \y in {-0.5, 3.5}
{
    \draw[densely dashed] (-0.5, \y) -- (3.5, \y);
}

\foreach \x in {0, ..., 3}
{
    \foreach \y in {-0.5, 3.5}
    {
        \draw (\x, \y) node[bluedot] {};
    }
}
\foreach \y in {0, ..., 3}
{
    \foreach \x in {-0.5, 3.5}
    {
        \draw (\x, \y) node[reddot] {};
    }
}

\foreach \x in {0, ..., 3}
{
    \draw (\x,0) -- (\x,3);
}
\foreach \y in {0, ..., 3}
{
    \draw (0,\y) -- (3,\y);
}

\foreach \x in {0, ..., 3}
{
    \foreach \y in {0, ..., 3}
    {
        \draw (\x, \y) node[dot] {};
    }
}

\foreach \x in {0.5, ..., 2.5}
{
    \foreach \y in {0, ..., 3}
    {
        \draw (\x, \y) node[redcross] {};
    }
}

\foreach \x in {0, ..., 3}
{
    \foreach \y in {0.5, ..., 2.5}
    {
        \draw (\x, \y) node[bluecross] {};
    }
}

\matrix [draw,below left] at (-1,2.5) {
  \node [dot,label=right:{$\mathbf{u}$}] {}; \\
  \node [redcross,label=right:{$\mathbf{v}_x$}] {}; \\
  \node [bluecross,label=right:{$\mathbf{v}_y$}] {}; \\
  \node [reddot,label=right:{$\mathbf{v}_x = 0$}] {}; \\
  \node [bluedot,label=right:{$\mathbf{v}_y = 0$}] {}; \\
  \begin{scope}
    \node (B) at (1,0) {$\partial \Omega$};
    \draw[densely dashed] (B) -- ++(-0.8, 0);
  \end{scope} \\
};

\end{tikzpicture}

\caption{2D staggered grid with Neumann BCs on pressure for all $\partial \Omega$.}
\label{fig: sg-neumann}
\end{subfigure}
\hfill
\begin{subfigure}[b]{0.49\textwidth}
\centering

\begin{tikzpicture}

\tikzstyle{dot}=[circle, fill, inner sep=0pt, minimum size=5pt]
\tikzstyle{blackdot}=[circle, draw=black, inner sep=0pt, minimum size=5pt]
\tikzstyle{reddot}=[circle, draw=red, inner sep=0pt, minimum size=5pt]
\tikzstyle{bluedot}=[circle, draw=blue, inner sep=0pt, minimum size=5pt]
\tikzstyle{redcross}=[cross out, draw=red, minimum size=5pt, inner sep=0pt, outer sep=0pt]
\tikzstyle{bluecross}=[cross out, draw=blue, minimum size=5pt, inner sep=0pt, outer sep=0pt]

\coordinate (O) at (-0.5,3.5);
\draw[thick,->] (O) -- ++(.5,0) node[anchor=south east]{$x$};
\draw[thick,->] (O) -- ++(0,-.5) node[anchor=south east]{$y$};

\foreach \x in {0, 3}
{
    \draw[densely dashed] (\x, -0) -- (\x, 3);
}
\foreach \y in {0, 3}
{
    \draw[densely dashed] (0, \y) -- (3, \y);
}

\foreach \x in {0, ..., 3}
{
    \foreach \y in {0, 3}
    {
        \draw (\x, \y) node[blackdot] {};
    }
}
\foreach \y in {0, ..., 3}
{
    \foreach \x in {0, 3}
    {
        \draw (\x, \y) node[blackdot] {};
    }
}

\foreach \x in {1, 2}
{
    \draw (\x,0) -- (\x,3);
}
\foreach \y in {1, 2}
{
    \draw (0,\y) -- (3,\y);
}

\foreach \x in {1, 2}
{
    \foreach \y in {1, 2}
    {
        \draw (\x, \y) node[dot] {};
    }
}

\foreach \x in {0.5, ..., 2.5}
{
    \foreach \y in {0, ..., 3}
    {
        \draw (\x, \y) node[redcross] {};
    }
}

\foreach \x in {0, ..., 3}
{
    \foreach \y in {0.5, ..., 2.5}
    {
        \draw (\x, \y) node[bluecross] {};
    }
}

\matrix [draw, below left] at (-1,2.5) {
  \node [dot,label=right:{$\mathbf{u}$}] {}; \\
  \node [redcross,label=right:{$\mathbf{v}_x$}] {}; \\
  \node [bluecross,label=right:{$\mathbf{v}_y$}] {}; \\
  \node [blackdot,label=right:{$\mathbf{u} = 0$}] {}; \\
  \begin{scope}
    \node (B) at (1,0) {$\partial \Omega$};
    \draw[densely dashed] (B) -- ++(-0.8, 0);
  \end{scope} \\
};

\end{tikzpicture}

\caption{2D staggered grid with Dirichlet BCs on pressure for all $\partial \Omega$.}
\label{fig: sg-dirichlet}
\end{subfigure}
\hfill
\begin{subfigure}[b]{0.49\textwidth}
\centering

\begin{tikzpicture}

\tikzstyle{dot}=[circle, fill, inner sep=0pt, minimum size=5pt]
\tikzstyle{blackdot}=[circle, draw=black, inner sep=0pt, minimum size=5pt]
\tikzstyle{reddot}=[circle, draw=red, inner sep=0pt, minimum size=5pt]
\tikzstyle{bluedot}=[circle, draw=blue, inner sep=0pt, minimum size=5pt]
\tikzstyle{redcross}=[cross out, draw=red, minimum size=5pt, inner sep=0pt, outer sep=0pt]
\tikzstyle{bluecross}=[cross out, draw=blue, minimum size=5pt, inner sep=0pt, outer sep=0pt]

\coordinate (O) at (-1,3.5);
\draw[thick,->] (O) -- ++(.5,0) node[anchor=south east]{$x$};
\draw[thick,->] (O) -- ++(0,-.5) node[anchor=south east]{$y$};

\foreach \x in {-0.5, 3}
{
    \draw[densely dashed] (\x, -0.5) -- (\x, 3);
}
\foreach \y in {-0.5, 3}
{
    \draw[densely dashed] (-0.5, \y) -- (3, \y);
}

\foreach \x in {0, ..., 3}
{
    \foreach \y in {-0.5}
    {
        \draw (\x, \y) node[bluedot] {};
    }
}
\foreach \y in {0, ..., 3}
{
    \foreach \x in {-0.5}
    {
        \draw (\x, \y) node[reddot] {};
    }
}

\foreach \x in {0, ..., 2}
{
    \draw (\x,0) -- (\x,3);
}
\foreach \y in {0, ..., 2}
{
    \draw (0,\y) -- (3,\y);
}

\foreach \x in {0, ..., 2}
{
    \foreach \y in {0, ..., 2}
    {
        \draw (\x, \y) node[dot] {};
    }
}
\foreach \x in {0, ..., 3}
{
    \draw (\x, 3) node[blackdot] {};
}
\foreach \y in {0, ..., 3}
{
    \draw (3, \y) node[blackdot] {};
}

\foreach \x in {0.5, ..., 2.5}
{
    \foreach \y in {0, ..., 3}
    {
        \draw (\x, \y) node[redcross] {};
    }
}

\foreach \x in {0, ..., 3}
{
    \foreach \y in {0.5, ..., 2.5}
    {
        \draw (\x, \y) node[bluecross] {};
    }
}

\matrix [draw,below left] at (-1,2.5) {
  \node [dot,label=right:{$\mathbf{u}$}] {}; \\
  \node [redcross,label=right:{$\mathbf{v}_x$}] {}; \\
  \node [bluecross,label=right:{$\mathbf{v}_y$}] {}; \\
  \node [blackdot,label=right:{$\mathbf{u} = 0$}] {}; \\
  \node [reddot,label=right:{$\mathbf{v}_x = 0$}] {}; \\
  \node [bluedot,label=right:{$\mathbf{v}_y = 0$}] {}; \\
  \begin{scope}
    \node (B) at (1,0) {$\partial \Omega$};
    \draw[densely dashed] (B) -- ++(-0.8, 0);
  \end{scope} \\
};
\end{tikzpicture}

\caption{2D staggered grid with Dirichlet BCs on pressure for top and right boundaries, Neumann BCs for bottom and left boundaries.}
\label{fig: sg-mixed}
\end{subfigure}

\caption{Examples of different boundary condition implementations.}
\end{figure*}
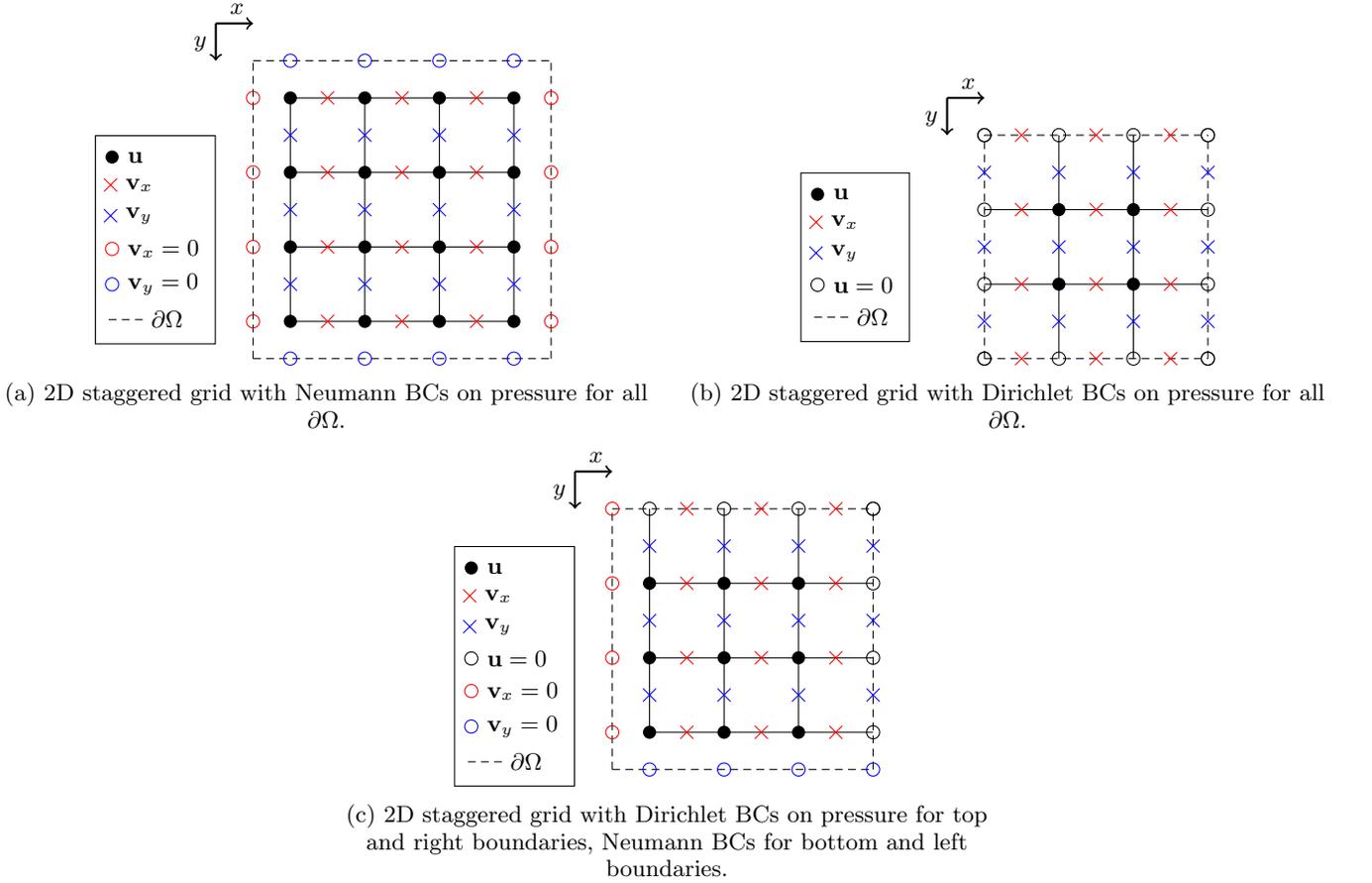

The staggered grid described in the previous section naturally implements Neumann BCs, as the velocities in the direction perpendicular to the domain's boundaries are assumed to be vanishing, i.e., $\mathbf{v}(\mathbf{x}) \cdot \mathbf{n}(\mathbf{x}) = 0$ for all $\mathbf{x} \in \partial \Omega$, where $\mathbf{n}(\mathbf{x})$ is the normal vector at position $\mathbf{x}$ on the boundary $\partial \Omega$. A visual representation is shown in \Cref{fig: sg-neumann}, where the actual boundaries of the domain are depicted with a dashed line.
To implement Dirichlet BCs, i.e., $\mathbf{u}(\mathbf{x}) = 0$ for all $\mathbf{x} \in \partial \Omega$, constraints on pressure points with indices in the set $\{\ell \mid \Phi_u(\ell) \in \partial \Omega\}$ must be added. \Cref{fig: sg-dirichlet} provides a visual representation of these constraints.
Using the method presented in \Cref{s: Constraints} we can employ Dirichlet or Neumann BCs along arbitrary boundaries that align with the grid. An example for a rectangular grid with changing boundary conditions is displayed in \Cref{fig: sg-mixed}.

Since these added constraints result in a system where $\mathbf{R}_f = 0$ (see \Cref{s: Constraints}), i.e., the unconstrained DOFs do not appear in the constraint matrix $\mathbf{R}$, the modified matrices will be $\underline{\mathbf{B}} = \mathbf{B}_{ff}$ and $\underline{\mathbf{A}} = \mathbf{A}_{ff}$, which are the original with some rows and columns removed corresponding to the newly constrained DOFs.

Because the matrices $\mathbf{B}$ and $\mathbf{A}$ were Hermitian and anti-Hermitian by construction, respectively, the modified matrices $\underline{\mathbf{B}}$ and $\underline{\mathbf{A}}$ will remain Hermitian and anti-Hermitian, since deleting one row and column corresponding to the same index of a Hermitian (or anti-Hermitian) matrix results in a matrix that is still Hermitian (or anti-Hermitian).


\section{Generalizing Boundary Conditions to Linear Constraints}\label{s: Constraints}

In \Cref{A: bcs implementation} we showed that in the acoustic case and for rectangular domains Neumann or Dirichlet BCs emerge naturally. In this section we introduce a general method to deal with BCs.
The method operates directly on the discretized but general wave equation
\begin{equation}
    \mathbf{B} \frac{d\mathbf{w}(t)}{dt}  = \mathbf{A}\mathbf{w}(t),
\end{equation}
where we aim to impose BCs as constraints. 
These constraints modify the wave equation and can introduce source terms. 
We assume that these constraints can be expressed in the following form:
\begin{equation}\label{eq: constraint matrix}
\mathbf{R}\mathbf{w} = [\mathbf{R}_f, \mathbf{R}_c] \begin{bmatrix}
  \mathbf{w}_f \\
  \mathbf{w}_c
\end{bmatrix} = \mathbf{b}(t),
\end{equation}
where $\mathbf{R} \in \mathbb{R}^{N_c \times N}$ is the constraint matrix, $N_c$ is the number of constraints, $\mathbf{w}_f \in \mathbb{R}^{N-N_c}$ represents the free (unconstrained) DOFs, and $\mathbf{w}_c \in \mathbb{R}^{N_c}$ represents the constrained DOFs, which will be eliminated from the equations of motion. 
The vector $\mathbf{b}(t)$ is non-zero if the constraints are non-homogeneous. 
For example, in a 2D acoustic wave problem with FD discretization, homogeneous Dirichlet (Neumann) BCs along any boundary orthogonal to the $x$ or $y$ velocity components can be enforced by setting $\mathbf{b}_i = 0$,  $\mathbf{R}_f = \mathbf{0}$ and $\mathbf{R}_c = \mathds{1}$ for all pressure (velocity) values on the boundary.

To incorporate the BCs (or constraints) into the wave equation, we follow the approach in \citep{hoepffner2007implementation}. 
We solve for the constrained DOFs
\begin{equation} \label{eq: constraint eqs solved}
    \mathbf{w}_c = -\mathbf{R}_c^{-1}\mathbf{R}_f \mathbf{w}_f + \mathbf{R}_c^{-1}\mathbf{b}(t).
\end{equation}
Next, we rearrange the equations of motion to separate the free and constrained components
\begin{equation}
 \begin{bmatrix}
  \mathbf{B}_{ff} & \mathbf{B}_{fc} \\
  \mathbf{B}_{cf} & \mathbf{B}_{cc}
\end{bmatrix}
\frac{d}{dt}
\begin{bmatrix}
  \mathbf{w}_f \\
  \mathbf{w}_c
\end{bmatrix}
=
\begin{bmatrix}
  \mathbf{A}_{ff} & \mathbf{A}_{fc} \\
  \mathbf{A}_{cf} & \mathbf{A}_{cc}
\end{bmatrix}
\begin{bmatrix}
  \mathbf{w}_f \\
  \mathbf{w}_c
\end{bmatrix}.   
\end{equation}
Rearranging the equations in this way preserves the Hermitian or anti-Hermitian nature of the matrices. 
Substituting Eq. \eqref{eq: constraint eqs solved} into the above equation and solving for the free DOFs yields
\begin{equation}
    \underline{\mathbf{B}} \frac{d\mathbf{w}_f(t)}{dt} = \underline{\mathbf{A}}\mathbf{w}_f(t) + \underline{\mathbf{s}}_{c}(t),
\end{equation}
where
\begin{equation}
    \underline{\mathbf{B}} = \mathbf{B}_{ff} - \mathbf{B}_{fc}\mathbf{R}_c^{-1}\mathbf{R}_f,
\end{equation}
and
\begin{equation}
    \underline{\mathbf{A}} = \mathbf{A}_{ff} - \mathbf{A}_{fc}\mathbf{R}_c^{-1}\mathbf{R}_f.
\end{equation}
For the constraints to be compatible with this approach, $\underline{\mathbf{B}}$ and $\underline{\mathbf{A}}$ must be Hermitian and anti-Hermitian, respectively. 
This condition is automatically satisfied when $\mathbf{R}_f = \mathbf{0}$. 
In other cases, these conditions can be used to assess whether the constraints are implementable using a HS scheme, which relies on the Hermitian or anti-Hermitian properties of the matrices. 
The source term is given by
\begin{equation}
    \underline{\mathbf{s}}_{c}(t) = \mathbf{A}_{fc}\mathbf{R}_c^{-1}\mathbf{b}(t)-\mathbf{B}_{fc}\mathbf{R}_c^{-1}\frac{d\mathbf{b}(t)}{dt},
\end{equation}
In general, this source cannot be implemented within our framework as the resulting wave field that needs to be initialized is neither radially symmetric nor populates a volume that is independent of $N$. However, if $\mathbf{b}(t)$ is pulse-like and a point source in space, then the logic of \Cref{s: n-independent} applies, giving rise to a quantum initialization that is independent on $N$. Hence, certain boundary or displacement control problems can be implemented within the presented framework.

In summary, this scheme offers a practical approach to implementing BCs: First, discretize the equations, ensuring that the operators maintain the necessary Hermitian or anti-Hermitian properties, then introduce the BCs along the chosen boundaries as described.

Finally, absorbing boundaries or perfectly matched layers have played a crucial role in computational wave physics \citep{higdon1992absorbing, berenger1996perfectly, fichtner2010full}. Due to their fundamentally non-energy conserving nature it is not straightforward to implement such BCs on a QC within the HS framework. However, they can approximately be implemented using random boundaries that diffuse the wavefield, thereby minimizing the energy that is back-scattered into the volume of interest \citep{clapp2009reverse, shen2015random}.


\bibliography{apssamp}

\begin{thebibliography}{86}%
\makeatletter
\providecommand \@ifxundefined [1]{%
 \@ifx{#1\undefined}
}%
\providecommand \@ifnum [1]{%
 \ifnum #1\expandafter \@firstoftwo
 \else \expandafter \@secondoftwo
 \fi
}%
\providecommand \@ifx [1]{%
 \ifx #1\expandafter \@firstoftwo
 \else \expandafter \@secondoftwo
 \fi
}%
\providecommand \natexlab [1]{#1}%
\providecommand \enquote  [1]{``#1''}%
\providecommand \bibnamefont  [1]{#1}%
\providecommand \bibfnamefont [1]{#1}%
\providecommand \citenamefont [1]{#1}%
\providecommand \href@noop [0]{\@secondoftwo}%
\providecommand \href [0]{\begingroup \@sanitize@url \@href}%
\providecommand \@href[1]{\@@startlink{#1}\@@href}%
\providecommand \@@href[1]{\endgroup#1\@@endlink}%
\providecommand \@sanitize@url [0]{\catcode `\\12\catcode `\$12\catcode `\&12\catcode `\#12\catcode `\^12\catcode `\_12\catcode `\%12\relax}%
\providecommand \@@startlink[1]{}%
\providecommand \@@endlink[0]{}%
\providecommand \url  [0]{\begingroup\@sanitize@url \@url }%
\providecommand \@url [1]{\endgroup\@href {#1}{\urlprefix }}%
\providecommand \urlprefix  [0]{URL }%
\providecommand \Eprint [0]{\href }%
\providecommand \doibase [0]{https://doi.org/}%
\providecommand \selectlanguage [0]{\@gobble}%
\providecommand \bibinfo  [0]{\@secondoftwo}%
\providecommand \bibfield  [0]{\@secondoftwo}%
\providecommand \translation [1]{[#1]}%
\providecommand \BibitemOpen [0]{}%
\providecommand \bibitemStop [0]{}%
\providecommand \bibitemNoStop [0]{.\EOS\space}%
\providecommand \EOS [0]{\spacefactor3000\relax}%
\providecommand \BibitemShut  [1]{\csname bibitem#1\endcsname}%
\let\auto@bib@innerbib\@empty
\bibitem [{\citenamefont {Fichtner}(2011)}]{fichtner2010full}%
  \BibitemOpen
  \bibfield  {author} {\bibinfo {author} {\bibfnamefont {A.}~\bibnamefont {Fichtner}},\ }\href {https://doi.org/10.1007/978-3-642-15807-0} {\emph {\bibinfo {title} {Full {Seismic} {Waveform} {Modelling} and {Inversion}}}},\ Advances in {Geophysical} and {Environmental} {Mechanics} and {Mathematics}\ (\bibinfo  {publisher} {Springer},\ \bibinfo {year} {2011})\BibitemShut {NoStop}%
\bibitem [{\citenamefont {Cesca}\ and\ \citenamefont {Grigoli}(2015)}]{cesca2015full}%
  \BibitemOpen
  \bibfield  {author} {\bibinfo {author} {\bibfnamefont {S.}~\bibnamefont {Cesca}}\ and\ \bibinfo {author} {\bibfnamefont {F.}~\bibnamefont {Grigoli}},\ }in\ \href {https://doi.org/10.1016/bs.agph.2014.12.002} {\emph {\bibinfo {booktitle} {Advances in {Geophysics}}}},\ Vol.~\bibinfo {volume} {56},\ \bibinfo {editor} {edited by\ \bibinfo {editor} {\bibfnamefont {R.}~\bibnamefont {Dmowska}}}\ (\bibinfo  {publisher} {Elsevier},\ \bibinfo {year} {2015})\ pp.\ \bibinfo {pages} {169--228}\BibitemShut {NoStop}%
\bibitem [{\citenamefont {Igel}(2017)}]{igel2017computational}%
  \BibitemOpen
  \bibfield  {author} {\bibinfo {author} {\bibfnamefont {H.}~\bibnamefont {Igel}},\ }\href {https://books.google.com/books?id=zslLDQAAQBAJ} {\emph {\bibinfo {title} {Computational {Seismology}: {A} {Practical} {Introduction}}}}\ (\bibinfo  {publisher} {Oxford University Press},\ \bibinfo {year} {2017})\BibitemShut {NoStop}%
\bibitem [{\citenamefont {Arrowsmith}\ \emph {et~al.}(2022)\citenamefont {Arrowsmith}, \citenamefont {Trugman}, \citenamefont {MacCarthy}, \citenamefont {Bergen}, \citenamefont {Lumley},\ and\ \citenamefont {Magnani}}]{arrowsmith2022big}%
  \BibitemOpen
  \bibfield  {author} {\bibinfo {author} {\bibfnamefont {S.~J.}\ \bibnamefont {Arrowsmith}}, \bibinfo {author} {\bibfnamefont {D.~T.}\ \bibnamefont {Trugman}}, \bibinfo {author} {\bibfnamefont {J.}~\bibnamefont {MacCarthy}}, \bibinfo {author} {\bibfnamefont {K.~J.}\ \bibnamefont {Bergen}}, \bibinfo {author} {\bibfnamefont {D.}~\bibnamefont {Lumley}},\ and\ \bibinfo {author} {\bibfnamefont {M.~B.}\ \bibnamefont {Magnani}},\ }\href {https://doi.org/10.1029/2021RG000769} {\bibfield  {journal} {\bibinfo  {journal} {Reviews of Geophysics}\ }\textbf {\bibinfo {volume} {60}},\ \bibinfo {pages} {e2021RG000769} (\bibinfo {year} {2022})}\BibitemShut {NoStop}%
\bibitem [{\citenamefont {Pratt}\ \emph {et~al.}(2007)\citenamefont {Pratt}, \citenamefont {Huang}, \citenamefont {Duric},\ and\ \citenamefont {Littrup}}]{pratt2007sound}%
  \BibitemOpen
  \bibfield  {author} {\bibinfo {author} {\bibfnamefont {R.~G.}\ \bibnamefont {Pratt}}, \bibinfo {author} {\bibfnamefont {L.}~\bibnamefont {Huang}}, \bibinfo {author} {\bibfnamefont {N.}~\bibnamefont {Duric}},\ and\ \bibinfo {author} {\bibfnamefont {P.}~\bibnamefont {Littrup}},\ }in\ \href {https://doi.org/10.1117/12.708789} {\emph {\bibinfo {booktitle} {Medical {Imaging} 2007: {Physics} of {Medical} {Imaging}}}},\ Vol.\ \bibinfo {volume} {6510}\ (\bibinfo  {publisher} {SPIE},\ \bibinfo {year} {2007})\ pp.\ \bibinfo {pages} {1523--1534}\BibitemShut {NoStop}%
\bibitem [{\citenamefont {Gemmeke}\ and\ \citenamefont {Ruiter}(2007)}]{gemmeke20073d}%
  \BibitemOpen
  \bibfield  {author} {\bibinfo {author} {\bibfnamefont {H.}~\bibnamefont {Gemmeke}}\ and\ \bibinfo {author} {\bibfnamefont {N.~V.}\ \bibnamefont {Ruiter}},\ }\href {https://doi.org/10.1016/j.nima.2007.06.116} {\bibfield  {journal} {\bibinfo  {journal} {Nuclear Instruments and Methods in Physics Research Section A: Accelerators, Spectrometers, Detectors and Associated Equipment}\ }\textbf {\bibinfo {volume} {580}},\ \bibinfo {pages} {1057} (\bibinfo {year} {2007})}\BibitemShut {NoStop}%
\bibitem [{\citenamefont {Sanches}\ \emph {et~al.}(2012)\citenamefont {Sanches}, \citenamefont {Laine},\ and\ \citenamefont {Suri}}]{sanches2012ultrasound}%
  \BibitemOpen
  \bibinfo {editor} {\bibfnamefont {J.~M.}\ \bibnamefont {Sanches}}, \bibinfo {editor} {\bibfnamefont {A.~F.}\ \bibnamefont {Laine}},\ and\ \bibinfo {editor} {\bibfnamefont {J.~S.}\ \bibnamefont {Suri}},\ eds.,\ \href {https://doi.org/10.1007/978-1-4614-1180-2} {\emph {\bibinfo {title} {Ultrasound {Imaging}: {Advances} and {Applications}}}}\ (\bibinfo  {publisher} {Springer},\ \bibinfo {year} {2012})\BibitemShut {NoStop}%
\bibitem [{\citenamefont {Ould~Naffa}\ \emph {et~al.}(2002)\citenamefont {Ould~Naffa}, \citenamefont {Goueygou}, \citenamefont {Piwakowski},\ and\ \citenamefont {Buyle-Bodin}}]{naffa2002detection}%
  \BibitemOpen
  \bibfield  {author} {\bibinfo {author} {\bibfnamefont {S.}~\bibnamefont {Ould~Naffa}}, \bibinfo {author} {\bibfnamefont {M.}~\bibnamefont {Goueygou}}, \bibinfo {author} {\bibfnamefont {B.}~\bibnamefont {Piwakowski}},\ and\ \bibinfo {author} {\bibfnamefont {F.}~\bibnamefont {Buyle-Bodin}},\ }\href {https://doi.org/10.1016/S0041-624X(02)00146-4} {\bibfield  {journal} {\bibinfo  {journal} {Ultrasonics}\ }\textbf {\bibinfo {volume} {40}},\ \bibinfo {pages} {247} (\bibinfo {year} {2002})}\BibitemShut {NoStop}%
\bibitem [{\citenamefont {Kim}\ and\ \citenamefont {Wagner}(2016)}]{kim2016non}%
  \BibitemOpen
  \bibfield  {author} {\bibinfo {author} {\bibfnamefont {K.}~\bibnamefont {Kim}}\ and\ \bibinfo {author} {\bibfnamefont {W.~R.}\ \bibnamefont {Wagner}},\ }\href {https://doi.org/10.1007/s10439-015-1495-0} {\bibfield  {journal} {\bibinfo  {journal} {Annals of Biomedical Engineering}\ }\textbf {\bibinfo {volume} {44}},\ \bibinfo {pages} {621} (\bibinfo {year} {2016})}\BibitemShut {NoStop}%
\bibitem [{\citenamefont {Lopez}\ \emph {et~al.}(2018)\citenamefont {Lopez}, \citenamefont {Bacelar}, \citenamefont {Pires}, \citenamefont {Santos}, \citenamefont {Sousa},\ and\ \citenamefont {Quintino}}]{lopez2018non}%
  \BibitemOpen
  \bibfield  {author} {\bibinfo {author} {\bibfnamefont {A.}~\bibnamefont {Lopez}}, \bibinfo {author} {\bibfnamefont {R.}~\bibnamefont {Bacelar}}, \bibinfo {author} {\bibfnamefont {I.}~\bibnamefont {Pires}}, \bibinfo {author} {\bibfnamefont {T.~G.}\ \bibnamefont {Santos}}, \bibinfo {author} {\bibfnamefont {J.~P.}\ \bibnamefont {Sousa}},\ and\ \bibinfo {author} {\bibfnamefont {L.}~\bibnamefont {Quintino}},\ }\href {https://doi.org/10.1016/j.addma.2018.03.020} {\bibfield  {journal} {\bibinfo  {journal} {Additive Manufacturing}\ }\textbf {\bibinfo {volume} {21}},\ \bibinfo {pages} {298} (\bibinfo {year} {2018})}\BibitemShut {NoStop}%
\bibitem [{\citenamefont {Bendsoe}\ and\ \citenamefont {Sigmund}(2013)}]{bendsoe2013topology}%
  \BibitemOpen
  \bibfield  {author} {\bibinfo {author} {\bibfnamefont {M.~P.}\ \bibnamefont {Bendsoe}}\ and\ \bibinfo {author} {\bibfnamefont {O.}~\bibnamefont {Sigmund}},\ }\href {https://books.google.ch/books?id=ZCjsCAAAQBAJ} {\emph {\bibinfo {title} {Topology {Optimization}: {Theory}, {Methods}, and {Applications}}}}\ (\bibinfo  {publisher} {Springer Science \& Business Media},\ \bibinfo {year} {2013})\BibitemShut {NoStop}%
\bibitem [{\citenamefont {Christiansen}\ and\ \citenamefont {Sigmund}(2021)}]{christiansen2021inverse}%
  \BibitemOpen
  \bibfield  {author} {\bibinfo {author} {\bibfnamefont {R.~E.}\ \bibnamefont {Christiansen}}\ and\ \bibinfo {author} {\bibfnamefont {O.}~\bibnamefont {Sigmund}},\ }\href {https://doi.org/10.1364/JOSAB.406048} {\bibfield  {journal} {\bibinfo  {journal} {JOSA B}\ }\textbf {\bibinfo {volume} {38}},\ \bibinfo {pages} {496} (\bibinfo {year} {2021})}\BibitemShut {NoStop}%
\bibitem [{\citenamefont {Sharma}\ \emph {et~al.}(2022)\citenamefont {Sharma}, \citenamefont {Kosta}, \citenamefont {Shmuel},\ and\ \citenamefont {Amir}}]{sharma2022gradient}%
  \BibitemOpen
  \bibfield  {author} {\bibinfo {author} {\bibfnamefont {A.~K.}\ \bibnamefont {Sharma}}, \bibinfo {author} {\bibfnamefont {M.}~\bibnamefont {Kosta}}, \bibinfo {author} {\bibfnamefont {G.}~\bibnamefont {Shmuel}},\ and\ \bibinfo {author} {\bibfnamefont {O.}~\bibnamefont {Amir}},\ }\href {https://doi.org/10.1016/j.compstruct.2021.114846} {\bibfield  {journal} {\bibinfo  {journal} {Composite Structures}\ }\textbf {\bibinfo {volume} {280}},\ \bibinfo {pages} {114846} (\bibinfo {year} {2022})}\BibitemShut {NoStop}%
\bibitem [{\citenamefont {Bordiga}\ \emph {et~al.}(2024)\citenamefont {Bordiga}, \citenamefont {Medina}, \citenamefont {Jafarzadeh}, \citenamefont {Bösch}, \citenamefont {Adams}, \citenamefont {Tournat},\ and\ \citenamefont {Bertoldi}}]{bordiga2024automated}%
  \BibitemOpen
  \bibfield  {author} {\bibinfo {author} {\bibfnamefont {G.}~\bibnamefont {Bordiga}}, \bibinfo {author} {\bibfnamefont {E.}~\bibnamefont {Medina}}, \bibinfo {author} {\bibfnamefont {S.}~\bibnamefont {Jafarzadeh}}, \bibinfo {author} {\bibfnamefont {C.}~\bibnamefont {Bösch}}, \bibinfo {author} {\bibfnamefont {R.~P.}\ \bibnamefont {Adams}}, \bibinfo {author} {\bibfnamefont {V.}~\bibnamefont {Tournat}},\ and\ \bibinfo {author} {\bibfnamefont {K.}~\bibnamefont {Bertoldi}},\ }\href {https://doi.org/10.1038/s41563-024-02008-6} {\bibfield  {journal} {\bibinfo  {journal} {Nature Materials}\ ,\ \bibinfo {pages} {1}} (\bibinfo {year} {2024})}\BibitemShut {NoStop}%
\bibitem [{\citenamefont {Yilmaz}(2001)}]{yilmaz2001seismic}%
  \BibitemOpen
  \bibfield  {author} {\bibinfo {author} {\bibfnamefont {{\"O}.}~\bibnamefont {Yilmaz}},\ }\href {https://doi.org/https://doi.org/10.1190/1.9781560801580} {\emph {\bibinfo {title} {Seismic {Data} {Analysis}}}}\ (\bibinfo  {publisher} {Society of Exploration Geophysicists},\ \bibinfo {year} {2001})\BibitemShut {NoStop}%
\bibitem [{\citenamefont {Fichtner}\ and\ \citenamefont {Trampert}(2011)}]{fichtner2011resolution}%
  \BibitemOpen
  \bibfield  {author} {\bibinfo {author} {\bibfnamefont {A.}~\bibnamefont {Fichtner}}\ and\ \bibinfo {author} {\bibfnamefont {J.}~\bibnamefont {Trampert}},\ }\href {https://doi.org/10.1111/j.1365-246X.2011.05218.x} {\bibfield  {journal} {\bibinfo  {journal} {Geophysical Journal International}\ }\textbf {\bibinfo {volume} {187}},\ \bibinfo {pages} {1604} (\bibinfo {year} {2011})}\BibitemShut {NoStop}%
\bibitem [{\citenamefont {Rawlinson}\ \emph {et~al.}(2014)\citenamefont {Rawlinson}, \citenamefont {Fichtner}, \citenamefont {Sambridge},\ and\ \citenamefont {Young}}]{rawlinson2014seismic}%
  \BibitemOpen
  \bibfield  {author} {\bibinfo {author} {\bibfnamefont {N.}~\bibnamefont {Rawlinson}}, \bibinfo {author} {\bibfnamefont {A.}~\bibnamefont {Fichtner}}, \bibinfo {author} {\bibfnamefont {M.}~\bibnamefont {Sambridge}},\ and\ \bibinfo {author} {\bibfnamefont {M.~K.}\ \bibnamefont {Young}},\ }in\ \href {https://doi.org/10.1016/bs.agph.2014.08.001} {\emph {\bibinfo {booktitle} {Advances in {Geophysics}}}},\ Vol.~\bibinfo {volume} {55},\ \bibinfo {editor} {edited by\ \bibinfo {editor} {\bibfnamefont {R.}~\bibnamefont {Dmowska}}}\ (\bibinfo  {publisher} {Elsevier},\ \bibinfo {year} {2014})\ pp.\ \bibinfo {pages} {1--76}\BibitemShut {NoStop}%
\bibitem [{\citenamefont {Betancourt}(2018)}]{betancourt2017conceptual}%
  \BibitemOpen
  \bibfield  {author} {\bibinfo {author} {\bibfnamefont {M.}~\bibnamefont {Betancourt}}} (\bibinfo {year} {2018}),\ \bibinfo {note} {arXiv:1701.02434}\BibitemShut {NoStop}%
\bibitem [{\citenamefont {Sen}\ and\ \citenamefont {Biswas}(2017)}]{sen2017transdimensional}%
  \BibitemOpen
  \bibfield  {author} {\bibinfo {author} {\bibfnamefont {M.~K.}\ \bibnamefont {Sen}}\ and\ \bibinfo {author} {\bibfnamefont {R.}~\bibnamefont {Biswas}},\ }\href {https://doi.org/10.1190/geo2016-0010.1} {\bibfield  {journal} {\bibinfo  {journal} {GEOPHYSICS}\ }\textbf {\bibinfo {volume} {82}},\ \bibinfo {pages} {R119} (\bibinfo {year} {2017})}\BibitemShut {NoStop}%
\bibitem [{\citenamefont {Fichtner}\ \emph {et~al.}(2019)\citenamefont {Fichtner}, \citenamefont {Zunino},\ and\ \citenamefont {Gebraad}}]{fichtner2019hamiltonian}%
  \BibitemOpen
  \bibfield  {author} {\bibinfo {author} {\bibfnamefont {A.}~\bibnamefont {Fichtner}}, \bibinfo {author} {\bibfnamefont {A.}~\bibnamefont {Zunino}},\ and\ \bibinfo {author} {\bibfnamefont {L.}~\bibnamefont {Gebraad}},\ }\href {https://doi.org/10.1093/gji/ggy496} {\bibfield  {journal} {\bibinfo  {journal} {Geophysical Journal International}\ }\textbf {\bibinfo {volume} {216}},\ \bibinfo {pages} {1344} (\bibinfo {year} {2019})}\BibitemShut {NoStop}%
\bibitem [{\citenamefont {Ulrich}\ \emph {et~al.}(2022)\citenamefont {Ulrich}, \citenamefont {Boehm}, \citenamefont {Zunino},\ and\ \citenamefont {Fichtner}}]{ulrich2022analyzing}%
  \BibitemOpen
  \bibfield  {author} {\bibinfo {author} {\bibfnamefont {I.~E.}\ \bibnamefont {Ulrich}}, \bibinfo {author} {\bibfnamefont {C.}~\bibnamefont {Boehm}}, \bibinfo {author} {\bibfnamefont {A.}~\bibnamefont {Zunino}},\ and\ \bibinfo {author} {\bibfnamefont {A.}~\bibnamefont {Fichtner}},\ }in\ \href {https://doi.org/10.1117/12.2608546} {\emph {\bibinfo {booktitle} {Medical {Imaging} 2022: {Ultrasonic} {Imaging} and {Tomography}}}},\ Vol.\ \bibinfo {volume} {12038}\ (\bibinfo  {publisher} {SPIE},\ \bibinfo {year} {2022})\ pp.\ \bibinfo {pages} {48--60}\BibitemShut {NoStop}%
\bibitem [{\citenamefont {Nielsen}\ and\ \citenamefont {Chuang}(2010)}]{nielsen2010quantum}%
  \BibitemOpen
  \bibfield  {author} {\bibinfo {author} {\bibfnamefont {M.~A.}\ \bibnamefont {Nielsen}}\ and\ \bibinfo {author} {\bibfnamefont {I.~L.}\ \bibnamefont {Chuang}},\ }\href {https://www.google.com/books/edition/Quantum_Computation_and_Quantum_Informat/-s4DEy7o-a0C} {\emph {\bibinfo {title} {Quantum computation and quantum information}}},\ \bibinfo {edition} {10th}\ ed.\ (\bibinfo  {publisher} {Cambridge university press},\ \bibinfo {year} {2010})\BibitemShut {NoStop}%
\bibitem [{\citenamefont {Montanaro}(2016)}]{montanaro2016quantum}%
  \BibitemOpen
  \bibfield  {author} {\bibinfo {author} {\bibfnamefont {A.}~\bibnamefont {Montanaro}},\ }\href {https://doi.org/10.1038/npjqi.2015.23} {\bibfield  {journal} {\bibinfo  {journal} {npj Quantum Information}\ }\textbf {\bibinfo {volume} {2}},\ \bibinfo {pages} {1} (\bibinfo {year} {2016})}\BibitemShut {NoStop}%
\bibitem [{\citenamefont {Cerezo}\ \emph {et~al.}(2021)\citenamefont {Cerezo}, \citenamefont {Arrasmith}, \citenamefont {Babbush}, \citenamefont {Benjamin}, \citenamefont {Endo}, \citenamefont {Fujii}, \citenamefont {McClean}, \citenamefont {Mitarai}, \citenamefont {Yuan}, \citenamefont {Cincio},\ and\ \citenamefont {Coles}}]{cerezo2021variational}%
  \BibitemOpen
  \bibfield  {author} {\bibinfo {author} {\bibfnamefont {M.}~\bibnamefont {Cerezo}}, \bibinfo {author} {\bibfnamefont {A.}~\bibnamefont {Arrasmith}}, \bibinfo {author} {\bibfnamefont {R.}~\bibnamefont {Babbush}}, \bibinfo {author} {\bibfnamefont {S.~C.}\ \bibnamefont {Benjamin}}, \bibinfo {author} {\bibfnamefont {S.}~\bibnamefont {Endo}}, \bibinfo {author} {\bibfnamefont {K.}~\bibnamefont {Fujii}}, \bibinfo {author} {\bibfnamefont {J.~R.}\ \bibnamefont {McClean}}, \bibinfo {author} {\bibfnamefont {K.}~\bibnamefont {Mitarai}}, \bibinfo {author} {\bibfnamefont {X.}~\bibnamefont {Yuan}}, \bibinfo {author} {\bibfnamefont {L.}~\bibnamefont {Cincio}},\ and\ \bibinfo {author} {\bibfnamefont {P.~J.}\ \bibnamefont {Coles}},\ }\href {https://doi.org/10.1038/s42254-021-00348-9} {\bibfield  {journal} {\bibinfo  {journal} {Nature Reviews Physics}\ }\textbf {\bibinfo {volume} {3}},\ \bibinfo {pages} {625} (\bibinfo {year} {2021})}\BibitemShut {NoStop}%
\bibitem [{\citenamefont {Martyn}\ \emph {et~al.}(2021)\citenamefont {Martyn}, \citenamefont {Rossi}, \citenamefont {Tan},\ and\ \citenamefont {Chuang}}]{PRXQuantum.2.040203}%
  \BibitemOpen
  \bibfield  {author} {\bibinfo {author} {\bibfnamefont {J.~M.}\ \bibnamefont {Martyn}}, \bibinfo {author} {\bibfnamefont {Z.~M.}\ \bibnamefont {Rossi}}, \bibinfo {author} {\bibfnamefont {A.~K.}\ \bibnamefont {Tan}},\ and\ \bibinfo {author} {\bibfnamefont {I.~L.}\ \bibnamefont {Chuang}},\ }\href {https://doi.org/10.1103/PRXQuantum.2.040203} {\bibfield  {journal} {\bibinfo  {journal} {PRX Quantum}\ }\textbf {\bibinfo {volume} {2}},\ \bibinfo {pages} {040203} (\bibinfo {year} {2021})}\BibitemShut {NoStop}%
\bibitem [{\citenamefont {Childs}\ and\ \citenamefont {Wiebe}(2012)}]{childs2012hamiltonian}%
  \BibitemOpen
  \bibfield  {author} {\bibinfo {author} {\bibfnamefont {A.~M.}\ \bibnamefont {Childs}}\ and\ \bibinfo {author} {\bibfnamefont {N.}~\bibnamefont {Wiebe}}} (\bibinfo {year} {2012}),\ \bibinfo {note} {arXiv:1202.5822}\BibitemShut {NoStop}%
\bibitem [{\citenamefont {Low}\ and\ \citenamefont {Chuang}(2017)}]{low2017optimal}%
  \BibitemOpen
  \bibfield  {author} {\bibinfo {author} {\bibfnamefont {G.~H.}\ \bibnamefont {Low}}\ and\ \bibinfo {author} {\bibfnamefont {I.~L.}\ \bibnamefont {Chuang}},\ }\href {https://doi.org/10.1103/PhysRevLett.118.010501} {\bibfield  {journal} {\bibinfo  {journal} {Phys. Rev. Lett.}\ }\textbf {\bibinfo {volume} {118}},\ \bibinfo {pages} {010501} (\bibinfo {year} {2017})}\BibitemShut {NoStop}%
\bibitem [{\citenamefont {Low}\ and\ \citenamefont {Chuang}(2019)}]{low2019hamiltonian}%
  \BibitemOpen
  \bibfield  {author} {\bibinfo {author} {\bibfnamefont {G.~H.}\ \bibnamefont {Low}}\ and\ \bibinfo {author} {\bibfnamefont {I.~L.}\ \bibnamefont {Chuang}},\ }\href {https://doi.org/10.22331/q-2019-07-12-163} {\bibfield  {journal} {\bibinfo  {journal} {Quantum}\ }\textbf {\bibinfo {volume} {3}},\ \bibinfo {pages} {163} (\bibinfo {year} {2019})}\BibitemShut {NoStop}%
\bibitem [{\citenamefont {Low}\ \emph {et~al.}(2023)\citenamefont {Low}, \citenamefont {Su}, \citenamefont {Tong},\ and\ \citenamefont {Tran}}]{low2023complexity}%
  \BibitemOpen
  \bibfield  {author} {\bibinfo {author} {\bibfnamefont {G.~H.}\ \bibnamefont {Low}}, \bibinfo {author} {\bibfnamefont {Y.}~\bibnamefont {Su}}, \bibinfo {author} {\bibfnamefont {Y.}~\bibnamefont {Tong}},\ and\ \bibinfo {author} {\bibfnamefont {M.~C.}\ \bibnamefont {Tran}},\ }\href {https://doi.org/10.1103/PRXQuantum.4.020323} {\bibfield  {journal} {\bibinfo  {journal} {PRX Quantum}\ }\textbf {\bibinfo {volume} {4}},\ \bibinfo {pages} {020323} (\bibinfo {year} {2023})}\BibitemShut {NoStop}%
\bibitem [{\citenamefont {Costa}\ \emph {et~al.}(2019)\citenamefont {Costa}, \citenamefont {Jordan},\ and\ \citenamefont {Ostrander}}]{costa2019quantum}%
  \BibitemOpen
  \bibfield  {author} {\bibinfo {author} {\bibfnamefont {P.~C.~S.}\ \bibnamefont {Costa}}, \bibinfo {author} {\bibfnamefont {S.}~\bibnamefont {Jordan}},\ and\ \bibinfo {author} {\bibfnamefont {A.}~\bibnamefont {Ostrander}},\ }\href {https://doi.org/10.1103/PhysRevA.99.012323} {\bibfield  {journal} {\bibinfo  {journal} {Physical Review A}\ }\textbf {\bibinfo {volume} {99}},\ \bibinfo {pages} {012323} (\bibinfo {year} {2019})}\BibitemShut {NoStop}%
\bibitem [{\citenamefont {Suau}\ \emph {et~al.}(2021)\citenamefont {Suau}, \citenamefont {Staffelbach},\ and\ \citenamefont {Calandra}}]{suau2021practical}%
  \BibitemOpen
  \bibfield  {author} {\bibinfo {author} {\bibfnamefont {A.}~\bibnamefont {Suau}}, \bibinfo {author} {\bibfnamefont {G.}~\bibnamefont {Staffelbach}},\ and\ \bibinfo {author} {\bibfnamefont {H.}~\bibnamefont {Calandra}},\ }\href {https://doi.org/10.1145/3430030} {\bibfield  {journal} {\bibinfo  {journal} {ACM Transactions on Quantum Computing}\ }\textbf {\bibinfo {volume} {2}},\ \bibinfo {pages} {2:1} (\bibinfo {year} {2021})}\BibitemShut {NoStop}%
\bibitem [{\citenamefont {Koukoutsis}\ \emph {et~al.}(2023)\citenamefont {Koukoutsis}, \citenamefont {Hizanidis}, \citenamefont {Ram},\ and\ \citenamefont {Vahala}}]{koukoutsis2023}%
  \BibitemOpen
  \bibfield  {author} {\bibinfo {author} {\bibfnamefont {E.}~\bibnamefont {Koukoutsis}}, \bibinfo {author} {\bibfnamefont {K.}~\bibnamefont {Hizanidis}}, \bibinfo {author} {\bibfnamefont {A.~K.}\ \bibnamefont {Ram}},\ and\ \bibinfo {author} {\bibfnamefont {G.}~\bibnamefont {Vahala}},\ }\href {https://doi.org/10.1103/PhysRevA.107.042215} {\bibfield  {journal} {\bibinfo  {journal} {Physical Review A}\ }\textbf {\bibinfo {volume} {107}},\ \bibinfo {pages} {042215} (\bibinfo {year} {2023})}\BibitemShut {NoStop}%
\bibitem [{\citenamefont {Sato}\ \emph {et~al.}(2024{\natexlab{a}})\citenamefont {Sato}, \citenamefont {Kondo}, \citenamefont {Hamamura}, \citenamefont {Onodera},\ and\ \citenamefont {Yamamoto}}]{sato2024hamiltonian}%
  \BibitemOpen
  \bibfield  {author} {\bibinfo {author} {\bibfnamefont {Y.}~\bibnamefont {Sato}}, \bibinfo {author} {\bibfnamefont {R.}~\bibnamefont {Kondo}}, \bibinfo {author} {\bibfnamefont {I.}~\bibnamefont {Hamamura}}, \bibinfo {author} {\bibfnamefont {T.}~\bibnamefont {Onodera}},\ and\ \bibinfo {author} {\bibfnamefont {N.}~\bibnamefont {Yamamoto}}} (\bibinfo {year} {2024}{\natexlab{a}}),\ \bibinfo {note} {arXiv:2402.18398}\BibitemShut {NoStop}%
\bibitem [{\citenamefont {Sato}\ \emph {et~al.}(2024{\natexlab{b}})\citenamefont {Sato}, \citenamefont {Tezuka}, \citenamefont {Kondo},\ and\ \citenamefont {Yamamoto}}]{sato2024quantum}%
  \BibitemOpen
  \bibfield  {author} {\bibinfo {author} {\bibfnamefont {Y.}~\bibnamefont {Sato}}, \bibinfo {author} {\bibfnamefont {H.}~\bibnamefont {Tezuka}}, \bibinfo {author} {\bibfnamefont {R.}~\bibnamefont {Kondo}},\ and\ \bibinfo {author} {\bibfnamefont {N.}~\bibnamefont {Yamamoto}}} (\bibinfo {year} {2024}{\natexlab{b}}),\ \bibinfo {note} {arXiv:2407.05019}\BibitemShut {NoStop}%
\bibitem [{\citenamefont {Schade}\ \emph {et~al.}(2024)\citenamefont {Schade}, \citenamefont {Bösch}, \citenamefont {Hapla},\ and\ \citenamefont {Fichtner}}]{schade2024quantum}%
  \BibitemOpen
  \bibfield  {author} {\bibinfo {author} {\bibfnamefont {M.}~\bibnamefont {Schade}}, \bibinfo {author} {\bibfnamefont {C.}~\bibnamefont {Bösch}}, \bibinfo {author} {\bibfnamefont {V.}~\bibnamefont {Hapla}},\ and\ \bibinfo {author} {\bibfnamefont {A.}~\bibnamefont {Fichtner}},\ }\href {https://doi.org/10.1093/gji/ggae160} {\bibfield  {journal} {\bibinfo  {journal} {Geophysical Journal International}\ }\textbf {\bibinfo {volume} {238}},\ \bibinfo {pages} {321} (\bibinfo {year} {2024})}\BibitemShut {NoStop}%
\bibitem [{\citenamefont {Babbush}\ \emph {et~al.}(2023)\citenamefont {Babbush}, \citenamefont {Berry}, \citenamefont {Kothari}, \citenamefont {Somma},\ and\ \citenamefont {Wiebe}}]{babbush2023exponential}%
  \BibitemOpen
  \bibfield  {author} {\bibinfo {author} {\bibfnamefont {R.}~\bibnamefont {Babbush}}, \bibinfo {author} {\bibfnamefont {D.~W.}\ \bibnamefont {Berry}}, \bibinfo {author} {\bibfnamefont {R.}~\bibnamefont {Kothari}}, \bibinfo {author} {\bibfnamefont {R.~D.}\ \bibnamefont {Somma}},\ and\ \bibinfo {author} {\bibfnamefont {N.}~\bibnamefont {Wiebe}},\ }\href {https://doi.org/10.1103/PhysRevX.13.041041} {\bibfield  {journal} {\bibinfo  {journal} {Physical Review X}\ }\textbf {\bibinfo {volume} {13}},\ \bibinfo {pages} {041041} (\bibinfo {year} {2023})}\BibitemShut {NoStop}%
\bibitem [{\citenamefont {Ito}\ \emph {et~al.}(2023)\citenamefont {Ito}, \citenamefont {Tanaka},\ and\ \citenamefont {Fujii}}]{ito2023map}%
  \BibitemOpen
  \bibfield  {author} {\bibinfo {author} {\bibfnamefont {Y.}~\bibnamefont {Ito}}, \bibinfo {author} {\bibfnamefont {Y.}~\bibnamefont {Tanaka}},\ and\ \bibinfo {author} {\bibfnamefont {K.}~\bibnamefont {Fujii}}} (\bibinfo {year} {2023}),\ \bibinfo {note} {arXiv:2311.15628}\BibitemShut {NoStop}%
\bibitem [{\citenamefont {Danz}\ \emph {et~al.}(2024)\citenamefont {Danz}, \citenamefont {Berta}, \citenamefont {Schröder}, \citenamefont {Kienast}, \citenamefont {Wilhelm},\ and\ \citenamefont {Ciani}}]{danz2024calculating}%
  \BibitemOpen
  \bibfield  {author} {\bibinfo {author} {\bibfnamefont {S.}~\bibnamefont {Danz}}, \bibinfo {author} {\bibfnamefont {M.}~\bibnamefont {Berta}}, \bibinfo {author} {\bibfnamefont {S.}~\bibnamefont {Schröder}}, \bibinfo {author} {\bibfnamefont {P.}~\bibnamefont {Kienast}}, \bibinfo {author} {\bibfnamefont {F.~K.}\ \bibnamefont {Wilhelm}},\ and\ \bibinfo {author} {\bibfnamefont {A.}~\bibnamefont {Ciani}}} (\bibinfo {year} {2024}),\ \bibinfo {note} {arXiv:2405.08694}\BibitemShut {NoStop}%
\bibitem [{\citenamefont {Jin}\ \emph {et~al.}(2024)\citenamefont {Jin}, \citenamefont {Liu},\ and\ \citenamefont {Ma}}]{jin2023quantum}%
  \BibitemOpen
  \bibfield  {author} {\bibinfo {author} {\bibfnamefont {S.}~\bibnamefont {Jin}}, \bibinfo {author} {\bibfnamefont {N.}~\bibnamefont {Liu}},\ and\ \bibinfo {author} {\bibfnamefont {C.}~\bibnamefont {Ma}},\ }\href {https://doi.org/10.1051/m2an/2024046} {\bibfield  {journal} {\bibinfo  {journal} {ESAIM: Mathematical Modelling and Numerical Analysis}\ }\textbf {\bibinfo {volume} {58}},\ \bibinfo {pages} {1853} (\bibinfo {year} {2024})}\BibitemShut {NoStop}%
\bibitem [{\citenamefont {Jin}\ \emph {et~al.}(2023)\citenamefont {Jin}, \citenamefont {Liu},\ and\ \citenamefont {Yu}}]{jin2022quantum}%
  \BibitemOpen
  \bibfield  {author} {\bibinfo {author} {\bibfnamefont {S.}~\bibnamefont {Jin}}, \bibinfo {author} {\bibfnamefont {N.}~\bibnamefont {Liu}},\ and\ \bibinfo {author} {\bibfnamefont {Y.}~\bibnamefont {Yu}},\ }\href {https://doi.org/10.1103/PhysRevA.108.032603} {\bibfield  {journal} {\bibinfo  {journal} {Physical Review A}\ }\textbf {\bibinfo {volume} {108}},\ \bibinfo {pages} {032603} (\bibinfo {year} {2023})}\BibitemShut {NoStop}%
\bibitem [{\citenamefont {Ma}\ \emph {et~al.}(2024)\citenamefont {Ma}, \citenamefont {Jin}, \citenamefont {Liu}, \citenamefont {Wang},\ and\ \citenamefont {Zhang}}]{ma2024schr}%
  \BibitemOpen
  \bibfield  {author} {\bibinfo {author} {\bibfnamefont {C.}~\bibnamefont {Ma}}, \bibinfo {author} {\bibfnamefont {S.}~\bibnamefont {Jin}}, \bibinfo {author} {\bibfnamefont {N.}~\bibnamefont {Liu}}, \bibinfo {author} {\bibfnamefont {K.}~\bibnamefont {Wang}},\ and\ \bibinfo {author} {\bibfnamefont {L.}~\bibnamefont {Zhang}}} (\bibinfo {year} {2024}),\ \bibinfo {note} {arXiv:2411.10999}\BibitemShut {NoStop}%
\bibitem [{\citenamefont {Johnson}(2007)}]{johnson2007notes}%
  \BibitemOpen
  \bibfield  {author} {\bibinfo {author} {\bibfnamefont {S.~G.}\ \bibnamefont {Johnson}},\ }\href {https://math.mit.edu/~stevenj/18.369/spring09/wave-equations.pdf} {\bibinfo {title} {Notes on the algebraic structure of wave equations}} (\bibinfo {year} {2007})\BibitemShut {NoStop}%
\bibitem [{\citenamefont {Mostafazadeh}(2010)}]{mostafazadeh2010pseudo}%
  \BibitemOpen
  \bibfield  {author} {\bibinfo {author} {\bibfnamefont {A.}~\bibnamefont {Mostafazadeh}},\ }\href {https://doi.org/10.1142/S0219887810004816} {\bibfield  {journal} {\bibinfo  {journal} {International Journal of Geometric Methods in Modern Physics}\ }\textbf {\bibinfo {volume} {07}},\ \bibinfo {pages} {1191} (\bibinfo {year} {2010})}\BibitemShut {NoStop}%
\bibitem [{\citenamefont {Kaltenbacher}(2018)}]{Kaltenbacher2018}%
  \BibitemOpen
  \bibfield  {author} {\bibinfo {author} {\bibfnamefont {M.}~\bibnamefont {Kaltenbacher}},\ }\bibinfo {title} {Fundamental equations of acoustics},\ in\ \href {https://doi.org/10.1007/978-3-319-59038-7_1} {\emph {\bibinfo {booktitle} {Computational Acoustics}}}\ (\bibinfo  {publisher} {Springer International Publishing},\ \bibinfo {year} {2018})\ p.~\bibinfo {pages} {12}\BibitemShut {NoStop}%
\bibitem [{\citenamefont {Griffiths}(2023)}]{griffiths2023introduction}%
  \BibitemOpen
  \bibfield  {author} {\bibinfo {author} {\bibfnamefont {D.~J.}\ \bibnamefont {Griffiths}},\ }\href {https://www.google.com/books/edition/Introduction_to_Electrodynamics/I9jbEAAAQBAJ} {\emph {\bibinfo {title} {Introduction to {Electrodynamics}}}}\ (\bibinfo  {publisher} {Cambridge University Press},\ \bibinfo {year} {2023})\BibitemShut {NoStop}%
\bibitem [{\citenamefont {Johnson}\ and\ \citenamefont {Joannopoulos}(2001)}]{johnson2001photonic}%
  \BibitemOpen
  \bibfield  {author} {\bibinfo {author} {\bibfnamefont {S.~G.}\ \bibnamefont {Johnson}}\ and\ \bibinfo {author} {\bibfnamefont {J.~D.}\ \bibnamefont {Joannopoulos}},\ }\href {https://www.google.com/books/edition/Photonic_Crystals/LOZAsek9y7oC} {\emph {\bibinfo {title} {Photonic {Crystals}: {The} {Road} from {Theory} to {Practice}}}}\ (\bibinfo  {publisher} {Springer Science \& Business Media},\ \bibinfo {year} {2001})\BibitemShut {NoStop}%
\bibitem [{\citenamefont {Jackson}(2021)}]{jackson1999classical}%
  \BibitemOpen
  \bibfield  {author} {\bibinfo {author} {\bibfnamefont {J.}~\bibnamefont {Jackson}},\ }\href {https://books.google.ch/books?id=6VV-EAAAQBAJ} {\emph {\bibinfo {title} {Classical Electrodynamics}}}\ (\bibinfo  {publisher} {Wiley},\ \bibinfo {year} {2021})\BibitemShut {NoStop}%
\bibitem [{\citenamefont {Voigt}(1966)}]{voigt1910lehrbuch}%
  \BibitemOpen
  \bibfield  {author} {\bibinfo {author} {\bibfnamefont {W.}~\bibnamefont {Voigt}},\ }\href {https://doi.org/10.1007/978-3-663-15884-4} {\emph {\bibinfo {title} {Lehrbuch der {Kristallphysik}}}}\ (\bibinfo  {publisher} {Vieweg und Teubner Verlag},\ \bibinfo {year} {1966})\BibitemShut {NoStop}%
\bibitem [{\citenamefont {Scheibner}\ \emph {et~al.}(2020)\citenamefont {Scheibner}, \citenamefont {Souslov}, \citenamefont {Banerjee}, \citenamefont {Surówka}, \citenamefont {Irvine},\ and\ \citenamefont {Vitelli}}]{scheibner2020odd}%
  \BibitemOpen
  \bibfield  {author} {\bibinfo {author} {\bibfnamefont {C.}~\bibnamefont {Scheibner}}, \bibinfo {author} {\bibfnamefont {A.}~\bibnamefont {Souslov}}, \bibinfo {author} {\bibfnamefont {D.}~\bibnamefont {Banerjee}}, \bibinfo {author} {\bibfnamefont {P.}~\bibnamefont {Surówka}}, \bibinfo {author} {\bibfnamefont {W.~T.~M.}\ \bibnamefont {Irvine}},\ and\ \bibinfo {author} {\bibfnamefont {V.}~\bibnamefont {Vitelli}},\ }\href {https://doi.org/10.1038/s41567-020-0795-y} {\bibfield  {journal} {\bibinfo  {journal} {Nature Physics}\ }\textbf {\bibinfo {volume} {16}},\ \bibinfo {pages} {475} (\bibinfo {year} {2020})}\BibitemShut {NoStop}%
\bibitem [{\citenamefont {Aki}\ and\ \citenamefont {Richards}(2002)}]{aki2002quantitative}%
  \BibitemOpen
  \bibfield  {author} {\bibinfo {author} {\bibfnamefont {K.}~\bibnamefont {Aki}}\ and\ \bibinfo {author} {\bibfnamefont {P.~G.}\ \bibnamefont {Richards}},\ }\href {https://books.google.com/books/about/Quantitative_Seismology.html?id=sRhawFG5_EcC} {\emph {\bibinfo {title} {Quantitative {Seismology}, 2nd {Ed}.}}}\ (\bibinfo  {publisher} {University Science Books},\ \bibinfo {year} {2002})\BibitemShut {NoStop}%
\bibitem [{\citenamefont {Landau}\ \emph {et~al.}(2012)\citenamefont {Landau}, \citenamefont {Pitaevskii}, \citenamefont {Kosevich},\ and\ \citenamefont {Lifshitz}}]{landau2012theory}%
  \BibitemOpen
  \bibfield  {author} {\bibinfo {author} {\bibfnamefont {L.}~\bibnamefont {Landau}}, \bibinfo {author} {\bibfnamefont {L.}~\bibnamefont {Pitaevskii}}, \bibinfo {author} {\bibfnamefont {A.}~\bibnamefont {Kosevich}},\ and\ \bibinfo {author} {\bibfnamefont {E.}~\bibnamefont {Lifshitz}},\ }\href {https://books.google.ch/books?id=NXRaWJb4HdkC} {\emph {\bibinfo {title} {Theory of Elasticity: Volume 7}}},\ \bibinfo {number} {Bd. 7}\ (\bibinfo  {publisher} {Butterworth-Heinemann},\ \bibinfo {year} {2012})\BibitemShut {NoStop}%
\bibitem [{\citenamefont {Hoepffner}(2007)}]{hoepffner2007implementation}%
  \BibitemOpen
  \bibfield  {author} {\bibinfo {author} {\bibfnamefont {J.}~\bibnamefont {Hoepffner}},\ }\href {http://www.ida.upmc.fr/~hoepffner/boundarycondition.pdf} {\bibinfo {title} {Implementation of boundary conditions}} (\bibinfo {year} {2007})\BibitemShut {NoStop}%
\bibitem [{\citenamefont {Banjai}\ and\ \citenamefont {Sauter}(2009)}]{banjai2009rapid}%
  \BibitemOpen
  \bibfield  {author} {\bibinfo {author} {\bibfnamefont {L.}~\bibnamefont {Banjai}}\ and\ \bibinfo {author} {\bibfnamefont {S.}~\bibnamefont {Sauter}},\ }\href {https://doi.org/10.1137/070690754} {\bibfield  {journal} {\bibinfo  {journal} {SIAM Journal on Numerical Analysis}\ }\textbf {\bibinfo {volume} {47}},\ \bibinfo {pages} {227} (\bibinfo {year} {2009})}\BibitemShut {NoStop}%
\bibitem [{\citenamefont {Qian}\ and\ \citenamefont {Chew}(2009)}]{qian2009fast}%
  \BibitemOpen
  \bibfield  {author} {\bibinfo {author} {\bibfnamefont {Z.-G.}\ \bibnamefont {Qian}}\ and\ \bibinfo {author} {\bibfnamefont {W.~C.}\ \bibnamefont {Chew}},\ }\href {https://doi.org/10.1109/TAP.2009.2023629} {\bibfield  {journal} {\bibinfo  {journal} {IEEE Transactions on Antennas and Propagation}\ }\textbf {\bibinfo {volume} {57}},\ \bibinfo {pages} {3594} (\bibinfo {year} {2009})}\BibitemShut {NoStop}%
\bibitem [{\citenamefont {Wei}\ \emph {et~al.}(2012)\citenamefont {Wei}, \citenamefont {Peng},\ and\ \citenamefont {Lee}}]{wei2012fast}%
  \BibitemOpen
  \bibfield  {author} {\bibinfo {author} {\bibfnamefont {J.-G.}\ \bibnamefont {Wei}}, \bibinfo {author} {\bibfnamefont {Z.}~\bibnamefont {Peng}},\ and\ \bibinfo {author} {\bibfnamefont {J.-F.}\ \bibnamefont {Lee}},\ }\href {https://doi.org/10.1029/2012RS004988} {\bibfield  {journal} {\bibinfo  {journal} {Radio Science}\ }\textbf {\bibinfo {volume} {47}},\ \bibinfo {pages} {1} (\bibinfo {year} {2012})}\BibitemShut {NoStop}%
\bibitem [{\citenamefont {Omar}\ and\ \citenamefont {Jiao}(2015)}]{omar2015linear}%
  \BibitemOpen
  \bibfield  {author} {\bibinfo {author} {\bibfnamefont {S.}~\bibnamefont {Omar}}\ and\ \bibinfo {author} {\bibfnamefont {D.}~\bibnamefont {Jiao}},\ }\href {https://doi.org/10.1109/TMTT.2015.2396494} {\bibfield  {journal} {\bibinfo  {journal} {IEEE Transactions on Microwave Theory and Techniques}\ }\textbf {\bibinfo {volume} {63}},\ \bibinfo {pages} {897} (\bibinfo {year} {2015})}\BibitemShut {NoStop}%
\bibitem [{\citenamefont {Smith}\ \emph {et~al.}(2019)\citenamefont {Smith}, \citenamefont {Tsynkov},\ and\ \citenamefont {Turkel}}]{smith2019compact}%
  \BibitemOpen
  \bibfield  {author} {\bibinfo {author} {\bibfnamefont {F.}~\bibnamefont {Smith}}, \bibinfo {author} {\bibfnamefont {S.}~\bibnamefont {Tsynkov}},\ and\ \bibinfo {author} {\bibfnamefont {E.}~\bibnamefont {Turkel}},\ }\href {https://doi.org/10.1007/s10915-019-00970-x} {\bibfield  {journal} {\bibinfo  {journal} {Journal of Scientific Computing}\ }\textbf {\bibinfo {volume} {81}},\ \bibinfo {pages} {1181} (\bibinfo {year} {2019})}\BibitemShut {NoStop}%
\bibitem [{\citenamefont {Harrow}\ and\ \citenamefont {Wei}(2019)}]{harrow2020adaptive}%
  \BibitemOpen
  \bibfield  {author} {\bibinfo {author} {\bibfnamefont {A.~W.}\ \bibnamefont {Harrow}}\ and\ \bibinfo {author} {\bibfnamefont {A.~Y.}\ \bibnamefont {Wei}},\ }in\ \href {https://doi.org/10.1137/1.9781611975994.12} {\emph {\bibinfo {booktitle} {Proceedings of the 2020 {ACM}-{SIAM} {Symposium} on {Discrete} {Algorithms} ({SODA})}}},\ \bibinfo {series and number} {Proceedings}\ (\bibinfo  {publisher} {Society for Industrial and Applied Mathematics},\ \bibinfo {year} {2019})\ pp.\ \bibinfo {pages} {193--212}\BibitemShut {NoStop}%
\bibitem [{\citenamefont {Rall}\ and\ \citenamefont {Fuller}(2023)}]{rall2023amplitude}%
  \BibitemOpen
  \bibfield  {author} {\bibinfo {author} {\bibfnamefont {P.}~\bibnamefont {Rall}}\ and\ \bibinfo {author} {\bibfnamefont {B.}~\bibnamefont {Fuller}},\ }\href {https://doi.org/10.22331/q-2023-03-02-937} {\bibfield  {journal} {\bibinfo  {journal} {Quantum}\ }\textbf {\bibinfo {volume} {7}},\ \bibinfo {pages} {937} (\bibinfo {year} {2023})}\BibitemShut {NoStop}%
\bibitem [{\citenamefont {Krovi}(2024)}]{krovi2024quantum}%
  \BibitemOpen
  \bibfield  {author} {\bibinfo {author} {\bibfnamefont {H.}~\bibnamefont {Krovi}}} (\bibinfo {year} {2024}),\ \bibinfo {note} {arXiv:2404.07303}\BibitemShut {NoStop}%
\bibitem [{\citenamefont {Koukoutsis}\ \emph {et~al.}(2024)\citenamefont {Koukoutsis}, \citenamefont {Hizanidis}, \citenamefont {Ram},\ and\ \citenamefont {Vahala}}]{koukoutsis2024quantum}%
  \BibitemOpen
  \bibfield  {author} {\bibinfo {author} {\bibfnamefont {E.}~\bibnamefont {Koukoutsis}}, \bibinfo {author} {\bibfnamefont {K.}~\bibnamefont {Hizanidis}}, \bibinfo {author} {\bibfnamefont {A.~K.}\ \bibnamefont {Ram}},\ and\ \bibinfo {author} {\bibfnamefont {G.}~\bibnamefont {Vahala}},\ }\href {https://doi.org/10.1016/j.future.2024.05.028} {\bibfield  {journal} {\bibinfo  {journal} {Future Generation Computer Systems}\ }\textbf {\bibinfo {volume} {159}},\ \bibinfo {pages} {221} (\bibinfo {year} {2024})}\BibitemShut {NoStop}%
\bibitem [{\citenamefont {Hu}\ \emph {et~al.}(2020)\citenamefont {Hu}, \citenamefont {Xia},\ and\ \citenamefont {Kais}}]{hu2020quantum}%
  \BibitemOpen
  \bibfield  {author} {\bibinfo {author} {\bibfnamefont {Z.}~\bibnamefont {Hu}}, \bibinfo {author} {\bibfnamefont {R.}~\bibnamefont {Xia}},\ and\ \bibinfo {author} {\bibfnamefont {S.}~\bibnamefont {Kais}},\ }\href {https://doi.org/10.1038/s41598-020-60321-x} {\bibfield  {journal} {\bibinfo  {journal} {Scientific Reports}\ }\textbf {\bibinfo {volume} {10}},\ \bibinfo {pages} {3301} (\bibinfo {year} {2020})}\BibitemShut {NoStop}%
\bibitem [{\citenamefont {Nassar}\ \emph {et~al.}(2018)\citenamefont {Nassar}, \citenamefont {Chen}, \citenamefont {Norris},\ and\ \citenamefont {Huang}}]{nassar2018quantization}%
  \BibitemOpen
  \bibfield  {author} {\bibinfo {author} {\bibfnamefont {H.}~\bibnamefont {Nassar}}, \bibinfo {author} {\bibfnamefont {H.}~\bibnamefont {Chen}}, \bibinfo {author} {\bibfnamefont {A.~N.}\ \bibnamefont {Norris}},\ and\ \bibinfo {author} {\bibfnamefont {G.~L.}\ \bibnamefont {Huang}},\ }\href {https://doi.org/10.1103/PhysRevB.97.014305} {\bibfield  {journal} {\bibinfo  {journal} {Physical Review B}\ }\textbf {\bibinfo {volume} {97}},\ \bibinfo {pages} {014305} (\bibinfo {year} {2018})}\BibitemShut {NoStop}%
\bibitem [{\citenamefont {Chen}\ \emph {et~al.}(2019)\citenamefont {Chen}, \citenamefont {Yao}, \citenamefont {Nassar},\ and\ \citenamefont {Huang}}]{chen2019mechanical}%
  \BibitemOpen
  \bibfield  {author} {\bibinfo {author} {\bibfnamefont {H.}~\bibnamefont {Chen}}, \bibinfo {author} {\bibfnamefont {L.}~\bibnamefont {Yao}}, \bibinfo {author} {\bibfnamefont {H.}~\bibnamefont {Nassar}},\ and\ \bibinfo {author} {\bibfnamefont {G.}~\bibnamefont {Huang}},\ }\href {https://doi.org/10.1103/PhysRevApplied.11.044029} {\bibfield  {journal} {\bibinfo  {journal} {Physical Review Applied}\ }\textbf {\bibinfo {volume} {11}},\ \bibinfo {pages} {044029} (\bibinfo {year} {2019})}\BibitemShut {NoStop}%
\bibitem [{\citenamefont {Galiffi}\ \emph {et~al.}(2022)\citenamefont {Galiffi}, \citenamefont {Tirole}, \citenamefont {Yin}, \citenamefont {Li}, \citenamefont {Vezzoli}, \citenamefont {Huidobro}, \citenamefont {Silveirinha}, \citenamefont {Sapienza}, \citenamefont {Alù},\ and\ \citenamefont {Pendry}}]{galiffi2022photonics}%
  \BibitemOpen
  \bibfield  {author} {\bibinfo {author} {\bibfnamefont {E.}~\bibnamefont {Galiffi}}, \bibinfo {author} {\bibfnamefont {R.}~\bibnamefont {Tirole}}, \bibinfo {author} {\bibfnamefont {S.}~\bibnamefont {Yin}}, \bibinfo {author} {\bibfnamefont {H.}~\bibnamefont {Li}}, \bibinfo {author} {\bibfnamefont {S.}~\bibnamefont {Vezzoli}}, \bibinfo {author} {\bibfnamefont {P.~A.}\ \bibnamefont {Huidobro}}, \bibinfo {author} {\bibfnamefont {M.~G.}\ \bibnamefont {Silveirinha}}, \bibinfo {author} {\bibfnamefont {R.}~\bibnamefont {Sapienza}}, \bibinfo {author} {\bibfnamefont {A.}~\bibnamefont {Alù}},\ and\ \bibinfo {author} {\bibfnamefont {J.~B.}\ \bibnamefont {Pendry}},\ }\href {https://doi.org/10.1117/1.AP.4.1.014002} {\bibfield  {journal} {\bibinfo  {journal} {Advanced Photonics}\ }\textbf {\bibinfo {volume} {4}},\ \bibinfo {pages} {014002} (\bibinfo {year} {2022})}\BibitemShut {NoStop}%
\bibitem [{\citenamefont {Bösch}\ \emph {et~al.}(2024)\citenamefont {Bösch}, \citenamefont {Fichtner},\ and\ \citenamefont {Serra-Garcia}}]{bosch2023differences}%
  \BibitemOpen
  \bibfield  {author} {\bibinfo {author} {\bibfnamefont {C.}~\bibnamefont {Bösch}}, \bibinfo {author} {\bibfnamefont {A.}~\bibnamefont {Fichtner}},\ and\ \bibinfo {author} {\bibfnamefont {M.}~\bibnamefont {Serra-Garcia}},\ }\href {https://doi.org/10.1103/PhysRevB.110.064307} {\bibfield  {journal} {\bibinfo  {journal} {Physical Review B}\ }\textbf {\bibinfo {volume} {110}},\ \bibinfo {pages} {064307} (\bibinfo {year} {2024})}\BibitemShut {NoStop}%
\bibitem [{\citenamefont {Kikuchi}\ \emph {et~al.}(2023)\citenamefont {Kikuchi}, \citenamefont {Mc~Keever}, \citenamefont {Coopmans}, \citenamefont {Lubasch},\ and\ \citenamefont {Benedetti}}]{kikuchi2023realization}%
  \BibitemOpen
  \bibfield  {author} {\bibinfo {author} {\bibfnamefont {Y.}~\bibnamefont {Kikuchi}}, \bibinfo {author} {\bibfnamefont {C.}~\bibnamefont {Mc~Keever}}, \bibinfo {author} {\bibfnamefont {L.}~\bibnamefont {Coopmans}}, \bibinfo {author} {\bibfnamefont {M.}~\bibnamefont {Lubasch}},\ and\ \bibinfo {author} {\bibfnamefont {M.}~\bibnamefont {Benedetti}},\ }\href {https://doi.org/10.1038/s41534-023-00762-0} {\bibfield  {journal} {\bibinfo  {journal} {npj Quantum Information}\ }\textbf {\bibinfo {volume} {9}},\ \bibinfo {pages} {1} (\bibinfo {year} {2023})}\BibitemShut {NoStop}%
\bibitem [{\citenamefont {Wright}\ \emph {et~al.}(2024)\citenamefont {Wright}, \citenamefont {Mc~Keever}, \citenamefont {First}, \citenamefont {Johnston}, \citenamefont {Tillay}, \citenamefont {Chaney}, \citenamefont {Rosenkranz},\ and\ \citenamefont {Lubasch}}]{wright2024noisy}%
  \BibitemOpen
  \bibfield  {author} {\bibinfo {author} {\bibfnamefont {L.}~\bibnamefont {Wright}}, \bibinfo {author} {\bibfnamefont {C.}~\bibnamefont {Mc~Keever}}, \bibinfo {author} {\bibfnamefont {J.~T.}\ \bibnamefont {First}}, \bibinfo {author} {\bibfnamefont {R.}~\bibnamefont {Johnston}}, \bibinfo {author} {\bibfnamefont {J.}~\bibnamefont {Tillay}}, \bibinfo {author} {\bibfnamefont {S.}~\bibnamefont {Chaney}}, \bibinfo {author} {\bibfnamefont {M.}~\bibnamefont {Rosenkranz}},\ and\ \bibinfo {author} {\bibfnamefont {M.}~\bibnamefont {Lubasch}},\ }\href {https://doi.org/10.1103/PhysRevResearch.6.043169} {\bibfield  {journal} {\bibinfo  {journal} {Physical Review Research}\ }\textbf {\bibinfo {volume} {6}},\ \bibinfo {pages} {043169} (\bibinfo {year} {2024})}\BibitemShut {NoStop}%
\bibitem [{\citenamefont {Acharya}(2024)}]{acharya2024quantum}%
  \BibitemOpen
  \bibfield  {author} {\bibinfo {author} {\bibfnamefont {R.}~\bibnamefont {Acharya}}} (\bibinfo {year} {2024}),\ \bibinfo {note} {arXiv:2408.13687}\BibitemShut {NoStop}%
\bibitem [{\citenamefont {Iten}\ \emph {et~al.}(2016)\citenamefont {Iten}, \citenamefont {Colbeck}, \citenamefont {Kukuljan}, \citenamefont {Home},\ and\ \citenamefont {Christandl}}]{iten2016quantum}%
  \BibitemOpen
  \bibfield  {author} {\bibinfo {author} {\bibfnamefont {R.}~\bibnamefont {Iten}}, \bibinfo {author} {\bibfnamefont {R.}~\bibnamefont {Colbeck}}, \bibinfo {author} {\bibfnamefont {I.}~\bibnamefont {Kukuljan}}, \bibinfo {author} {\bibfnamefont {J.}~\bibnamefont {Home}},\ and\ \bibinfo {author} {\bibfnamefont {M.}~\bibnamefont {Christandl}},\ }\href {https://doi.org/10.1103/PhysRevA.93.032318} {\bibfield  {journal} {\bibinfo  {journal} {Physical Review A}\ }\textbf {\bibinfo {volume} {93}},\ \bibinfo {pages} {032318} (\bibinfo {year} {2016})}\BibitemShut {NoStop}%
\bibitem [{\citenamefont {He}\ \emph {et~al.}(2017)\citenamefont {He}, \citenamefont {Luo}, \citenamefont {Zhang}, \citenamefont {Wang},\ and\ \citenamefont {Wang}}]{he2017decompositions}%
  \BibitemOpen
  \bibfield  {author} {\bibinfo {author} {\bibfnamefont {Y.}~\bibnamefont {He}}, \bibinfo {author} {\bibfnamefont {M.-X.}\ \bibnamefont {Luo}}, \bibinfo {author} {\bibfnamefont {E.}~\bibnamefont {Zhang}}, \bibinfo {author} {\bibfnamefont {H.-K.}\ \bibnamefont {Wang}},\ and\ \bibinfo {author} {\bibfnamefont {X.-F.}\ \bibnamefont {Wang}},\ }\href {https://doi.org/10.1007/s10773-017-3389-4} {\bibfield  {journal} {\bibinfo  {journal} {International Journal of Theoretical Physics}\ }\textbf {\bibinfo {volume} {56}},\ \bibinfo {pages} {2350} (\bibinfo {year} {2017})}\BibitemShut {NoStop}%
\bibitem [{\citenamefont {Zindorf}\ and\ \citenamefont {Bose}(2024)}]{zindorf2024efficient}%
  \BibitemOpen
  \bibfield  {author} {\bibinfo {author} {\bibfnamefont {B.}~\bibnamefont {Zindorf}}\ and\ \bibinfo {author} {\bibfnamefont {S.}~\bibnamefont {Bose}}} (\bibinfo {year} {2024}),\ \bibinfo {note} {arXiv:2404.02279}\BibitemShut {NoStop}%
\bibitem [{\citenamefont {Martinez}\ \emph {et~al.}(2016)\citenamefont {Martinez}, \citenamefont {Monz}, \citenamefont {Nigg}, \citenamefont {Schindler},\ and\ \citenamefont {Blatt}}]{martinez2016compiling}%
  \BibitemOpen
  \bibfield  {author} {\bibinfo {author} {\bibfnamefont {E.~A.}\ \bibnamefont {Martinez}}, \bibinfo {author} {\bibfnamefont {T.}~\bibnamefont {Monz}}, \bibinfo {author} {\bibfnamefont {D.}~\bibnamefont {Nigg}}, \bibinfo {author} {\bibfnamefont {P.}~\bibnamefont {Schindler}},\ and\ \bibinfo {author} {\bibfnamefont {R.}~\bibnamefont {Blatt}},\ }\href {https://doi.org/10.1088/1367-2630/18/6/063029} {\bibfield  {journal} {\bibinfo  {journal} {New Journal of Physics}\ }\textbf {\bibinfo {volume} {18}},\ \bibinfo {pages} {063029} (\bibinfo {year} {2016})}\BibitemShut {NoStop}%
\bibitem [{\citenamefont {Goel}\ and\ \citenamefont {Freericks}(2021)}]{goel2021native}%
  \BibitemOpen
  \bibfield  {author} {\bibinfo {author} {\bibfnamefont {N.}~\bibnamefont {Goel}}\ and\ \bibinfo {author} {\bibfnamefont {J.~K.}\ \bibnamefont {Freericks}},\ }\href {https://doi.org/10.1088/2058-9565/ac1e02} {\bibfield  {journal} {\bibinfo  {journal} {Quantum Science and Technology}\ }\textbf {\bibinfo {volume} {6}},\ \bibinfo {pages} {044010} (\bibinfo {year} {2021})}\BibitemShut {NoStop}%
\bibitem [{\citenamefont {Kim}\ \emph {et~al.}(2022)\citenamefont {Kim}, \citenamefont {Morvan}, \citenamefont {Nguyen}, \citenamefont {Naik}, \citenamefont {Jünger}, \citenamefont {Chen}, \citenamefont {Kreikebaum}, \citenamefont {Santiago},\ and\ \citenamefont {Siddiqi}}]{kim2022high}%
  \BibitemOpen
  \bibfield  {author} {\bibinfo {author} {\bibfnamefont {Y.}~\bibnamefont {Kim}}, \bibinfo {author} {\bibfnamefont {A.}~\bibnamefont {Morvan}}, \bibinfo {author} {\bibfnamefont {L.~B.}\ \bibnamefont {Nguyen}}, \bibinfo {author} {\bibfnamefont {R.~K.}\ \bibnamefont {Naik}}, \bibinfo {author} {\bibfnamefont {C.}~\bibnamefont {Jünger}}, \bibinfo {author} {\bibfnamefont {L.}~\bibnamefont {Chen}}, \bibinfo {author} {\bibfnamefont {J.~M.}\ \bibnamefont {Kreikebaum}}, \bibinfo {author} {\bibfnamefont {D.~I.}\ \bibnamefont {Santiago}},\ and\ \bibinfo {author} {\bibfnamefont {I.}~\bibnamefont {Siddiqi}},\ }\href {https://doi.org/10.1038/s41567-022-01590-3} {\bibfield  {journal} {\bibinfo  {journal} {Nature Physics}\ }\textbf {\bibinfo {volume} {18}},\ \bibinfo {pages} {783} (\bibinfo {year} {2022})}\BibitemShut {NoStop}%
\bibitem [{\citenamefont {Knill}\ \emph {et~al.}(2007)\citenamefont {Knill}, \citenamefont {Ortiz},\ and\ \citenamefont {Somma}}]{knill2007optimal}%
  \BibitemOpen
  \bibfield  {author} {\bibinfo {author} {\bibfnamefont {E.}~\bibnamefont {Knill}}, \bibinfo {author} {\bibfnamefont {G.}~\bibnamefont {Ortiz}},\ and\ \bibinfo {author} {\bibfnamefont {R.~D.}\ \bibnamefont {Somma}},\ }\href {https://doi.org/10.1103/PhysRevA.75.012328} {\bibfield  {journal} {\bibinfo  {journal} {Physical Review A}\ }\textbf {\bibinfo {volume} {75}},\ \bibinfo {pages} {012328} (\bibinfo {year} {2007})}\BibitemShut {NoStop}%
\bibitem [{\citenamefont {Gleinig}\ and\ \citenamefont {Hoefler}(2021)}]{gleinig2021efficient}%
  \BibitemOpen
  \bibfield  {author} {\bibinfo {author} {\bibfnamefont {N.}~\bibnamefont {Gleinig}}\ and\ \bibinfo {author} {\bibfnamefont {T.}~\bibnamefont {Hoefler}},\ }in\ \href {https://doi.org/10.1109/DAC18074.2021.9586240} {\emph {\bibinfo {booktitle} {2021 58th {ACM}/{IEEE} {Design} {Automation} {Conference} ({DAC})}}}\ (\bibinfo {year} {2021})\ pp.\ \bibinfo {pages} {433--438}\BibitemShut {NoStop}%
\bibitem [{\citenamefont {Malvetti}\ \emph {et~al.}(2021)\citenamefont {Malvetti}, \citenamefont {Iten},\ and\ \citenamefont {Colbeck}}]{malvetti2021quantum}%
  \BibitemOpen
  \bibfield  {author} {\bibinfo {author} {\bibfnamefont {E.}~\bibnamefont {Malvetti}}, \bibinfo {author} {\bibfnamefont {R.}~\bibnamefont {Iten}},\ and\ \bibinfo {author} {\bibfnamefont {R.}~\bibnamefont {Colbeck}},\ }\href {https://doi.org/10.22331/q-2021-03-15-412} {\bibfield  {journal} {\bibinfo  {journal} {Quantum}\ }\textbf {\bibinfo {volume} {5}},\ \bibinfo {pages} {412} (\bibinfo {year} {2021})}\BibitemShut {NoStop}%
\bibitem [{\citenamefont {de~Veras}\ \emph {et~al.}(2022)\citenamefont {de~Veras}, \citenamefont {da~Silva},\ and\ \citenamefont {da~Silva}}]{de2022double}%
  \BibitemOpen
  \bibfield  {author} {\bibinfo {author} {\bibfnamefont {T.~M.~L.}\ \bibnamefont {de~Veras}}, \bibinfo {author} {\bibfnamefont {L.~D.}\ \bibnamefont {da~Silva}},\ and\ \bibinfo {author} {\bibfnamefont {A.~J.}\ \bibnamefont {da~Silva}},\ }\href {https://doi.org/10.1007/s11128-022-03549-y} {\bibfield  {journal} {\bibinfo  {journal} {Quantum Information Processing}\ }\textbf {\bibinfo {volume} {21}},\ \bibinfo {pages} {204} (\bibinfo {year} {2022})}\BibitemShut {NoStop}%
\bibitem [{\citenamefont {Zhang}\ \emph {et~al.}(2022)\citenamefont {Zhang}, \citenamefont {Li},\ and\ \citenamefont {Yuan}}]{zhang2022quantum}%
  \BibitemOpen
  \bibfield  {author} {\bibinfo {author} {\bibfnamefont {X.-M.}\ \bibnamefont {Zhang}}, \bibinfo {author} {\bibfnamefont {T.}~\bibnamefont {Li}},\ and\ \bibinfo {author} {\bibfnamefont {X.}~\bibnamefont {Yuan}},\ }\href {https://doi.org/10.1103/PhysRevLett.129.230504} {\bibfield  {journal} {\bibinfo  {journal} {Physical Review Letters}\ }\textbf {\bibinfo {volume} {129}},\ \bibinfo {pages} {230504} (\bibinfo {year} {2022})}\BibitemShut {NoStop}%
\bibitem [{\citenamefont {Vale}\ \emph {et~al.}(2024)\citenamefont {Vale}, \citenamefont {Azevedo}, \citenamefont {Araújo}, \citenamefont {Araujo},\ and\ \citenamefont {da~Silva}}]{vale2023circuit}%
  \BibitemOpen
  \bibfield  {author} {\bibinfo {author} {\bibfnamefont {R.}~\bibnamefont {Vale}}, \bibinfo {author} {\bibfnamefont {T.~M.~D.}\ \bibnamefont {Azevedo}}, \bibinfo {author} {\bibfnamefont {I.~C.~S.}\ \bibnamefont {Araújo}}, \bibinfo {author} {\bibfnamefont {I.~F.}\ \bibnamefont {Araujo}},\ and\ \bibinfo {author} {\bibfnamefont {A.~J.}\ \bibnamefont {da~Silva}},\ }\href {https://doi.org/10.1109/TCAD.2023.3327102} {\bibfield  {journal} {\bibinfo  {journal} {IEEE Transactions on Computer-Aided Design of Integrated Circuits and Systems}\ }\textbf {\bibinfo {volume} {43}},\ \bibinfo {pages} {802} (\bibinfo {year} {2024})}\BibitemShut {NoStop}%
\bibitem [{\citenamefont {Virieux}(1984)}]{virieux1984}%
  \BibitemOpen
  \bibfield  {author} {\bibinfo {author} {\bibfnamefont {J.}~\bibnamefont {Virieux}},\ }\href {https://doi.org/10.1190/1.1441605} {\bibfield  {journal} {\bibinfo  {journal} {Geophysics}\ }\textbf {\bibinfo {volume} {49}},\ \bibinfo {pages} {1933} (\bibinfo {year} {1984})}\BibitemShut {NoStop}%
\bibitem [{\citenamefont {Higdon}(1992)}]{higdon1992absorbing}%
  \BibitemOpen
  \bibfield  {author} {\bibinfo {author} {\bibfnamefont {R.~L.}\ \bibnamefont {Higdon}},\ }\href {https://doi.org/10.1016/0021-9991(92)90016-R} {\bibfield  {journal} {\bibinfo  {journal} {Journal of Computational Physics}\ }\textbf {\bibinfo {volume} {101}},\ \bibinfo {pages} {386} (\bibinfo {year} {1992})}\BibitemShut {NoStop}%
\bibitem [{\citenamefont {Berenger}(1994)}]{berenger1996perfectly}%
  \BibitemOpen
  \bibfield  {author} {\bibinfo {author} {\bibfnamefont {J.-P.}\ \bibnamefont {Berenger}},\ }\href {https://doi.org/10.1006/jcph.1994.1159} {\bibfield  {journal} {\bibinfo  {journal} {Journal of Computational Physics}\ }\textbf {\bibinfo {volume} {114}},\ \bibinfo {pages} {185} (\bibinfo {year} {1994})}\BibitemShut {NoStop}%
\bibitem [{\citenamefont {Clapp}(2009)}]{clapp2009reverse}%
  \BibitemOpen
  \bibfield  {author} {\bibinfo {author} {\bibfnamefont {R.~G.}\ \bibnamefont {Clapp}},\ }\bibinfo {title} {Reverse time migration with random boundaries},\ in\ \href {https://doi.org/10.1190/1.3255432} {\emph {\bibinfo {booktitle} {SEG Technical Program Expanded Abstracts 2009}}}\ (\bibinfo  {publisher} {Society of Exploration Geophysicists},\ \bibinfo {year} {2009})\ pp.\ \bibinfo {pages} {2809--2813}\BibitemShut {NoStop}%
\bibitem [{\citenamefont {Shen}\ and\ \citenamefont {Clapp}(2015)}]{shen2015random}%
  \BibitemOpen
  \bibfield  {author} {\bibinfo {author} {\bibfnamefont {X.}~\bibnamefont {Shen}}\ and\ \bibinfo {author} {\bibfnamefont {R.~G.}\ \bibnamefont {Clapp}},\ }\href {https://doi.org/10.1190/geo2014-0542.1} {\bibfield  {journal} {\bibinfo  {journal} {Geophysics}\ }\textbf {\bibinfo {volume} {80}},\ \bibinfo {pages} {R351} (\bibinfo {year} {2015})}\BibitemShut {NoStop}%
\end{thebibliography}%
\end{document}